\documentclass[11pt,a4paper,twoside]{book}
\usepackage{latexsym, amsmath}
\usepackage{framed}
\usepackage{amssymb}
\usepackage{amscd}
\usepackage{graphicx}
\usepackage{wrapfig}                                                                 
\usepackage{fancyhdr}
\usepackage{color}
\usepackage{pst-plot, pstricks}
\definecolor{darkblue}{rgb}{0,0,0.5}
\definecolor{darkgray}{rgb}{0.2,0.2,0.2}

\newcommand{\cn}{{\hr_n}}
\newcommand{\cnq}{{\hr_n^\mathbb{Q}}}
\newcommand{\hr}{{\cal H}}

\newcommand{\Fock}{{\{0,1\}^*}}
\newcommand{\C}{{\mathbb C}}
\newcommand{\R}{{\mathbb R}}
\newcommand{\N}{{\mathbb N}}
\newcommand{\idn}{\mathbf{1}}
\newcommand{\Z}{{\mathbb Z}}

\newcommand{\eps}{{\varepsilon}}

\newcommand{\s}{{\{0,1\}^*}}

\newcommand{\qka}{{\overline{QK}}}
\newcommand{\proof}{{\bf Proof.} }
\newtheorem{theorem}{Theorem}[section]
\newtheorem{lemma}[theorem]{Lemma}
\newtheorem{corollary}[theorem]{Corollary}
\newtheorem{definition}[theorem]{Definition}
\newtheorem{proposition}[theorem]{Proposition}
\newtheorem{conjecture}[theorem]{Conjecture}
\newtheorem{example}[theorem]{Example}
\newcommand{\nix}{{\rule{0pt}{2pt}}}
\newcommand{\qedd}{{\nix\nolinebreak\hfill\hfill\nolinebreak$\Box$}}
\newcommand{\qed}{{\qedd\par\medskip\noindent}}
\newcommand{\lineclear}{{\rule{0pt}{0pt}\nopagebreak\par\nopagebreak\noindent}}

\definecolor{shadecolor}{rgb}{.95,.95,.95}

\newenvironment{fshaded}{%
\MakeFramed {\FrameRestore}}%
{\endMakeFramed}

\newenvironment{flemma}[1][]{\definecolor{shadecolor}{rgb}{0.95,0.95,0.95}%
\definecolor{framecolor}{rgb}{0,0,0}%
\begin{fshaded}\begin{lemma}[#1]}{\end{lemma}\end{fshaded}}
\newenvironment{ftheorem}[1][]{\definecolor{shadecolor}{rgb}{.95,.95,0.95}%
\definecolor{framecolor}{rgb}{0,0,0}%
\begin{fshaded}\begin{theorem}[#1]}{\end{theorem}\end{fshaded}}
\newenvironment{fcorollary}[1][]{\definecolor{shadecolor}{rgb}{.95,.95,0.95}%
\definecolor{framecolor}{rgb}{0,0,0}%
\begin{fshaded}\begin{corollary}[#1]}{\end{corollary}\end{fshaded}}
\newenvironment{fproposition}[1][]{\definecolor{shadecolor}{rgb}{.95,.95,0.95}%
\definecolor{framecolor}{rgb}{0,0,0}%
\begin{fshaded}\begin{proposition}[#1]}{\end{proposition}\end{fshaded}}

\newenvironment{emptyflemma}[1][]{\definecolor{shadecolor}{rgb}{0.95,0.95,0.95}%
\definecolor{framecolor}{rgb}{0,0,0}%
\begin{fshaded}\begin{lemma}}{\end{lemma}\end{fshaded}}
\newenvironment{emptyftheorem}[1][]{\definecolor{shadecolor}{rgb}{.95,.95,0.95}%
\definecolor{framecolor}{rgb}{0,0,0}%
\begin{fshaded}\begin{theorem}}{\end{theorem}\end{fshaded}}

\makeindex
\makeglossary
\newcommand{\helv}{
  \rmfamily\scshape}
\begin{document}
\frontmatter

\thispagestyle{empty}

\begin{center}
  \rule{0mm}{10mm}

  {\center{\LARGE \bf Quantum Kolmogorov Complexity  \\[3mm] and the Quantum Turing Machine } \vspace{1mm} \\[18mm]}

\end{center}

\begin{center}
  vorgelegt von \\[2mm]
  Diplom-Physiker \\[2mm]
  Markus~M\"uller \\[2mm]
  Berlin \\[1cm]        
  von der Fakult\"at II -- Mathematik- und Naturwissenschaften \\[2mm]
  der Technischen Universit\"at Berlin\\[2mm]
  zur Erlangung des akademischen Grades \\[2mm]
  Doktor der Naturwissenschaften\\[2mm]
  Dr.~rer.~nat.\\[1cm]
  genehmigte Dissertation\\[1cm] 
\end{center}
\vspace{-5mm}

\begin{flushleft}
Promotionsausschuss:\\[6mm]
  \begin{tabular}{ll}
    Vorsitzender: & Prof.~Dr.~R.~H.~M\"ohring\\[2mm]
    Berichter: & Prof.~Dr.~R.~Seiler\\[2mm]
    Berichter: & Prof.~Dr.~A.~Knauf\\[2mm]
    Zus\"atzlicher Gutachter: & Prof.~Dr.~F.~Benatti
  \end{tabular}\\[6mm]
 
Tag der wissenschaftlichen Aussprache:\enspace\enspace 31. 8. 2007\\[10mm]
\end{flushleft}

\begin{center}
  Berlin, 2007
\end{center}

\begin{center}
  {\bf D 83}
\end{center}

\pagebreak
\section*{Acknowledgements}
I would like to express my deep and sincere gratitude to my supervisor, Prof. Ruedi Seiler, for his help
and confidence during the time of this work. His enthusiasm about mathematical physics and his friendly 
way of supporting me was always an important motivation for me.

I am deeply grateful to my former colleagues, co-authors, and friends Arleta Szko\l a, Rainer Siegmund-Schultze, Tyll Kr\"uger
and Fabio Benatti. Without their deep knowledge, support and many interesting discussions, this thesis would
not have been possible.
I was greatly enjoying the time that we were doing research together, and I hope that many further joint projects
will follow.

Sincere thanks go to Rainer W\"ust. During the time that we were teaching mathematics to undergraduates together,
I was benefiting greatly from his help in many respects. His support, the friendly working atmosphere and his confidence
made teaching for me a great experience. Without that, it would never have been possible to travel to conferences
and finish this thesis in such a short amount of time.

There are many friends and colleagues that supported me in many important ways. I would like to say thank you to
Igor Bjelakovi\'c, Sabine Jansen, Juliane Rama, Till Micheler, Christopher Witte, Franz-Josef Schmitt, Dierk Schleicher,
David Gross, Stephan Fischer, and Caroline Rogers, as well as the audiences of the DFG workshops in Leipzig, Berlin,
Erlangen, Darmstadt, and Braunschweig, in particular to Nihat Ay, Andreas Knauf, and Burkhard K\"ummerer.
Especially, I want to thank Andreas Knauf for being a reviewer of my thesis, and for giving me support and motivation
in many ways since the time I was starting to study physics in Erlangen.

Finally, I would like to express my deep love and gratitude to my family, especially to my parents, my grand-parents and my little sister
Janina. Without you, this thesis and nothing else would ever have been possible.

\addtocontents{toc}{\vskip -1.2cm}

\chapter{Abstract}
The purpose of this thesis is to give a formal definition of quantum Kolmogorov complexity and
rigorous mathematical proofs of its basic properties.

Classical Kolmogorov complexity is a well-known and useful measure of randomness for binary strings.
In recent years, several different quantum generalizations of Kolmogorov complexity have been proposed.
The most natural generalization is due to Berthiaume et al. \cite{Berthiaume},
defining the complexity of a quantum bit (qubit) string as the length of the shortest quantum input
for a universal quantum computer that outputs the desired string. Except for slight modifications,
it is this definition of quantum Kolmogorov complexity that we study in this thesis.

We start by analyzing certain aspects of the underlying quantum Turing machine (QTM) model
in a more detailed formal rigour than was done previously.
Afterwards, we apply these results to quantum Kolmogorov complexity.

Our first result, based on work by Bernstein and Vazirani \cite{BernsteinVazirani},
is a proof of the existence of a universal QTM which simulates every other QTM for
an arbitrary number of time steps and than halts with probability one.
In addition, we show that every input that makes a QTM {\em almost} halt can be modified to make the universal QTM
halt entirely, by adding at most a constant number of qubits.

It follows that quantum Kolmogorov complexity has the invariance property, i.e.
it depends on the choice of the universal QTM only up to an additive constant. Moreover,
the quantum complexity of classical strings agrees with classical complexity, again up to an additive constant.
The proofs are based on several analytic estimates.

Furthermore, we prove several incompressibility theorems for quantum Kolmogorov complexity. Finally,
we show that for ergodic quantum information sources, complexity rate and entropy rate
coincide with probability one.

The thesis is finished with an outlook on a possible application of quantum Kolmogorov complexity
in statistical mechanics.

\chapter{Zusammenfassung}
\vskip -0.5cm
Ziel dieser Arbeit ist es, den Begriff der Quanten-Kolmogorov-Komplexit\"at formal zu definieren
und seine wichtigsten Eigenschaften rigoros zu beweisen.

Die klassische Kolmogorov-Komplexit\"at ist ein bekanntes und n\"utzliches Ma\ss\enspace f\"ur
die Zuf\"alligkeit endlicher W\"orter. In den letzten Jahren wurden unterschiedliche Quantenverallgemeinerungen
der Kolmogorov-Komplexit\"at vorgeschlagen. Die nat\"urlichste Art der Verallgemeinerung stammt von
Berthiaume u.a. \cite{Berthiaume}, die die Komplexit\"at eines Quantenwortes definieren als die L\"ange
der k\"urzesten Quanteneingabe f\"ur einen universellen Quantencomputer, die als Ausgabe das entsprechende
Wort produziert. Abgesehen von kleinen \"Anderungen soll dieser Komplexit\"atsbegriff in der hier
vorliegenden Arbeit untersucht werden.

Zun\"achst untersuchen wir verschiedene Aspekte des zugrunde liegenden Modells der Quantenturingmaschine (QTM),
und zwar mit gr\"o\ss erer formaler Genauigkeit als in bisherigen Arbeiten. Anschlie\ss end wenden wir
diese Resultate auf die Quanten-Kolmogorov-Komplexit\"at an.

Unser erstes Ergebnis, basierend auf der Arbeit von Bernstein und Vazirani~\cite{BernsteinVazirani},
ist ein Beweis f\"ur die Existenz einer universellen QTM, die jede andere QTM f\"ur eine beliebige
Anzahl von Zeitschritten simulieren kann, und dann selbst mit Wahrscheinlichkeit eins h\"alt.
Weiterhin zeigen wir, dass jede Eingabe, die eine QTM {\em beinahe} halten l\"asst, modifiziert werden
kann, um eine Eingabe zu erhalten, die die universelle QTM vollst\"andig halten l\"asst, wobei
sich die Eingabel\"ange h\"ochstens um eine konstante Anzahl von Qubits vergr\"o\ss ert.

Daraus folgt, dass die Quanten-Kolmogorov-Komplexit\"at die Invarianzeigenschaft besitzt, d.h.
sie ist bis auf eine additive Konstante unabh\"angig von der Wahl der universellen QTM.
Au\ss erdem stimmt die Quantenkomplexit\"at klassischer W\"orter mit deren klassischer
Komplexit\"at \"uberein, wieder bis auf eine additive Konstante. Die entsprechenden
Beweise beruhen auf verschiedenen analytischen Absch\"atzungen.

Weiterhin beweisen wir mehrere S\"atze, die zeigen, dass nur wenige Quantenw\"orter kleine
Quantenkomplexit\"at besitzen k\"onnen. Schlie\ss lich zeigen wir, dass bei ergodischen
Quantendatenquellen Komplexit\"atsrate und Entropierate mit Wahrscheinlichkeit eins \"ubereinstimmen.

Den Abschluss der Arbeit bildet ein Ausblick auf eine m\"ogliche Anwendung der Quanten-Kolmogorov-Komplexit\"at
in der statistischen Mechanik.

\tableofcontents 
\mainmatter
\chapter{Introduction}
Kolmogorov complexity is an important measure of the information content of single binary strings.
It is motivated by the fact that regular objects tend to have short descriptions. Consider for example
two binary strings $s$ and $t$, both consisting of a million bits, namely
\[
   s=101010101010101010101010\ldots,\quad t=1101011101000000010110101\ldots
\]
The string $s$ is purely repetitive, while the string $t$ looks quite irregular;
in fact, it has been recorded during a physics experiment with some radioactive source.

So why does $t$ look more irregular than $s$?
We can easily describe $s$ by saying that $s$ consists of $500.000$ repetitions of $10$, while we need
a lot more words and effort to specify the exact value of $t$. Thus, it makes sense to measure the irregularity
or randomness of a binary string as the length of its shortest description. To avoid problems, we have to beware of
self-contradictory descriptions like the following:
\begin{center}
{\em ``Let $n$ be the smallest integer that cannot be described in less than a hundred words.''}
\end{center}
This statement is the well-known {\em Berry Paradox}, cf. \cite{Vitanyibook}. So we should only accept
descriptions that are explicit enough to give instructions for constructing the corresponding string unambiguously
and purely mechanically. This requirement is definitely fulfilled by computer programs that make a predefined computer halt
and output some string in a finite amount of time.
So we choose some universal computer
$U$ and measure the irregularity, or {\em Kolmogorov complexity $C$}, of some binary string $s$ as the length $\ell$ of
the shortest program that makes the universal computer output $s$:
\[
   C(s):=\min\{\ell(p)\,\,|\,\,U(p)=s\}.
\]
For regular strings like $s$ (even if they have some large length $n$), we can find short computer programs like {\em ``print
$n$ times the string $10$''}, while for strings like $t$, there seems to be no obvious way to compress the binary
digits into a short computer program (although there might be one which we do not know). To encode some integer $n$,
we need about $\log n$ bits, where $\log=\log_2$ here and in the remainder of the thesis denotes the binary logarithm. Thus,
\[
   C(\underbrace{1111\ldots 1}_{\mbox{\em n}}) \leq \log n+\mathcal{O}(1),\,\,\mbox{ while }\,\,C(\underbrace{10110100\ldots 1}_{\mbox{{\em n} random bits}})
   \approx n.
\]

The mathematical theory of Kolmogorov complexity and some related notions like algorithmic probability
is called algorithmic information theory. It has been developed since the 1960's by Kolmogorov \cite{Kolmogorov65}, Solomonoff
\cite{Solomonoff}, Chaitin \cite{Chaitin}, and others, and is still a lively field of research.

In recent years, there has been extensive study on how the extraordinary world of quantum mechanics changes the way that information
can be transmitted, stored and processed in our universe. In this field of research, called quantum information theory, many aspects
of classical information theory have already been extended and generalized to the quantum situation.
It is thus natural to ask whether also some quantum counterpart of Kolmogorov complexity can be found.
It is tempting to try so for several reasons:
\begin{itemize}
\item Kolmogorov complexity has applications in many areas, including classical computer science, information theory and
statistical mechanics. Thus, one may hope that its quantum counterpart is similarly useful in areas like
quantum information theory or quantum statistical mechanics.
\item Quantum Kolmogorov complexity promises to unite two different kinds of randomness in a single theory:
quantum randomness, originating from measurements in quantum theory, and algorithmic randomness, corresponding
to incompressibility.
\item Every quantum system in our universe that behaves according to some computable time evolution
is a quantum computer, in the sense that it can in principle be simulated by a quantum Turing machine.
By definition, the corresponding computation cannot change the complexity of
the system's state too much. In this case, quantum Kolmogorov complexity might turn out to be a useful invariant.
\end{itemize}

In the next section, we briefly describe previous work on quantum Kolmogorov complexity,
while in Section~\ref{SecSyllabus}, we describe what is done in this thesis, why it is done, and in what way.

\section{Previous Work on Quantum Kolmogorov Complexity}
\label{SecQC}
While classical
information theory deals with finite binary strings\footnote{Note that $\lambda$ denotes the empty string of length zero.}
\[
   \s=\{\lambda,0,1,00,01,10,11,000,001,\ldots\},
\]
quantum information theory allows arbitrary superpositions of classical strings like
\[
   |\psi\rangle=\frac 1 {\sqrt{2}}\left(|001\rangle+|110\rangle\right).
\]
The idea of quantum Kolmogorov complexity is to assign some complexity measure $C(|\psi\rangle)$ to every such
quantum state $|\psi\rangle$, namely the length of the shortest program for a universal quantum computer
to produce the state $|\psi\rangle$.

Yet, in contrast to the classical situation, it is not clear at the outset what the details of such a definition
should look like.
What, for example, is exactly meant by ``universal quantum computer''? Then, what is a proper ``program'' or ``input''
for a quantum computer - is it a classical bit string, or some quantum state itself? In the second case, what is
the ``length'' of such a quantum state? Moreover, do we demand that the quantum computer produces the state $|\psi\rangle$
exactly, or do we allow some error tolerance in the continuum of quantum states?

In recent years, there have been several attempts to define and study quantum Kolmogorov complexity.
Most of them seem to be inequivalent, reflecting the different possibilities mentioned above. In the remainder of this
section, we will briefly discuss some of them. The definition which is used in this thesis can be found in
Section~\ref{SecDefQComplexity}.

The first definition of quantum Kolmogorov complexity is due to Svozil~\cite{SvozilQC}. He defines the algorithmic complexity $H$
of a vector $s\in\mathfrak{H}$ in some Hilbert space $\mathfrak H$ as the length of the shortest {\em classical} program $p$
for a universal quantum
computer $C$ to output that element,
\[
   H(s):=\min_{C(p)=s} \ell(p).
\]
Since there are countably many classical binary strings, but uncountably many quantum states, this definition has
the disadvantage that it is undefined (or infinite) for many states $s\in\mathfrak{H}$.

Later, a similar definition was given by Vit\'anyi \cite{Vitanyi}. He also allows only classical inputs,
but circumvents the aforementioned problem by allowing some error and introducing some penalty term for non-perfect output.
His definition reads
\[
   K(|x\rangle)=\min\{\ell(p)+\lceil -\log \|\langle z|x\rangle\|^2\rceil:Q(p)=|z\rangle\},
\]
where $Q$ is some universal quantum Turing machine. In this case, the output $Q(p)$ of the machine $Q$ on input $p$
does not have to be exactly equal to $|x\rangle$, but can differ by a small amount. Nevertheless, if $Q(p)$ and
the desired state $|x\rangle$ differ too much, then the penalty term $\lceil -\log \|\langle z|x\rangle\|^2\rceil$
gets large, and the minimum is attained at another argument, not at $p$.

Mora and Briegel \cite{Briegel1, Briegel2} define the quantum Kolmogorov complexity of some quantum state
as the length of the shortest classical description of some quantum circuit that prepares that state.
Maybe this approach is related to the ones mentioned before. In any case, it seems to have the advantage
to be more utilizable for applications than other definitions of complexity.

The first purely quantum definition has been given by Berthiaume, van Dam, and Laplante \cite{Berthiaume}.
They explicitly allow inputs that are themselves quantum, i.e. superpositions of classical strings of some
common length. They define
\[
   QC^\alpha(|\psi\rangle)=\min\{\ell(|\varphi\rangle)\,\,|\,\,\langle\psi|U(|\varphi\rangle)|\psi\rangle\geq\alpha\},
\]
that is, the complexity of $|\psi\rangle$ is the length of the shortest quantum input $|\varphi\rangle$ that produces
$|\psi\rangle$ with some fidelity which is larger than $\alpha$. Thus, for $\alpha=1$, $U(|\varphi\rangle)$ must be
equal to $|\psi\rangle$, while for $\alpha<1$, some inaccuracy in the output of the universal quantum computer $U$ is allowed.
Moreover, they define a similar notion of complexity by means of an approximation scheme, which will be described later
on in Section~\ref{SecDefQComplexity}.

We argue that this kind of definition is in some sense the most natural quantum extension of Kolmogorov complexity,
since inputs and outputs are treated symmetrically. In a quantum world, quantum computers can have quantum inputs.
Our definition in Section~\ref{SecDefQComplexity} is thus very similar; we basically use the definition by Berthiaume et al.,
except for slight modifications (e.g. we also allow superpositions of strings of different lengths).

We give some evidence why this kind of definition is natural in Section~\ref{SecQBrudno}, where we prove an intimate connection between
von Neumann entropy and this kind of complexity, which seems to be impossible for all definitions of
quantum complexity that are restricted to classical inputs.

A quite different idea of how to define quantum Kolmogorov complexity has been elaborated by G\'acs \cite{Gacs}.
His approach is motivated by Levin's coding theorem from classical Kolmogorov complexity. Levin's coding theorem is about
so-called {\em semimeasures}, i.e. ``probability distributions'' $p$ on the strings such that the sum $\sum_{x\in\s}p(x)$
may be less than one. A semimeasure is called {\em semicomputable} if there is a monotonically increasing, computable
sequence of functions converging to it. There is a theorem stating that there exists a so-called {\em universal}
semicomputable semimeasure $\mu$, where universal means that $\mu(x)\geq \nu(x)\cdot c_\nu$ for every other
semicomputable semimeasure $\nu$, and $c_\nu$ is a constant not depending on $x$.

Levin's coding theorem says that the Kolmogorov complexity of some string $x$ equals $-\log \mu(x)$ within some
additive constant. Contrariwise, this means that Kolmogorov complexity can also be defined as the negative
logarithm of some universal semicomputable semimeasure without reference to program lengths.

Consequently, G\'acs showed the existence of a universal semicomputable semi-density matrix, and then defined its logarithm
as the quantum Kolmogorov complexity of some quantum state. It is not clear how this approach is related to the other
definitions, although he shows some interesting estimations among the different definitions in his paper.
Moreover, the fact that his definition works without reference to any model of a quantum computer is a striking
feature, but may also make it different to relate his notion to concrete program lengths in quantum computation.
A similar and more general idea has been elaborated by Tadaki \cite{Tadaki}, but for different purpose.

It is an open problem whether all these definitions are unrelated or some of them are equivalent.
The aim of this thesis is not
to solve this problem, but rather to give a rigorous analysis of the definition given by Berthiaume et al. \cite{Berthiaume},
although some of the results on this thesis might in the end contribute to the classification of the different
complexity notions.

\section{Synopsis and Main Results}
\label{SecSyllabus}
In this section, we describe how this thesis is organized. This thesis consists of two parts.
The first part is about quantum Turing machines, the second part is about quantum Kolmogorov complexity.

As the purpose of this thesis is to develop
the basics of quantum Kolmogorov complexity in full mathematical rigour, it is necessary to study in detail
the underlying model of quantum computation, which is the quantum Turing machine \label{AbbrQTM}(QTM).
There is nothing special about the QTM model; other models
of quantum computation like the circuit model (cf. \cite{NielsenChuang}) or
measurement-based quantum computers \cite{Perdrix} are equivalent in their computational power
(see, for example, \cite{Nishimura}). We chose this model as there is a large volume of existing literature
discussing various aspects of QTMs. Also, the model seems interesting in itself, as it is
a direct quantization of the popular model of classical computation, the Turing machine (TM).

It will be shown in Chapter~\ref{ChapterQuantumKolmogorovComplexity} that
many important properties of quantum Kolmogorov complexity, like the invariance property,
are sensitive to the details of quantum computation itself. Most of the previous
work studied QTMs with the purpose to analyze {\em computational complexity}, i.e. to answer
questions like how efficient (fast) quantum algorithms can be, and how efficiently different quantum
computers can simulate each other.
As quantum Kolmogorov complexity is insensitive to execution times of algorithms, but instead
studies the program lengths, different aspects of quantum computation become important.
In more detail, in Chapter~\ref{ChapterQTM}, we proceed in the following way:
\begin{itemize}
\item In Section~\ref{SecDefQTM}, we start by defining the notion of a {\em qubit string} and give
two different ways to quantify its length. Then, we give a mathematical framework for QTMs, based on
the work by Bernstein and Vazirani \cite{BernsteinVazirani}; we define a QTM as a special kind of partial map
on the qubit strings.
\item In Section~\ref{SecHaltingUniv}, we discuss the problem of defining when a QTM halts.
We argue that the most natural and useful definition of halting, at least in the context
of quantum Kolmogorov complexity, is to demand {\em perfect halting} and to dismiss
any input which brings the QTM into some superposition or mixture of halting and non-halting.

Moreover, we discuss the notion of {\em universality} of a QTM. We show that the previous
definition of a universal QTM by Bernstein and Vazirani is perfectly suitable for the study
of computational complexity, but is not sufficient for studying quantum Kolmogorov complexity.
This is due to the restriction that in the previous approach, the halting time has to
be specified in advance.

\item Consequently, in Section~\ref{SecConstruction}, we give a full proof that there exists
a universal QTM which simulates every other QTM without knowing the halting time in advance,
and then halts perfectly. This result is necessary to show in Chapter~\ref{ChapterQuantumKolmogorovComplexity}
that quantum Kolmogorov complexity depends on the choice of the universal QTM only up to an
additive constant.

The construction of this ``strongly universal'' QTM is based on the observation that the
valid inputs are organized in mutually orthogonal {\em halting spaces}. Moreover, these
halting spaces can be computably approximated. We define these {\em approximate halting spaces}
and show several properties, based on analytic estimates.

Some slightly different universality results are needed for the different notions of quantum
Kolmogorov complexity (e.g. with or without a second parameter) that we study in Chapter~\ref{ChapterQuantumKolmogorovComplexity}.
Thus, we also describe how the proof can be modified to obtain the various different universality results.

\item In Section~\ref{SecHaltingStability}, we show a stability result for the halting scheme of QTMs:
every input which makes a QTM {\em almost} halt can be modified to make the QTM halt {\em perfectly},
by adding at most a constant number of qubits.
This shows that the halting scheme defined before in Section~\ref{SecDefQTM} is not ``unphysical'',
since it has some inherent error tolerance that was not expected from the beginning. It also means
that we can to some extent use quantum programs
with probabilistic behaviour for estimates of quantum Kolmogorov complexity.
\end{itemize}

In Chapter~\ref{ChapterQuantumKolmogorovComplexity}, we then turn to the study of quantum Kolmogorov complexity.
\begin{itemize}
\item In Section~\ref{SecDefQComplexity}, we give four different definitions of quantum Kolmogorov complexity
($QC$, $QC^\delta$, $\qka$ and $\qka^\delta)$. They differ on the one hand by the way we quantify the length
of qubit strings (base length $\ell$ or average length $\bar\ell$), and on the other hand by the way we allow some error in
the QTM's output. Yet, they are similar enough to be studied all at the same time. Most of the time, we
will nevertheless restrict our analysis to the complexities $QC$ and $QC^\delta$, since they are in some sense easier to handle
than $\qka$ and $\qka^\delta$.
\item In Section~\ref{SecQCountingArgument}, we prove some ``quantum counting argument'', which allows to
derive an upper bound on the number of mutually orthogonal vectors that are reproduced by quantum operations within
some fixed error tolerance. Furthermore, we prove two incompressibility theorems for quantum Kolmogorov complexity.
\item We show that quantum Kolmogorov complexity is {\em invariant} in Section~\ref{SecInvariance}. This means
that it depends on the choice of the universal QTM only up to an additive constant. In the classical case,
the invariance theorem is the cornerstone for the whole theory of Kolmogorov complexity, and in the quantum case,
we expect that it will be of similar importance.
\item The aim of defining a quantum Kolmogorov complexity is to find a generalization of classical
Kolmogorov complexity to quantum systems.
In Section~\ref{SecQCofClass}, we show that this point of view is justified by proving that both
complexities closely coincide on the domain of classical strings. That is, the quantum complexity $QC$
of classical strings equals the classical complexity $C$ up to some constant. For the quantum complexity $QC^\delta$
with some fixed error tolerance $\delta$ for the QTM's output, we prove that both are equal up to some factor $1/(1-4\delta)$.
\item In Section~\ref{SecQBrudno}, we prove that the von Neumann entropy rate of an ergodic quantum information source
is arbitrarily close to its Kolmogorov complexity rate with probability one. This generalizes a classical theorem which has first
been conjectured by Zvonkin and Levin~\cite{Zvonkin} and was later proved by Brudno~\cite{Brudno}.

The case that is typically studied in quantum information theory is an i.i.d. source, that is,
many copies of a single density operator $\rho$. Ergodic sources generalize this model to
the case where the source is still stationary, but the different instances can be correlated in complicated ways.
The quantum Brudno's theorem shows that for such sources, the randomness (quantum Kolmogorov complexity) of single strings emitted
by the source typically equals the randomness of the source itself (its von Neumann entropy).

This part of the thesis is joint work with F. Benatti, T. Kr\"uger, Ra. Siegmund-Schultze, and A. Szko\l a.
\end{itemize}

Finally, in a summary and outlook, we discuss perspectives for further research and propose a concrete
application of quantum Kolmogorov complexity in quantum statistical mechanics.

\chapter{The Quantum Turing Machine}
\label{ChapterQTM}
The previous work on quantum Turing machines (QTMs) focused
on {\em computational} complexity, i.e. on questions like how efficient QTMs can perform certain
tasks or simulate other quantum computing machines. Since quantum Kolmogorov complexity does not
depend on the time of computation, but only focuses on the length of the input, we have to explore
different aspects of QTMs which have not been analyzed in this way before.

Note that the results on QTMs that we prove in this chapter may also be valid for other quantum
computing devices, as long as they map input quantum states to output quantum states, and may
or may not halt at some time step.

\section{Definition of Quantum Turing Machines}
\label{SecDefQTM}
In 1985, Deutsch~\cite{Deutsch} proposed
the first model of a quantum Turing machine (QTM), elaborating on an
even earlier idea by Feynman~\cite{Feynman}. Bernstein and Vazirani~\cite{BernsteinVazirani}
worked out the theory in more detail and proved that there exists an efficient
universal QTM (it will be discussed in Section~\ref{SecHaltingUniv} in what sense). A more compact presentation
of these results can be found in the book by Gruska~\cite{Gr}.
Ozawa and Nishimura~\cite{Ozawa}
gave necessary and sufficient conditions that a QTM's transition function results in
unitary time evolution. Benioff~\cite{Benioff} has worked out a slightly
different definition which is based on a local Hamiltonian instead of a local transition amplitude.

The definition of QTMs that we use in this thesis will be completely equivalent to that by Bernstein and Vazirani.
Yet, we will use some different kind of notation which makes it easier (or at least more clear) to derive
analytic estimates like ``{\em how much does the state of the control change at most, if the
input changes by some amount?}''. Also, we use the word QTM not only for the model itself, but also
for the partial function which it generates.

We start by defining the quantum analogue of a bit string.

\subsection{Indeterminate-Length Qubit Strings}
\label{SubsecQubitStrings}
The quantum analogue of a bit string, a so-called {\em qubit string}, is a superposition of several classical bit strings.
To be as general as possible, we would like to allow also superpositions of strings of {\em different} lengths like
\[
   |\varphi\rangle:=\frac 1 {\sqrt 2}\left( |00\rangle+|11011\rangle\right).
\]
Such quantum states are called {\em indeterminate-length qubit strings}. They have been studied
by Schumacher and Westmoreland \cite{SchumacherWestmoreland}, as well as by Bostr\"om and Felbinger \cite{Bostroem} in the context
of lossless quantum data compression.

Let \label{HK}$\hr_k:=\left(\C^{\{0,1\}}\right)^{\otimes k}$ be the Hilbert space
of $k$ qubits ($k\in\N_0$). We write $\C^{\{0,1\}}$ for $\C^2$ to
indicate that we fix two orthonormal {\em computational basis vectors}
$|0\rangle$ and $|1\rangle$. The Hilbert space $\hr_\s$ which contains indeterminate-length qubit strings like $|\varphi\rangle$
can be formally defined as the direct sum
\[
   \hr_{\s}:=\bigoplus_{k=0}^\infty \hr_k.
   \label{DefHrS}
\]
The classical finite binary strings $\{0,1\}^*$ are identified with
the computational basis
vectors in $\hr_\s$, i.e. $\hr_{\s}\simeq \ell^2(\{\lambda,
0,1,00,01,\ldots\})$, where $\lambda$ denotes the empty string.
We also use the notation
\[
   \hr_{\leq n}:=\bigoplus_{k=0}^n \hr_k
\]
and treat it as a subspace of $\hr_\s$.

To be as general as possible,
we do not only allow superpositions of strings of different lengths,
but also {\em mixtures}, i.e. our qubit strings are arbitrary density operators
on $\hr_\s$. It will become clear in the next sections that QTMs naturally
produce mixed qubit strings as outputs. Moreover, it will be a useful feature
that the result of applying the partial trace to segments of qubit strings
will itself be a qubit string.

Furthermore, we would like to say what the {\em length} of a qubit string is.
It was already noticed in \cite{SchumacherWestmoreland} and \cite{Bostroem} that
there are two different natural possibilities, which we will give in the next definition.

Before we state the definition of a qubit string, we fix some notation: if $\hr$ is a Hilbert space,
than we denote by \label{DefTraceClassOp}$\mathcal{T}(\hr)$ the trace-class operators on $\hr$. Moreover, $\mathcal{T}_1^+(\hr)$
shall denote the {\em density operators} on $\hr$, that is, the positive trace-class operators
with trace $1$.

\begin{definition}[Qubit Strings and their Length]
\label{DefLength}
\lineclear
An (indeterminate-length) {\em qubit string} $\sigma$
is a density operator on $\hr_\s$. Normalized vectors $|\psi\rangle\in\hr_\s$
will also be called qubit strings, identifying them with the corresponding
density operator $|\psi\rangle\langle\psi|$.

The {\em base length} (or just {\em length}) of a qubit string $\sigma\in\mathcal{T}_1^+(\hr_\s)$ is defined as
\[
   \ell(\sigma):=\max\{\ell(s)\,\,|\,\, \langle s|\sigma|s\rangle>0,\,\,s\in\s\}
\]
or as $\ell(\sigma)=\infty$ if the maximum does not exist. Moreover, we define the {\em average length}
$\bar\ell(\sigma)\in\R_0^+\cup\{\infty\}$ as
\[
   \bar\ell(\sigma):={\rm Tr}(\sigma\Lambda),\label{EqTrTrTr}
\]
where $\Lambda$ is the unbounded self-adjoint {\em length operator}. It is defined as
\[
   \Lambda=\sum_{n=0}^\infty n\cdot P_n
\]
on its obvious domain of definition, where $P_n$ denotes the projector onto the subspace $\hr_n$ of $\hr_\s$.
\end{definition}

For example, the qubit string $|\psi\rangle:=\frac 1 {\sqrt 2} \left(|0\rangle+|1101\rangle\right)$
has length $\ell(|\psi\rangle)=4$, i.e. the length of an indeterminate-length
qubit string equals the maximal length of any computational basis vector that has non-zero
coefficient in the superposition. This is motivated by the fact that a qubit string $\sigma$
needs at least $\ell(\sigma)$ cells on a QTM's tape to be stored perfectly (compare
Subsection~\ref{SubsecQTMs}).

On the other hand, we have $\bar\ell(|\psi\rangle)=\frac 1 2 1+\frac 1 2 4=\frac 5 2$.
Using either $\ell$ or $\bar\ell$ will give two different definitions of quantum Kolmogorov complexity.
The idea to use $\bar\ell$ in that definition has first been proposed by Rogers and Vedral \cite{Rogers}.

In contrast to classical bit strings, there are uncountably
many qubit strings that cannot be perfectly distinguished by means
of any quantum measurement. A good measure for the difference
between two quantum states is the trace distance (cf. \cite{NielsenChuang})
\begin{equation}
   \|\rho-\sigma\|_{\rm Tr}:=\frac 1 2 {\rm Tr} \left\vert \rho-\sigma
   \right\vert.
   \label{tr-dist}
\end{equation}
It has the nice operational meaning to be the maximum difference in probability for a yes-no-measurement
if either applied to $\rho$ or $\sigma$, cf. \cite{NielsenChuang}.

This distance measure on the qubit strings will be used in
our definition of quantum Kolmogorov complexity in Section~\ref{SecDefQComplexity}.

\subsection{Mathematical Framework for QTMs}
\label{SubsecQTMs}
To understand the notion of a quantum Turing machine (QTM), we first explain
how a classical Turing machine (TM) is defined.

We can think of a classical TM as consisting of three different parts: a control $\mathbf C$,
a head ${\mathbf H}$, and a tape ${\mathbf T}$. The tape consists of cells that are indexed
by the integers, and carry some symbol from a finite alphabet $\Sigma$. In the simplest
case, the alphabet consists of a zero, a one, and a special blank symbol $\#$. At the beginning
of the computation, all the cells are blank, i.e. carry the special symbol $\#$, except
for those cells that contain the input bit string.

The head points to one of the cells. It is connected to the control, which in every step
of the computation is in one ``internal state'' $q$ out of a finite set $Q$. At the
beginning of the computation, it is in the initial state $q_0\in Q$, while the end
of the computation (i.e. the halting of the TM) is attained if the control is in the so-called
final state $q_f\in Q$.

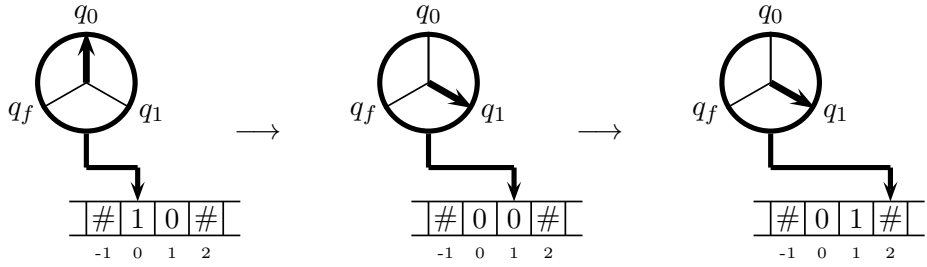
\begin{figure}[!hbt]
\psset{unit=0.45cm}
\begin{center}
\begin{pspicture}(-1,0.5)(26,8)
   {\psset{linewidth=0.05}\psline(0.5,1.5)(5.5,1.5)}
   {\psset{linewidth=0.05}\psline(0.5,2.5)(5.5,2.5)}
   {\psset{linewidth=0.05}\psline(1,1.5)(1,2.5)}
   {\psset{linewidth=0.05}\psline(2,1.5)(2,2.5)}
   {\psset{linewidth=0.05}\psline(3,1.5)(3,2.5)}
   {\psset{linewidth=0.05}\psline(4,1.5)(4,2.5)}
   {\psset{linewidth=0.05}\psline(5,1.5)(5,2.5)}
   {\psset{linewidth=0.15}\psline{->}(2.5,3.5)(2.5,2.5)}
   {\psset{linewidth=0.15}\psline(2.5,3.5)(1,3.5)}
   {\psset{linewidth=0.15}\psline(1,3.5)(1,4.5)}
   \pscircle[linewidth=0.15](1,6){1.5}
   {\psset{linewidth=0.05}\psline(1,6)(1,7.5)}
   {\psset{linewidth=0.2}\psline{->}(1,6)(1,7.5)}
   {\psset{linewidth=0.05}\psline(1,6)(2.3,5.25)}   
   {\psset{linewidth=0.05}\psline(1,6)(-0.3,5.25)}
   \rput(1.5,2){{$\#$}}
   \rput(2.5,2){{$1$}}
   \rput(3.5,2){{$0$}}
   \rput(4.5,2){{$\#$}}
   \rput(6,4.5){{$\longrightarrow$}}
   \rput(1.5,1){{\tiny{-1}}}
   \rput(2.5,1){{\tiny{0}}}
   \rput(3.5,1){{\tiny{1}}}
   \rput(4.5,1){{\tiny{2}}}
   \rput(1,8){{$q_0$}}   
   \rput(2.9,5.1){{$q_1$}}
   \rput(-0.9,5.1){{$q_f$}}

   {\psset{linewidth=0.05}\psline(10.5,1.5)(15.5,1.5)}
   {\psset{linewidth=0.05}\psline(10.5,2.5)(15.5,2.5)}
   {\psset{linewidth=0.05}\psline(11,1.5)(11,2.5)}
   {\psset{linewidth=0.05}\psline(12,1.5)(12,2.5)}
   {\psset{linewidth=0.05}\psline(13,1.5)(13,2.5)}
   {\psset{linewidth=0.05}\psline(14,1.5)(14,2.5)}
   {\psset{linewidth=0.05}\psline(15,1.5)(15,2.5)}
   {\psset{linewidth=0.15}\psline{->}(13.5,3.5)(13.5,2.5)}
   {\psset{linewidth=0.15}\psline(13.5,3.5)(11,3.5)}
   {\psset{linewidth=0.15}\psline(11,3.5)(11,4.5)}
   \pscircle[linewidth=0.15](11,6){1.5}
   {\psset{linewidth=0.05}\psline(11,6)(11,7.5)}
   {\psset{linewidth=0.05}\psline(11,6)(11,7.5)}
   {\psset{linewidth=0.2}\psline{->}(11,6)(12.3,5.25)}   
   {\psset{linewidth=0.05}\psline(11,6)(9.7,5.25)}
   \rput(11.5,2){{$\#$}}
   \rput(12.5,2){{$0$}}
   \rput(13.5,2){{$0$}}
   \rput(14.5,2){{$\#$}}
   \rput(16,4.5){{$\longrightarrow$}}
   \rput(11.5,1){{\tiny{-1}}}
   \rput(12.5,1){{\tiny{0}}}
   \rput(13.5,1){{\tiny{1}}}
   \rput(14.5,1){{\tiny{2}}}
   \rput(11,8){{$q_0$}}   
   \rput(12.9,5.1){{$q_1$}}
   \rput(9.1,5.1){{$q_f$}}

   {\psset{linewidth=0.05}\psline(20.5,1.5)(25.5,1.5)}
   {\psset{linewidth=0.05}\psline(20.5,2.5)(25.5,2.5)}
   {\psset{linewidth=0.05}\psline(21,1.5)(21,2.5)}
   {\psset{linewidth=0.05}\psline(22,1.5)(22,2.5)}
   {\psset{linewidth=0.05}\psline(23,1.5)(23,2.5)}
   {\psset{linewidth=0.05}\psline(24,1.5)(24,2.5)}
   {\psset{linewidth=0.05}\psline(25,1.5)(25,2.5)}
   {\psset{linewidth=0.15}\psline{->}(24.5,3.5)(24.5,2.5)}
   {\psset{linewidth=0.15}\psline(24.5,3.5)(21,3.5)}
   {\psset{linewidth=0.15}\psline(21,3.5)(21,4.5)}
   \pscircle[linewidth=0.15](21,6){1.5}
   {\psset{linewidth=0.05}\psline(21,6)(21,7.5)}
   {\psset{linewidth=0.05}\psline(21,6)(21,7.5)}
   {\psset{linewidth=0.2}\psline{->}(21,6)(22.3,5.25)}   
   {\psset{linewidth=0.05}\psline(21,6)(19.7,5.25)}
   \rput(21.5,2){{$\#$}}
   \rput(22.5,2){{$0$}}
   \rput(23.5,2){{$1$}}
   \rput(24.5,2){{$\#$}}
   \rput(21.5,1){{\tiny{-1}}}
   \rput(22.5,1){{\tiny{0}}}
   \rput(23.5,1){{\tiny{1}}}
   \rput(24.5,1){{\tiny{2}}}
   \rput(21,8){{$q_0$}}   
   \rput(22.9,5.1){{$q_1$}}
   \rput(19.1,5.1){{$q_f$}}      
\end{pspicture}
\caption{Time evolution of a Turing machine}
\label{AbbTM}
\end{center}
\end{figure}

The computation itself, i.e. the TM's time evolution, is determined by a so-called {\em transition function} $\delta$:
depending on the current state of the control $q\in Q$ and the symbol $\sigma\in\Sigma$ which is on the tape
cell where the head is pointing to, the TM turns into some new internal state $q'\in Q$, writes
some symbol $\sigma'\in\Sigma$ onto this tape cell, and then either turns left (L) or right (R). Thus, the transition
function $\delta$ is a map
\[
   \delta:Q\times\Sigma\to Q\times \Sigma\times \{L,R\}.
\]
As an example, we consider a TM with alphabet $\Sigma=\{0,1,\#\}$, internal states $Q=\{q_0,q_1,q_f\}$
and transition function $\delta$, given by
\begin{eqnarray*}
q_0,0&\stackrel\delta\mapsto& q_1,1,R \\
q_0,1&\stackrel\delta\mapsto& q_1,0,R \\
q_1,0&\stackrel\delta\mapsto& q_1,1,R \\
q_1,1&\stackrel\delta\mapsto& q_1,0,R \\
q_1,\#&\stackrel\delta\mapsto& q_f,\#,R.
\end{eqnarray*}
We have not defined $\delta(q_0,\#)$ and $\delta(q_f,\sigma)$ for any $\sigma$; we can define $\delta$
at these arguments in an arbitrary way. We imagine that this TM is started with some input bit
string $s$, which is written onto the tape segment $[0,\ell(s)-1]$. The head initially points to
cell number zero. The computation of the TM will then invert the string and halt.
As an example, in Figure~\ref{AbbTM}, we have depicted the first steps of the TM's time evolution
on input $s=10$.

A QTM is now defined analogously as a TM, but with the important difference that the
transition function is replaced by a {\em transition amplitude}.
That is, instead of having a single classical successor state for every internal state
and symbol on the tape, a QTM can evolve into a {\em superposition} of different classical
successor states.

For example, we may have a QTM that, if the control's internal state is $q_0\in Q$
and the tape symbol is a $0$, may turn into internal state $q_1$ and write a one and turn right,
as well as writing a zero and turning left, both at the same time in superposition, say
with complex amplitudes $\frac 1 {\sqrt 2}$ and $\frac {-i}{\sqrt 2}$.
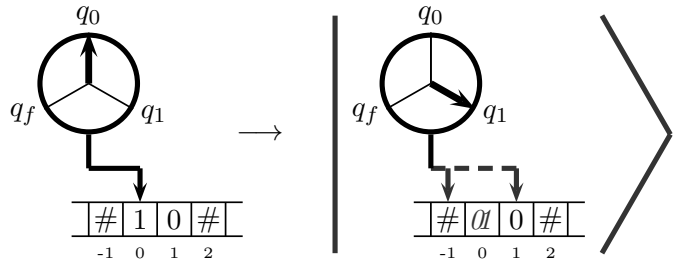
\begin{figure}[!hbt]
\psset{unit=0.45cm}
\begin{center}
\begin{pspicture}(-1,0.5)(14,8)
   {\psset{linewidth=0.05}\psline(0.5,1.5)(5.5,1.5)}
   {\psset{linewidth=0.05}\psline(0.5,2.5)(5.5,2.5)}
   {\psset{linewidth=0.05}\psline(1,1.5)(1,2.5)}
   {\psset{linewidth=0.05}\psline(2,1.5)(2,2.5)}
   {\psset{linewidth=0.05}\psline(3,1.5)(3,2.5)}
   {\psset{linewidth=0.05}\psline(4,1.5)(4,2.5)}
   {\psset{linewidth=0.05}\psline(5,1.5)(5,2.5)}
   {\psset{linewidth=0.15}\psline{->}(2.5,3.5)(2.5,2.5)}
   {\psset{linewidth=0.15}\psline(2.5,3.5)(1,3.5)}
   {\psset{linewidth=0.15}\psline(1,3.5)(1,4.5)}
   \pscircle[linewidth=0.15](1,6){1.5}
   {\psset{linewidth=0.05}\psline(1,6)(1,7.5)}
   {\psset{linewidth=0.2}\psline{->}(1,6)(1,7.5)}
   {\psset{linewidth=0.05}\psline(1,6)(2.3,5.25)}   
   {\psset{linewidth=0.05}\psline(1,6)(-0.3,5.25)}
   \rput(1.5,2){{$\#$}}
   \rput(2.5,2){{$1$}}
   \rput(3.5,2){{$0$}}
   \rput(4.5,2){{$\#$}}
   \rput(6,4.5){{$\longrightarrow$}}
   \rput(1.5,1){{\tiny{-1}}}
   \rput(2.5,1){{\tiny{0}}}
   \rput(3.5,1){{\tiny{1}}}
   \rput(4.5,1){{\tiny{2}}}
   \rput(1,8){{$q_0$}}   
   \rput(2.9,5.1){{$q_1$}}
   \rput(-0.9,5.1){{$q_f$}}
   {\psset{linewidth=0.05}\psline(10.5,1.5)(15.5,1.5)}
   {\psset{linewidth=0.05}\psline(10.5,2.5)(15.5,2.5)}
   {\psset{linewidth=0.05}\psline(11,1.5)(11,2.5)}
   {\psset{linewidth=0.05}\psline(12,1.5)(12,2.5)}
   {\psset{linewidth=0.05}\psline(13,1.5)(13,2.5)}
   {\psset{linewidth=0.05}\psline(14,1.5)(14,2.5)}
   {\psset{linewidth=0.05}\psline(15,1.5)(15,2.5)}
   {\psset{linewidth=0.15,linecolor=darkgray}\psline{->}(13.5,3.5)(13.5,2.5)}
   {\psset{linewidth=0.15,linecolor=darkgray}\psline{->}(11.5,3.5)(11.5,2.5)}
   {\psset{linewidth=0.15,linecolor=darkgray,linestyle=dashed}\psline(13.5,3.5)(11,3.5)}
   {\psset{linewidth=0.15}\psline(11,3.5)(11,4.5)}
   \pscircle[linewidth=0.15](11,6){1.5}
   {\psset{linewidth=0.05}\psline(11,6)(11,7.5)}
   {\psset{linewidth=0.05}\psline(11,6)(11,7.5)}
   {\psset{linewidth=0.2}\psline{->}(11,6)(12.3,5.25)}   
   {\psset{linewidth=0.05}\psline(11,6)(9.7,5.25)}
   \rput(11.5,2){{$\#$}}
   \rput(13.5,2){{$0$}}
   \rput(12.37,2){\textcolor{darkgray}{{$\it 0$}}}
   \rput(12.63,2){\textcolor{darkgray}{{$\it 1$}}}   
   \rput(14.5,2){{$\#$}}
   \rput(11.5,1){{\tiny{-1}}}
   \rput(12.5,1){{\tiny{0}}}
   \rput(13.5,1){{\tiny{1}}}
   \rput(14.5,1){{\tiny{2}}}
   \rput(11,8){{$q_0$}}   
   \rput(12.9,5.1){{$q_1$}}
   \rput(9.1,5.1){{$q_f$}}      
   {\psset{linewidth=0.15,linecolor=darkgray}\psline(8.2,1)(8.2,8)}
   {\psset{linewidth=0.15,linecolor=darkgray}\psline(16,1)(18,4.5)}
   {\psset{linewidth=0.15,linecolor=darkgray}\psline(16,8)(18,4.5)}   
\end{pspicture}
\caption{One step of time evolution of a quantum Turing machine}
\label{AbbQTM}
\end{center}
\end{figure}

A symbolic picture of this behaviour is depicted in Figure~\ref{AbbQTM}.
This can be written as
\[
   q_0,0\stackrel\delta\mapsto \underbrace{(q_1,1,R)}_{\frac 1 {\sqrt 2}},\underbrace{(q_1,0,L)}_{\frac{-i}{\sqrt 2}}.
\]
Formally, the transition amplitude $\delta$ is thus a mapping from $Q\times \Sigma$ to the {\em complex
functions} on $Q\times\Sigma\times\{L,R\}$. If the QTM as a whole is described by a Hilbert space $\hr_{QTM}$, then
we can linearly extend $\delta$ to define some global time evolution on $\hr_{QTM}$. We have to take care of two things:
\begin{itemize}
\item According to the postulates of quantum mechanics, we have to construct $\delta$ in such a way
that the resulting global time evolution on $\hr_{QTM}$ is {\em unitary}.
\item The complex amplitudes which are assigned to the successor states have to be
efficiently computable, which has the physical interpretation that we should be able
to efficiently prepare hardware (e.g. some quantum gate) which realizes the transitions specified by $\delta$.

Moreover, this requirement also guarantees that every QTM has a finite classical description,
that there is a universal QTM (see discussion below), and that we cannot ``hide'' information
(like the answer to infinitely many instances of the halting problem) in the transition amplitudes.
\end{itemize}

Consequently, Bernstein and Vazirani (\cite{BernsteinVazirani}, Def. 3.2.2) define
a quantum Turing machine $M$ as a triplet $(\Sigma,Q,\delta)$, where
$\Sigma$ is a finite alphabet with an identified blank symbol $\#$,
$Q$ is a finite set of states with an identified initial state $q_0$
and final state $q_f\neq q_0$, and $\delta:Q\times\Sigma\to\tilde\C^{Q \times \Sigma \times \{L,R\}}$
is the so-called the {\em quantum transition function}, determining the QTM's time evolution
in a way which is explained below.

Here, the symbol $\tilde\C$ denotes the set of complex numbers that are efficiently computable.
In more detail, $\alpha\in\tilde\C$ if and only if there is
a deterministic algorithm that computes the real and imaginary parts of $\alpha$ to within $2^{-n}$
in time polynomial in $n$.

Every QTM evolves in discrete, integer
time steps, where at every step, only a finite number of tape cells is non-blank.
For every QTM, there is a corresponding Hilbert space
\[
   \hr_{QTM}=\hr_{\mathbf{C}}\otimes \hr_{\mathbf{T}}
   \otimes \hr_{\mathbf{H}},
\]
where $\hr_{\mathbf C}=\C^Q$ is a finite-dimensional Hilbert space
spanned by the (orthonormal) control states $q\in Q$, while $\hr_{\mathbf T}=\ell^2(T)$
and $\hr_{\mathbf H}=\ell^2(\mathbb{Z})$
are separable Hilbert spaces describing the contents of the tape
and the position of the head.
In this definition, the symbol $T$ denotes the set of classical tape configurations with finitely many non-blank symbols,
i.e.
\begin{equation}
   T=\left\{(x_i)_{i\in\Z}\in\Sigma^\Z\,\,|\,\, x_i\neq\#
   \mbox{ for finitely many }i\in\Z\right\}.
   \label{EqDefT}
\end{equation}

For our purpose, it is useful to consider a special class of QTMs
with the property that their tape $\mathbf T$ consists of two different tracks
(cf. \cite[Def. 3.5.5]{BernsteinVazirani}),
an {\em input track} $\mathbf I$ and an {\em output track} $\mathbf O$. This can be achieved by
having an alphabet which is a Cartesian product of two alphabets,
in our case $\Sigma=\{0,1,\#\}\times \{0,1,\#\}$. Then, the tape Hilbert space
$\hr_{\mathbf T}$ can be written as $\hr_{\mathbf{T}}=\hr_{\mathbf{I}}\otimes
\hr_{\mathbf{O}}$, thus
\[
   \hr_{QTM}=\hr_{\mathbf C}\otimes\hr_{\mathbf I}\otimes\hr_{\mathbf O}\otimes\hr_{\mathbf H}.
\]

The transition
amplitude $\delta$ generates a linear operator $U_M$ on $\hr_{QTM}$ describing the
time evolution of the QTM $M$.
We identify $\sigma\in\mathcal{T}_1^+(\hr_\s)$ with the initial state
of $M$ on input $\sigma$, which is according to the definition in
\cite{BernsteinVazirani} a state on  $\hr_{QTM}$ where $\sigma$ is written on the input track
over the cell interval $[0, \ell(\sigma)-1]$, the empty symbol
$\#$ is written on the remaining cells of the input track and on the whole
output track, the control is in the initial state $q_0$ and the head is
in position $0$. By linearity, this e.g. means that the pure qubit string $|\psi\rangle=\frac 1 {\sqrt 2}
\left(|0\rangle+|11\rangle\right)$ is identified with the vector
$\frac 1 {\sqrt 2}\left(|0\#\rangle+|11\rangle\right)$
on input track cells number $0$ and $1$.

The global state $M^t(\sigma)\in\mathcal{T}_1^+(\hr_{QTM})$
of $M$ on input $\sigma$ at time $t\in\N_0$ is given by $M^t(\sigma)=\left(U_M\right)^t \sigma
\left(U_M^*\right)^t$. The state of the control at time $t$ is thus given by partial trace
over all the other parts of the machine, that is \label{MC}$M_{\mathbf C}^t(\sigma):={\rm Tr}_{\mathbf{T,H}}\left(
M^t(\sigma)\right)$ (similarly for the other parts of the QTM).
In accordance with \cite[Def. 3.5.1]{BernsteinVazirani}, we say that the
QTM $M$ {\em halts at time $t\in\N$ on input $\sigma\in\mathcal{T}_1^+(\hr_\s)$},
if and only if
\begin{equation}
   \langle q_f|M_{\rm\bf C}^t(\sigma)|q_f\rangle=1 \quad\mbox{ and }
   \quad\langle q_f|M_{\rm\bf C}^{t'}(\sigma)|q_f\rangle=0\quad
   \mbox{for every } t'<t,
   \label{EqHalting}
\end{equation}
where $q_f\in Q$ is the final state of the control (specified in the
definition of $M$) signaling the halting of the computation. 
See Subsection~\ref{SecHaltingUniv}
for a detailed discussion of this condition (Equation~(\ref{EqHalting})).

In this thesis, when we talk about a QTM, we do not mean the machine model itself,
but rather refer to the corresponding partial function on the qubit strings
which is computed by the QTM. Note
that this point of view is different from e.g. that of Ozawa~\cite{OzawaLocalTransition}
who describes a QTM as a map from $\Sigma^*$ to the set of probability distributions on $\Sigma^*$.

We still have to define what is meant by the output of a QTM $M$, once it has halted at some time $t$ on some input
qubit string $\sigma$.
We could take the state of the output tape $M_{\mathbf O}^t(\sigma)$ to be the output, but this is not
a qubit string, but instead a density operator on the Hilbert space $\hr_{\mathbf O}$. Hence, we define
a quantum operation $\mathcal{R}$ which maps the density operators on $\hr_{\mathbf O}$ to density operators
on $\hr_\s$, i.e. to the qubit strings. The operation $\mathcal{R}$ ``reads'' the output from the tape.

\begin{definition}[Reading Operation]
\label{DefReadingOp}
\lineclear
A quantum operation $\mathcal{R}:\mathcal{T}(\hr_{\mathbf O})\to\mathcal{T}(\hr_\s)$ is
called a {\em reading operation}, if for every finite set of classical strings $\{s_i\}_{i=1}^N\subset\s$, it holds that
\[
   \mathcal{R}\left(\mathbb{P}\left(
         \sum_{i=1}^N \alpha_i \left|\begin{array}{ccccccc}
               \ldots & \# & \# & s_i & \# & \# & \ldots \\
                & \mbox{\tiny -2} & \mbox{\tiny -1} & \mbox{\tiny 0} & \mbox{\tiny$\ell(s_i)$} & \mbox{\tiny $\ell(s_i)+1$} &
             \end{array}
         \right\rangle
      \right)
   \right)
   =\mathbb{P}\left(
      \sum_{i=1}^N \alpha_i |s_i\rangle
   \right)
\]
where $\mathbb{P}(|\varphi\rangle):=|\varphi\rangle\langle\varphi|$ denotes the projector onto $|\varphi\rangle$.
\end{definition}

The condition specified above does not determine $\mathcal{R}$ uniquely; there are many different reading operations.
For the remainder of this thesis, we fix the reading operation $\mathcal{R}$ which is specified in the following example.

\begin{example}
\label{ExR}
Let $T$ denote the classical output track configurations as defined in Equation~(\ref{EqDefT}), with
$\Sigma=\{0,1,\#\}$. Then, for every $t\in T$, let $R(t)$ be the classical string that consists of
the bits of $T$ from cell number zero to the last non-blank cell, i.e.
\begin{eqnarray*}
   R&:&T\to\s \\
   &&\left(\begin{array}{ccccccc}
               \ldots & ? & ? & s & \# & ? & \ldots \\
                & \mbox{\tiny -2} & \mbox{\tiny -1} & \mbox{\tiny 0} & \mbox{\tiny$\ell(s)$} & \mbox{\tiny $\ell(s)+1$} &
             \end{array}\right)
   \mapsto s.
\end{eqnarray*}
For every $s\in\s$, there is a countably-infinite number of $t\in T$ such that $R(t)=s$. Thus, to every $t\in T$,
we can assign a natural number $n(t)$ which is the number of $t$ in some enumeration of the set $\{t'\in T\,\,|\,\, R(t')=R(t)\}$;
we only demand that $n(t)=1$ if $t=\left(\begin{array}{ccccccc}
               \ldots & \# & \# & s & \# & \# & \ldots \\
                & \mbox{\tiny -2} & \mbox{\tiny -1} & \mbox{\tiny 0} & \mbox{\tiny$\ell(s)$} & \mbox{\tiny $\ell(s)+1$} &
             \end{array}\right)$.
Hence, if (as usual) $\ell^2\equiv \ell^2(\N)$ denotes the Hilbert space of square-summable sequences, then the map $U$,
defined by linear extension of
\begin{eqnarray*}
   U:\hr_{\mathbf O}&\to& \hr_\s\otimes \ell^2 \\
   |t\rangle&\mapsto& |R(t)\rangle\otimes |n(t)\rangle,
\end{eqnarray*}
is unitary. Then, the quantum operation
\begin{eqnarray*}
   \mathcal{R}:\mathcal{T}(\hr_{\mathbf O})&\to&\mathcal{T}(\hr_\s) \\
   \rho&\mapsto& {\rm Tr}_{\ell^2} \left( U\rho U^*\right)
\end{eqnarray*}
is a reading operation.
\end{example}

We are now ready to define QTMs as partial maps on the qubit strings.

\begin{definition}[Quantum Turing Machine (QTM)]
\label{DefQTM}
\lineclear
A partial map $M:\mathcal{T}_1^+(\hr_\s)\to\mathcal{T}_1^+(\hr_\s)$
will be called a QTM, if there is a Bernstein-Vazirani two-track QTM $M'=(\Sigma,Q,\delta)$
(see \cite{BernsteinVazirani}, Def. 3.5.5)
with the following properties:
\begin{itemize}
\item $\Sigma=\{0,1,\#\}\times \{0,1,\#\}$,
\item the corresponding time evolution operator $U_{M'}$ is unitary,
\item if $M'$ halts on input $\sigma$ at some time $t\in\N$, then
$M(\sigma)=\mathcal{R}\left({M'}_{\mathbf O}^t(\sigma)\right)$, where
$\mathcal{R}$ is the reading operation specified in Example~\ref{ExR} above.
Otherwise, $M(\sigma)$ is undefined.
\end{itemize}
A {\em fixed-length QTM} is the restriction of a QTM to the domain
$\bigcup_{n\in\N_0} \mathcal{T}_1^+(\hr_n)$
of length eigenstates. We denote the domain of definition of a QTM $M$ by ${\rm dom}\,M$.
\end{definition}

The definition of halting, given by Equation~(\ref{EqHalting}), is very important, as we will discuss
in Section~\ref{SecHaltingUniv}. On the other hand, changing certain details of a QTM's definition,
like the way to read the output or allowing a QTM's head to stay at its position instead of
turning left or right, should not change the results in this thesis.

A simple example of a fixed-length QTM is the identity map on the fixed-length qubit strings,
which corresponds to a machine that moves the contents of the input track to the output track.

\begin{example}
The identity map on the fixed-length qubit strings, i.e.
\begin{eqnarray*}
   {\rm id}:\bigcup_{n\in\N_0} \mathcal{T}_1^+(\hr_n)&\to&\bigcup_{n\in\N_0} \mathcal{T}_1^+(\hr_n)\\
   \rho&\mapsto&\rho
\end{eqnarray*}
is a fixed-length QTM.
\end{example}
\proof We start by defining a classical Turing machine that moves the content of the input
track to the output track and halts. Let $\Sigma:=\{0,1,\#\}^2$ and $Q=\{q_0,q_f\}$.
We look for a transition function $\delta:Q\times\Sigma\to Q\times\Sigma\times \{L,R\}$
such that
\begin{eqnarray*}
\left( q_0,\#\#\right)&\stackrel\delta\mapsto& \left(q_f,\#\#,R\right),\\
\left( q_0,0\#\right)&\stackrel\delta\mapsto& \left(q_0,\# 0,R\right),\\
\left( q_0,1\# \right)&\stackrel\delta\mapsto& \left(q_0,\# 1,R\right).
\end{eqnarray*}
This is not a complete definition, since we do not specify the action of $\delta$
on all the other configurations, but \cite[Corollary B.0.15]{BernsteinVazirani}
guarantees that $\delta$ can be extended to a total function on all the configurations
in some way (that we are not interested in) {\em such that the resulting TM $M$ is reversible}
as long as the following two conditions
are satisfied:
\begin{itemize}
\item[(1.)] Each state can be entered only from one direction, i.e. if $\delta(p_1,\sigma_1)=
(q,\tau_1,d_1)$ and $\delta(p_2,\sigma_2)=(q,\tau_2,d_2)$, then $d_1=d_2$.
\item[(2.)] The transition function $\delta$ is one-to-one when direction is ignored.
\end{itemize}
It is easily checked that both conditions are satisfied here. Moreover, it is not
difficult to see that the classical, reversible TM $M$ defined by the transition function $\delta$
moves the content of the input track bit by bit to the output track (while remaining in state $q_0$)
just until it detects the first blank symbol on the input track; in this case, it turns one more
step to the right and halts.

As $M$ is a reversible TM, $M$ is also a Bernstein-Vazirani QTM with unitary time
evolution, and thus, $M$ is a QTM in the sense of Definition~\ref{DefQTM}, one that
maps every classical binary string onto itself. Since the halting time and the
final position of the head of $M$ only depend on the length of the input, it follows
that superpositions of classical strings of common length are mapped to superpositions
(the same is true for mixtures). Thus, $M(\rho)=\rho$ for fixed-length qubit strings $\rho$.\qed

Given that an identity machine is simple to define on fixed-length inputs (it just moves
the contents of the input track to the output track), it is perhaps surprising that this
is not a QTM on indeterminate-length inputs. The reason is that if the input has indeterminate length,
there is no way to determine when the process of moving the contents to the other track
should halt: it halts at a superposition of different times if it is programmed as in the previous example,
and this contradicts the halting conditions of Equation~(\ref{EqHalting}).

\begin{example}
The identity map on the indeterminate-length qubit strings, i.e.
\begin{eqnarray*}
   {\rm id}:\mathcal{T}_1^+(\hr_\s)&\to&\mathcal{T}_1^+(\hr_\s)\\
   \rho&\mapsto&\rho
\end{eqnarray*}
 is {\em not} a QTM.
\end{example}
\proof Suppose the identity map on the indeterminate-length qubit strings
was a QTM. Let $\rho\in\mathcal{T}_1^+(\hr_\s)$
be an arbitrary indeterminate-length qubit string, and let $\tau\in\N$ denote the corresponding
halting time of the QTM $\rm id$ on input and output $\rho$.
Let $\sigma\in\mathcal{T}_1^+(\hr_\s)$ be another qubit string with
$\ell(\sigma)>\tau$.

For $\eps\in(0,1)$, let $\rho_\eps:=(1-\eps)\rho+\eps\sigma$. It follows that
$\ell(\rho_\eps)=\ell(\sigma)>\tau$. Since a QTM can only write one cell of the output tape at a time,
it follows that the halting time corresponding to $\rho_\eps$ must
be larger than $\tau$. Note that
\begin{equation}
   \|\rho-\rho_\eps\|_{\rm Tr}=\|\eps\rho-\eps\sigma\|_{\rm Tr}=\eps\|\rho-\sigma\|_{\rm Tr}\leq \eps.
   \label{eqTrDistAb}
\end{equation}
We know from the halting conditions in Equation~(\ref{EqHalting}) that
\[
   \langle q_f|{\rm id}_{\mathbf{C}}^\tau (\rho)|q_f\rangle=1\qquad\mbox{and}\\
   \qquad\langle q_f|{\rm id}_{\mathbf{C}}^\tau(\rho_\eps)|q_f\rangle=0.
\]
Thus, we get the inequality
\begin{eqnarray*}
\|\rho-\rho_\eps\|_{\rm Tr}&=&\left\| (U_{\rm id})^\tau \rho (U_{\rm id}^*)^\tau
-(U_{\rm id})^\tau \rho_\eps (U_{\rm id}^*)^\tau\right\|_{\rm Tr}\\
&=& \left\| {\rm id}^\tau(\rho)-{\rm id}^\tau(\rho_\eps)\right\|_{\rm Tr}\\
&\geq& \left\| {\rm id}_{\mathbf{C}}^\tau(\rho)-{\rm id}_{\mathbf{C}}^\tau(\rho_\eps)\right\|_{\rm Tr}=1
\end{eqnarray*}
which contradicts Equation~(\ref{eqTrDistAb}).\qed

For defining quantum Kolmogorov complexity, we will sometimes need to give
{\em two} inputs to a QTM, namely some qubit string $\sigma\in\mathcal{T}_1^+(\hr_\s)$ {\em and} an integer
$k\in\N$ both at the same time. Similarly as in the classical case, we can join $\sigma$
and a self-delimiting description $s_k\in\s$ of $k$ together by concatenation (which, in the quantum
case, is just the tensor product).

How can we do this? Since $\sigma$ may be a superposition or mixture of classical strings
of {\em different} lengths, it makes no sense to input $\sigma\otimes s_k$ into the QTM,
since the QTM cannot extract $s_k$ from the resulting qubit string.
But there is no problem with the other way round, i.e. to input $s_k\otimes \sigma$.
This leads to the following definition:

\begin{definition}[Parameter Encoding]
\label{DefEncoding}
Let $k\in\N$ and $\sigma\in\mathcal{T}_1^+(\hr_{\Fock})$. 
We define an encoding $\langle\cdot,\cdot\rangle:\N 
\times\mathcal{T}_1^+(\hr_{\s})\to \mathcal{T}_1^+(\hr_{\Fock})$
of a pair $(k,\sigma)$ into a single qubit string $\langle k,\sigma\rangle$ by
\[
   \langle k,\sigma\rangle:=|s_k\rangle\langle s_k|\otimes \sigma.
\]
Here, $s_k$ is the following self-delimiting description of $k$:
\begin{equation}
   s_k:=\underbrace{1111\ldots 1}_{\lfloor \log k\rfloor} 0 \underbrace{(\mbox{binary
   digits of }k)}_{\lfloor \log k\rfloor +1}.
   \label{EqSelfDelimiting}
\end{equation}
For every QTM $M$, we then set $M(k,\sigma):=M(\langle k,\sigma\rangle )$. Moreover,
if $\delta\in\mathbb{Q}^+$ is a rational number with $\delta=\frac l m$, and this fraction
cannot be reduced any further, then we define
\[
   M(\delta,\sigma):=M\left(\strut\langle l,\langle m,\sigma\rangle\rangle\right).
\]
\end{definition}

There are many other possibilities to encode an integer $k$ into some self-delimiting
binary string $s_k$. We chose this encoding since it is efficient enough
for our purpose (e.g. we can prove some relation like Lemma~\ref{LemRelation}),
but another choice of encoding will not change the results of this thesis.
See also the discussion after Lemma~\ref{LemRelation}. Also note that
\begin{eqnarray}
\label{encoding_length}
\ell(\langle k,\sigma\rangle)=2\lfloor \log k\rfloor +2+\ell(\sigma),
\end{eqnarray}
and the same equation holds true for average length $\bar\ell$.

In this thesis, we will sometimes consider the map $\sigma\to M(k,\sigma)$
for some QTM $M$ and some fixed integer $k$. We would like to apply everything that
we have learnt about QTMs to maps like this. Thus, the following lemma will be useful:
\begin{emptyflemma}
\label{LemIsAQTM}
For every QTM $M$ and $k\in\N$, the map $\sigma\mapsto M(k,\sigma)$ is itself a QTM.
\end{emptyflemma}
\proof Let $s_k$ be the self-delimiting description of $k$ as specified in
Equation~(\ref{EqSelfDelimiting}). Moreover,
let $T_k$ denote a classical reversible Turing machine that, ignoring its input,
prints the classical string $s_k\in\s$ onto its input track cells left
of the starting cell, i.e. onto the track segment $[-\ell(s_k),-1]$,
and then halts with the head pointing to the cell in position $-\ell(s_k)$.
As we know that these input track cells start with the empty symbol, this
can be done reversibly.

Since the reversible TM $T_k$ is also a QTM, there is a QTM that carries out the
computation of $T_k$, followed by the computation of $M$ (cf.
\cite[Dovetailing Lemma]{BernsteinVazirani}).
Nevertheless, the resulting QTM is not exactly what we want, since it will produce
$M$'s output on input $(k,\sigma)$ starting in output cell number $-\ell(s_k)$, not in cell $0$.

To circumvent this problem, we construct some modification $M'$ of $M$, which
then will give the correct output, if it is joined to $T_k$. To simplify the discussion,
we describe the solution for the special case that $s_k$ has length one.
Moreover, we restrict the proof to the situation that $M$ is a classical
reversible $TM$; the quantum generalization will be straightforward.

If $M$'s head points to some cell number $m\in\mathbb{Z}$, then $M$ reads and writes
cell number $m$ of the input track, and at the same time cell number $m$ of the output
track. The trick now is to program $M'$ in such a way that it effectively reads and writes
input track cell $m$, but output track cell $m+1$.
We choose the control state space $Q'$ of $M'$ to be three times as large
as $M$'s state space $Q$:
\[
   Q':=Q\times \{1,2,3\}.
\]
Now we construct some modified transition function $\delta'$ for the QTM $M'$
from $M$'s transition function $\delta$. Suppose that one of the transition
rules for $M$ is, for example,
\[
   q_5,(0,1)\stackrel\delta\mapsto q_6,(1,\#),L,
\]
which says that whenever $M$ is in state $q_5$ and reads the symbol $0$ on the
input track and $1$ on the output track, then it turns into state $q_6$, writes
a $1$ onto the input track and a blank symbol onto the output track and then turns left.

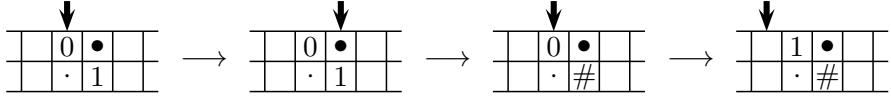
\begin{figure}[!hbt]
\psset{unit=0.4cm}
\begin{center}
\begin{pspicture}(0,0)(28,4)
   {\psset{linewidth=0.05}\psline(0.5,0.5)(5.5,0.5)}
   {\psset{linewidth=0.05}\psline(0.5,1.5)(5.5,1.5)}
   {\psset{linewidth=0.05}\psline(0.5,2.5)(5.5,2.5)}
   {\psset{linewidth=0.05}\psline(1,0.5)(1,2.5)}
   {\psset{linewidth=0.05}\psline(2,0.5)(2,2.5)}
   {\psset{linewidth=0.05}\psline(3,0.5)(3,2.5)}
   {\psset{linewidth=0.05}\psline(4,0.5)(4,2.5)}
   {\psset{linewidth=0.05}\psline(5,0.5)(5,2.5)}
   {\psset{linewidth=0.2}\psline{->}(2.5,3.5)(2.5,2.5)}
   \rput(2.5,2){{$0$}}
   \rput(2.5,1){{$\cdot$}}
   \rput(3.5,2){{$\bullet$}}
   \rput(3.5,1){{$1$}}
   \rput(7,1.5){{$\longrightarrow$}}
   {\psset{linewidth=0.05}\psline(8.5,0.5)(13.5,0.5)}
   {\psset{linewidth=0.05}\psline(8.5,1.5)(13.5,1.5)}
   {\psset{linewidth=0.05}\psline(8.5,2.5)(13.5,2.5)}
   {\psset{linewidth=0.05}\psline(9,0.5)(9,2.5)}
   {\psset{linewidth=0.05}\psline(10,0.5)(10,2.5)}
   {\psset{linewidth=0.05}\psline(11,0.5)(11,2.5)}
   {\psset{linewidth=0.05}\psline(12,0.5)(12,2.5)}
   {\psset{linewidth=0.05}\psline(13,0.5)(13,2.5)}
   {\psset{linewidth=0.2}\psline{->}(11.5,3.5)(11.5,2.5)}
   \rput(10.5,2){{$0$}}
   \rput(10.5,1){{$\cdot$}}
   \rput(11.5,2){{$\bullet$}}
   \rput(11.5,1){{$1$}}
   \rput(15,1.5){{$\longrightarrow$}}
   {\psset{linewidth=0.05}\psline(16.5,0.5)(21.5,0.5)}
   {\psset{linewidth=0.05}\psline(16.5,1.5)(21.5,1.5)}
   {\psset{linewidth=0.05}\psline(16.5,2.5)(21.5,2.5)}
   {\psset{linewidth=0.05}\psline(17,0.5)(17,2.5)}
   {\psset{linewidth=0.05}\psline(18,0.5)(18,2.5)}
   {\psset{linewidth=0.05}\psline(19,0.5)(19,2.5)}
   {\psset{linewidth=0.05}\psline(20,0.5)(20,2.5)}
   {\psset{linewidth=0.05}\psline(21,0.5)(21,2.5)}
   {\psset{linewidth=0.2}\psline{->}(18.5,3.5)(18.5,2.5)}
   \rput(18.5,2){{$0$}}
   \rput(18.5,1){{$\cdot$}}
   \rput(19.5,2){{$\bullet$}}
   \rput(19.5,1){{$\#$}}
   \rput(23,1.5){{$\longrightarrow$}}
   {\psset{linewidth=0.05}\psline(24.5,0.5)(29.5,0.5)}
   {\psset{linewidth=0.05}\psline(24.5,1.5)(29.5,1.5)}
   {\psset{linewidth=0.05}\psline(24.5,2.5)(29.5,2.5)}
   {\psset{linewidth=0.05}\psline(25,0.5)(25,2.5)}
   {\psset{linewidth=0.05}\psline(26,0.5)(26,2.5)}
   {\psset{linewidth=0.05}\psline(27,0.5)(27,2.5)}
   {\psset{linewidth=0.05}\psline(28,0.5)(28,2.5)}
   {\psset{linewidth=0.05}\psline(29,0.5)(29,2.5)}
   {\psset{linewidth=0.2}\psline{->}(25.5,3.5)(25.5,2.5)}
   \rput(26.5,2){{$1$}}
   \rput(26.5,1){{$\cdot$}}
   \rput(27.5,2){{$\bullet$}}
   \rput(27.5,1){{$\#$}}
\end{pspicture}
\caption{modified Turing machine}
\label{AbbModifiedTuringMachine}
\end{center}
\end{figure}

We decompose this step into three steps for $M'$:
\begin{eqnarray*}
   (q_5,1),(0,\cdot)&\stackrel{\delta'}\mapsto& (q_5,2),(0,\cdot),R \\
   (q_5,2),(\bullet,1)&\stackrel{\delta'}\mapsto& (q_5,3),(\bullet,\#),L \\
   (q_5,3),(0,\cdot)&\stackrel{\delta'}\mapsto& (q_6,1),(1,\cdot),L
\end{eqnarray*}
Here, $\cdot$ and $\bullet$ denote arbitrary symbols (zero, one, or blank).
The succession of steps that $M'$ performs with that transition function is
depicted in Figure~\ref{AbbModifiedTuringMachine}.

If the computations of $T_k$ are followed by the modified QTM $M'$,
then the output of the resulting QTM will thus be $M(k,\sigma)$.
\qed

\section{Halting and Universality of QTMs}
\label{SecHaltingUniv}
There has been a vivid discussion in the literature on the question when we
can consider a QTM as having {\em halted} on some input and how this is compatible
with unitary time evolution, see e.g. \cite{Myers, LindenPopescu, OzawaNondemolition, Shi, Miyadera}.
We will not get too deep into this discussion, but rather
analyze in detail the simple definition for halting by Bernstein and Vazirani \cite{BernsteinVazirani},
which we also use in this thesis, as specified in Equation~(\ref{EqHalting}). We argue below that
this definition is useful and natural, at least for the purpose to study quantum Kolmogorov complexity.

Note that whatever definition of ``halting'' we choose for a QTM, there is one problem which is
unavoidable in principle, originating from quantum theory itself. Suppose we are given some
classical string $s\in\s$, and we want to find out whether $s$ is halting for a given classical TM $T$ or not,
i.e. if $T$ halts on input $s$ or not.\footnote{In this discussion as well as in the remainder of
this thesis, we call some bit or qubit string $s$ halting for a TM or QTM $M$, if $M$ halts
on input $s$.} Then, we can always input $s$ into the TM $T$, and observe $T$'s computation for a long time.
Once we observe halting of $T$, we know for sure that $s$ is halting, of course.
If we have waited for a very long time and have not observed halting of $T$, we may believe
that $s$ is non-halting, although we can never be sure. Yet, if $T$ is a very simple TM for which
we can predict the time evolution completely, then we may find a proof
that $s$ is non-halting for $T$.

If we define some notion of ``halting'' for a QTM and qubit strings, this means that
we split the space of qubit strings into two parts: the halting qubit strings $H$ and
the non-halting qubit strings $N$.
\[
   \hr_\s=H\cup N\qquad\mbox{and}\qquad H\cap N=\emptyset.
\]
It follows immediately that $H$ and $N$ cannot be orthogonal, i.e.
\[
   H\not\perp N.
\]
Thus, if we have some unknown\footnote{``Unknown'' here means that we do not have a classical description of
$|\psi\rangle$, e.g. we do not know exactly how the state was created, and thus cannot obtain any copy of $|\psi\rangle$.}
quantum state $|\psi\rangle$,
and we are given the description of some QTM $M$,
then it is unavoidable that at least one of the following two problems occurs:
\begin{itemize}
\item[(a)] It may be true that $|\psi\rangle$ is halting for $M$, but we cannot find out with certainty by any possible
measurement that this is true.
\item[(b)] It may be true that $|\psi\rangle$ is non-halting for $M$, but we cannot prove this with certainty by any
possible measurement, even if $M$ is so simple that we can completely predict its time evolution.
\end{itemize}
It is impossible to get rid of both problems at once, but the definition of halting in this thesis
avoids problem (a), i.e. in principle, one can find out by measurement with certainty if
some input is halting for a QTM. Recall from Subsection~\ref{SubsecQTMs} how we have defined that a QTM $M$ halts on some
input $|\psi\rangle$ at time $t$: according to Equation~(\ref{EqHalting}), we demand that
\[
   \langle q_f|M_{\rm\bf C}^t(|\psi\rangle\langle\psi|)|q_f\rangle=1 \quad\mbox{ and }
   \quad\langle q_f|M_{\rm\bf C}^{t'}(|\psi\rangle\langle\psi|)|q_f\rangle=0\quad
   \mbox{for every } t'<t.
\]
Thus, given some unknown quantum state $|\psi\rangle$, {\em if} it is halting,
then we can find out for sure that it is, at least in principle, by supplying it as input to $M$
and periodically observing the control state. The aforementioned halting conditions guarantee
that projective measurements with respect to the projectors $|q_f\rangle\langle q_f|$ and
$\idn-|q_f\rangle\langle q_f|$ do not spoil the computation.

As the control state $M_{\mathbf{C}}^t(|\psi\rangle\langle\psi|)={\rm Tr}_{\mathbf{IOH}}\left(
U_M^t |\psi\rangle\langle\psi|(U_M^*)^t\right)$ is, in general, some mixed state on the control's Hilbert
space $\hr_{\mathbf{C}}$, the overlap with the final state $|q_f\rangle$ will generally be
some arbitrary number between zero and one. Hence, for most input qubit strings $|\psi\rangle$,
there will be no time $t\in\N$ such that the aforementioned halting conditions are satisfied. We call
those qubit strings {\em non-halting} in accordance with the discussion above, and otherwise
{\em $t$-halting}, where $t$ is the corresponding halting time.

In Subsection~\ref{SubsecHaltingSubspaces}, we analyze the resulting geometric structure
of the halting input qubit strings. We show that inputs $|\psi\rangle\in\cn$ with
some fixed length $n$ that make the QTM $M$ halt after $t$ steps form a linear
subspace $\hr_M^{(n)}(t)\subset\cn$. Moreover,
inputs with different halting times are mutually orthogonal, i.e.
$\hr_M^{(n)}(t)\perp \hr_M^{(n)}(t')$ if $t\neq t'$. According to the halting conditions given
above, this is almost obvious: Superpositions of $t$-halting inputs are again $t$-halting,
and inputs with different halting times can be perfectly distinguished, just
by observing their halting time.

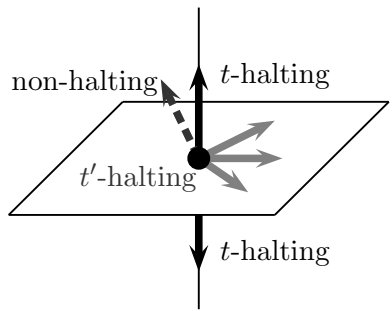
\begin{figure}[!hbt]
\psset{unit=0.5cm}
\begin{center}
\begin{pspicture}(0,0)(10,8.5)
   {\psset{linewidth=0.05}\psline(0,3)(7,3)}
   {\psset{linewidth=0.05}\psline(0,3)(3,6)}
   {\psset{linewidth=0.05}\psline(3,6)(10,6)}
   {\psset{linewidth=0.05}\psline(7,3)(10,6)}
   {\psset{linewidth=0.05}\psline(5,4.5)(5,8.5)}
   {\psset{linewidth=0.05}\psline(5,3)(5,0.5)}
   {\psset{linewidth=0.2}\psline{->}(5,4.5)(5,7)}
   {\psset{linewidth=0.2}\psline{->}(5,3)(5,1.5)}
   \rput(7,6.7){$t$-halting}
   \rput(7,2){$t$-halting}
   \rput(2,6.5){non-halting}
   {\psset{linewidth=0.2,linestyle=dashed,linecolor=darkgray}\psline{->}(5,4.5)(4,6.6)}
   {\psset{linewidth=0.2,linecolor=gray}\psline{->}(5,4.5)(7,5.5)}
   {\psset{linewidth=0.2,linecolor=gray}\psline{->}(5,4.5)(7.2,4.5)}
   {\psset{linewidth=0.2,linecolor=gray}\psline{->}(5,4.5)(6.3,3.6)}
   \rput(3.4,4){\textcolor{darkgray}{$t'$-halting}}
   \pscircle[linewidth=0.3](5,4.5){0.3}
\end{pspicture}
\caption{Mutually Orthogonal Halting Spaces}
\label{AbbHaltingSpaces}
\end{center}
\end{figure}

In Figure~\ref{AbbHaltingSpaces}, a geometrical
picture of the halting space structure is shown: The whole space $\R^3$ represents the space
of inputs of some fixed length $n$, i.e. $\cn$, while the plane and
the straight line represent two different halting spaces $\hr_M^{(n)}(t')$ and $\hr_M^{(n)}(t)$.
Every vector within these subspaces is perfectly halting, while every vector ``in between''
is non-halting and not considered a useful input for the QTM $M$.

At first, it seems that the halting conditions given above are far too restrictive. Don't we
loose a lot by dismissing every input which does not satisfy those conditions perfectly,
but, say, only approximately up to some small $\eps$? To see that it is not that bad, note that
\begin{itemize}
\item most (if not all) of the well-known quantum algorithms, like the quantum Fourier transform
or Shor's algorithm, have classically controlled halting. That is, the halting time is known
in advance, and can be controlled by a classical subprogram.
\item in Section~\ref{SecHaltingStability}, we show that every input that is {\em almost}
halting can be modified by adding at most a constant number of qubits to halt {\em perfectly},
i.e. to satisfy the aforementioned halting conditions. This can be interpreted as some kind
of ``stability result'', showing that the halting conditions are not ``unphysical'', but
have some kind of built-in error tolerance that was not expected from the beginning.
\end{itemize}

Moreover, this definition of halting is very useful. Given two QTMs $M_1$ and $M_2$, it enables
us to construct a QTM $M$ which carries out the computations of $M_1$, {\em followed by the
computations of $M_2$}, just by redirecting the final state $|q_f\rangle$ of $M_1$ to
the starting state $|q_0\rangle$ of $M_2$ (see \cite[Dovetailing Lemma 4.2.6]{BernsteinVazirani}).
In addition, it follows from this definition that QTMs are {\em quantum operations} (cf. Lemma~\ref{LemmaQTMsAreOperations}),
which is a very useful and plausible property.

Even more important, at each single time
step, an outside observer can make a measurement of the control state, described by
the operators $|q_f\rangle\langle q_f|$ and $\idn-|q_f\rangle\langle q_f|$ (thus observing
the halting time), without
spoiling the computation, as long as the input $|\psi\rangle$ is halting. As soon as
halting is detected, the observer can extract the output quantum state from the output track (tape)
and use it for further quantum information processing.
This is true even if the halting time
is very large, which typically happens in the study of Kolmogorov complexity.

Finally, if we instead introduced some probabilistic notion of halting (say, we demanded
that we observe halting of the QTM $M$ at some time $t$ with some large probability $p<1$),
then it would not be so clear how to define quantum Kolmogorov complexity correctly. Namely
if the halting probability is much less than one, it seems necessary to introduce some kind
of ``penalty term'' into the definition of quantum Kolmogorov complexity: there should
be some trade-off between program length and halting accuracy, and it is not so clear
what the correct trade-off should be. For example, what is the complexity of a qubit string
that has a program of length 100 which halts with probability $0.6$, and another program
of length 120 which halts with probability $0.9$? The definition of halting that we use
in this thesis avoids such questions.

\subsection{Different Notions of Universality for QTMs}
\label{SubsecUniversalities}
Bernstein and Vazirani~\cite{BernsteinVazirani} have shown that there exists a universal QTM
(UQTM) $\mathcal{U}$. It is important to understand what exactly they mean by ``universal''.
According to \cite[Thm. 7.0.2]{BernsteinVazirani}, this UQTM $\mathcal U$ has the property
that for every QTM $M$
there is some classical bit string $s_M\in\{0,1\}^*$ (containing a description of
the QTM $M$) such that
\begin{equation}
   \left\|\mathcal{U}(s_M,T,\delta,|\psi\rangle)-\mathcal{R}\circ M_{\mathbf O}^T(|\psi\rangle)\right\|_{\rm Tr}<\delta
   \label{EqWeakUniversality}
\end{equation}
for every input $|\psi\rangle$, accuracy $\delta>0$ and number of time steps $T\in\N$.

This means that the UQTM $\mathcal U$ simulates every other QTM $M$ within any desired accuracy
and outputs an approximation of the output track content of $M$ and halts,
as long as the number of time steps $T$ is given as input in advance.

Since the purpose of
Bernstein and Vazirani's work was to study the computational complexity of QTMs,
it was a reasonable assumption that the halting time $T$ is known in advance (and not too large)
and can be specified as additional input. The most important point for them was not to have short inputs, but
to prove that the simulation of $M$ by $\mathcal U$ is efficient, i.e. has only polynomial slowdown.

The situation is different if one is interested in studying quantum Kolmogorov complexity
instead. It will be explained in Subsection~\ref{SubsecInvariance}
below that the universality notion (\ref{EqWeakUniversality}) is not enough for proving
the important invariance property of quantum Kolmogorov complexity, which says that quantum
Kolmogorov complexity depends on the choice of the universal QTM only up to an additive constant.

To prove the invariance property, one needs a generalization 
of (\ref{EqWeakUniversality}), where the requirement to have the running time $T$ as
additional input is dropped. We show below in Subsection~\ref{SubsecStronglyUniversalQTMs}
that there exists a UQTM $\mathfrak U$ that satisfies such a generalized
universality property, i.e. that simulates every other QTM until that other
QTM has halted, without knowing that halting time in advance, and then halts itself.

Why is that so difficult to prove? At first, it seems that one can just program
the UQTM $\mathcal U$  mentioned in (\ref{EqWeakUniversality})
to simulate the other QTM $M$ for $T=1,2,3,\ldots$ time steps, and,
after every time step, to check if the simulation of $M$ has halted or not. If it has halted,
then $\mathcal U$ halts itself and prints out the output of $M$, otherwise it continues.

This approach works for classical TMs, but for QTMs, there is one problem: in general,
the UQTM $\mathcal U$ can simulate $M$ only approximately. The reason is the same as for the circuit
model, i.e. the set of basic unitary transformations that $\mathcal U$ can apply on its
tape may be algebraically independent from that of $M$, making a perfect simulation in principle
impossible. But if the simulation is only approximate, then the control state
of $M$ will also be simulated only approximately, which will force
$\mathcal U$ to halt only approximately. Thus, the restrictive halting conditions given
above in Equation~(\ref{EqHalting}) will inevitably be violated,
and the computation of $\mathcal U$ will be treated as invalid and be dismissed by definition.

This is a severe problem that cannot be circumvented easily. Many ideas for simple solutions
must fail, for example the idea to let $\mathcal U$ compute an upper bound on the halting time
$T$ of all inputs for $M$ of some length $n$ and just to proceed for $T$ time steps: upper bounds on
halting times are not computable. Another idea is that the computation of $\mathcal U$ should
somehow consist of a classical part that controls the computation and a quantum part that
does the unitary transformations on the data. But this idea is difficult to formalize.
Even for classical TMs, there is no general way to split the computation into ``program'' and
``data'' except for special cases, and for QTMs, by definition, global unitary time evolution
can entangle every part of a QTM with every other part.

Our proof idea rests instead on the observation that every {\em input} for a QTM which is
halting can be decomposed
into a classical and a quantum part, which is related to the mutual orthogonality of the
halting spaces. The proof is given in Section~\ref{SecConstruction}.
Note that we have already published the contents of this and the following section
in \cite{StronglyUniversal}.

\subsection{Quantum Complexity and its Supposed Invariance}
\label{SubsecInvariance}
As already explained in the introduction,
the classical Kolmogorov complexity $C_M(s)$ of a finite bit string
$s\in\{0,1\}^*$ is defined as the minimal length of any computer program $p$ that,
given as input into a TM $M$, outputs the string $s$ and makes $M$ halt:
\[
   C_M(s):=\min\left\{\ell(p)\,\,|\,\,M(p)=s\right\}.
\]
For this quantity, running times are not important; all that matters is the input length.
There is a crucial result that is the basis for the whole theory of Kolmogorov complexity
(see \cite{Vitanyibook}). Basically, it states that the choice of the computer $M$
is not important as long as $M$ is universal; choosing a different
universal computer will alter the complexity only up to some
additive constant. More specifically, there exists a universal computer $U$ such that
for every computer $M$ there is a constant $c_M\in\N$ such that
\begin{equation}
   C_U(s)\leq C_M(s)+c_M\qquad\mbox{for every }s\in\{0,1\}^*.
   \label{eqInvClass}
\end{equation}
This so-called ``invariance property'' follows easily from the following fact:
there exists a computer $U$ such
that for every computer $M$ and every input $s\in\{0,1\}^*$ there is an input $s'\in\{0,1\}^*$
such that $U(s')=M(s)$ and $\ell(s')\leq \ell(s)+c_M$, where $c_M\in\N$ is
a constant depending only on $M$. In short, there is a computer $U$ that produces every
output that is produced by any other computer,
while the length of the corresponding input blows up only by a constant summand.
One can think of the bit string $s'$ as consisting of the original bit string $s$
and of a description of the computer $M$ (of length $c_M$).

As the invariance property is so important for the theory of classical Kolmogorov complexity,
a study of quantum Kolmogorov complexity naturally asks for a quantum analogue of this property.
The notion of quantum complexity that we shall define in Chapter~\ref{ChapterQuantumKolmogorovComplexity}
is a slight modification of the definition given by Berthiaume et al. in \cite{Berthiaume}.
A closely related quantity has been considered recently by Rogers and Vedral~\cite{Rogers}.

In both cases \cite{Berthiaume} and \cite{Rogers}, it is claimed that quantum
Kolmogorov complexity is invariant up to an additive constant similar
to (\ref{eqInvClass}). Nevertheless, in \cite{Rogers} no proof is given and the proof 
in \cite{Berthiaume} is incomplete: in that proof, it is stated that
the existence of a universal QTM $\mathcal U$ in the sense
of Bernstein and Vazirani (see Equation~(\ref{EqWeakUniversality}))
makes it possible to mimic the classical proof and to conclude that the UQTM $\mathcal U$
outputs all that every other QTM outputs, implying invariance of quantum Kolmogorov complexity.

But this conclusion cannot be drawn so easily, because (\ref{EqWeakUniversality}) demands that the halting time $T$ is specified
as additional input, which can enlarge the input length dramatically, if $T$ is very large
(which typically happens in the study of Kolmogorov complexity).

As explained
above in Subsection~\ref{SubsecUniversalities}, it is not so easy to get rid of the halting time.
The main reason is that the UQTM $\mathcal U$ can simulate other QTMs only approximately.
Thus, it will also simulate the control state and the signaling of halting only approximately,
and cannot just ``halt whenever the simulation has halted'', because then, it will
violate the restrictive halting conditions given in Equation~(\ref{EqHalting}).
As we have chosen this definition of halting for good reasons (cf. the discussion at the beginning
of Section~\ref{SecHaltingUniv} above), we do not want to drop it. So what can we do?

The only way out is to give a proof that despite our restrictive definition of halting,
there still exists some UQTM $\mathfrak U$ that simulates every other QTM until that
other QTM has halted, even if it does not know the halting time in advance.
Yet, it is not enough to rely on the result (\ref{EqWeakUniversality}) by Bernstein and
Vazirani; we need another good idea how to do it. We describe our proof idea
in the next subsection, while the proof will be given below in Section~\ref{SecConstruction}.

\subsection{Strongly Universal QTMs}
\label{SubsecStronglyUniversalQTMs}
We are going to prove in Section~\ref{SecConstruction} below that there is
``strongly universal'' QTM that simulates
every other QTM until the other QTM has halted and then halts itself. Note that
the halting state is attained by $\mathfrak U$ {\em exactly} (with probability one) in accordance
with the strict halting definition given in Equation~(\ref{EqHalting}).
\begin{ftheorem}[Strongly Universal Quantum Turing Machine]
\label{MainTheorem1}
\lineclear
There is a fixed-length quantum Turing machine $\mathfrak{U}$ such that
for every QTM $M$ and
every qubit string $\sigma$
for which $M(\sigma)$ is defined, there is a qubit string $\sigma_M$
such that
\[
   \left\| \mathfrak{U}\,(\delta,\sigma_M)-M(\sigma)\right\|_{\rm Tr} <\delta
\]
for every $\delta\in\mathbb{Q}^+$, where the length of $\sigma_M$ is bounded
by $\ell(\sigma_M)\leq\ell(\sigma)+c_M$, and $c_M\in\N$ is a constant depending only
on $M$.
\end{ftheorem}

Note that $\sigma_M$ does not depend on $\delta$.

In Chapter~\ref{ChapterQuantumKolmogorovComplexity}, we study several notions of quantum
Kolmogorov complexity at once. To prove invariance for every single notion, we shall
also prove the following slight modifications of Theorem~\ref{MainTheorem1}:

\begin{fproposition}[Parameter Strongly Universal QTM]
\label{PropTwoParameter}
\lineclear
There is a fixed-length quantum Turing machine $\mathfrak{U}$ with the property
of Theorem~\ref{MainTheorem1} that additionally satisfies the following:
For every QTM $M$ and
every qubit string $\sigma\in\mathcal{T}_1^+\left(\hr_\s\right)$,
there is a qubit string $\sigma_M\in\mathcal{T}_1^+\left(\hr_\s\right)$
such that
\[
   \left\|\mathfrak{U}\left(k,\sigma_M\right)-M\left(
   2 k,\sigma\right)\right\|_{\rm Tr}<\frac 1 {2k}\qquad
   \mbox{for every }k\in\N
\]
if $M(2k,\sigma)$ is defined for every $k\in\N$, where the length of $\sigma_M$ is bounded
by $\ell(\sigma_M)\leq\ell(\sigma)+c_M$, and $c_M\in\N$ is a constant depending only
on $M$.
\end{fproposition}

It may first seem that this Proposition~\ref{PropTwoParameter} is a simple corollary
of Theorem~\ref{MainTheorem1}, but this is not true. The problem is that the computation
of $M(2k,\sigma)$ may take a different number of time steps $t_k$ for different $k$
(typically, $t_k\to\infty$ as $k\to\infty$). Just using the result of Theorem~\ref{MainTheorem1}
would give a corresponding qubit string $\sigma_M$ that depends on $k$, but here we demand
that the qubit string $\sigma_M$ is the {\em same} for every $k$, which will be important
for proving Theorem~\ref{MainTheorem2}.

We also sketch some proof idea for the following conjecture:

\begin{conjecture}[Average-Length Strongly Universal QTM]
\label{TheAvLengthUQTM}
\lineclear
There is a prefix QTM $\mathfrak V$ such that for every prefix QTM $M$ and every
qubit string $\sigma$ for which $M(\sigma)$ is defined, there is a qubit string
$\sigma_M$ such that
\[
   \|\mathfrak{V}(\delta,\sigma_M)-M(\sigma)\|_{\rm Tr}<\delta
\]
for every $\delta\in\mathbb{Q}^+$, where the average length of $\sigma_M$ is bounded
by $\bar\ell(\sigma_M)\leq\bar\ell(\sigma)+c_M$, and $c_M\in\N$ is a constant depending
only on $M$.
\end{conjecture}
We define the notion of a prefix QTM in Definition~\ref{DefPrefixQTM}. The reason why
we give a proof idea for this conjecture is that it explains why it seems that we
need the condition that $M$ has to be prefix-free. This supports the point of view that
average length $\bar\ell$ is intimately connected with the notion of prefix-free qubit strings.

We give a full proof of Theorem~\ref{MainTheorem1},
describing in every single detail how the corresponding UQTM $\mathfrak U$
works, below in Section~\ref{SecConstruction}. This involves many analytic estimates
to prove that certain numerical approximations made by $\mathfrak U$ are accurate enough.

Since the technical details are so similar, we will only sketch the proof
of Proposition~\ref{PropTwoParameter} in Section~\ref{SecConstruction}.
Although we have a proof sketch of Conjecture~\ref{TheAvLengthUQTM}, we do not think that
we have settled it completely (in contrast to Proposition~\ref{PropTwoParameter}) because
it depends heavily on the property that the domain of definition of the QTM is prefix-free, and it is not clear
that this fact survives the numerical approximations done by the QTM $\mathfrak V$.
In the remainder of this subsection, we describe the ideas of the proof of Theorem~\ref{MainTheorem1}.

The proof of Theorem~\ref{MainTheorem1} relies on the observation about the mutual orthogonality
of the halting spaces, as explained above at the beginning of Section~\ref{SecHaltingUniv}. Fix some QTM $M$, and
denote the set of vectors $|\psi\rangle\in\cn$
which cause $M$ to halt at time $t$ by $\hr_M^{(n)}(t)$. If $|\varphi\rangle\in
\cn$ is any halting input for $M$, then we can decompose $|\varphi\rangle$
in some sense into a classical and a quantum part. Namely, the information contained in $|\varphi\rangle$
can be split into a
\begin{itemize}
\item classical part: The vector $|\varphi\rangle$ is an element of
{\em which} of the subspaces $\hr_M^{(n)}(t)$?
\item quantum part: Given the halting time $\tau$ of $|\varphi\rangle$,
then {\em where} in the corresponding subspace $\hr_M^{(n)}(\tau)$
is $|\varphi\rangle$ situated?
\end{itemize}
Our goal is to find a QTM $\mathfrak U$ and an encoding
$|\tilde\varphi\rangle\in\hr_{n+1}$
of $|\varphi\rangle$
which is only one qubit longer and which makes the (cleverly programmed) QTM $\mathfrak U$ output a good
approximation of $M(|\varphi\rangle)$. First, we extract the quantum part out of $|\varphi\rangle$.
While $\dim \cn=2^n$, the halting space $\hr_M^{(n)}(\tau)$ that
contains $|\varphi\rangle$ is only a subspace and might have much smaller dimension $d<2^n$.
This means that we need less than $n$ qubits to describe the state $|\varphi\rangle$; indeed,
$\lceil \log d\rceil$ qubits are sufficient. In other words, there is some kind of ``standard
compression map'' $\mathcal C$ that maps every vector $|\psi\rangle\in\hr_M^{(n)}(\tau)$ into
the $\lceil \log d\rceil$-qubit-space $\left(\C^2\right)^{\otimes \lceil \log d\rceil}$.
Thus, the qubit string ${\mathcal C} |\varphi\rangle$ of length $\lceil \log d\rceil\leq n$
can be considered as the ``quantum part'' of $|\varphi\rangle$.

So how can the classical part of $|\varphi\rangle$ be encoded into a short classical binary string?
Our task is to specify
what halting space $\hr_M^{(n)}(\tau)$ corresponds to $|\varphi\rangle$. Unfortunately, it
is not possible to encode the halting time $\tau$ directly, since $\tau$ might be huge and
may not have a short description. Instead, we can encode the {\em halting number}.
Define the halting time sequence $\{t_i\}_{i=1}^N$ as the set of all integers
$t\in\N$ such that $\dim \hr_M^{(n)}(t)\geq 1$, ordered such that $t_i<t_{i+1}$
for every $i$, that is, the set of all halting times that can occur on inputs of length $n$.
Thus, there must be some $i\in\N$ such that $\tau=t_i$, and $i$ can be called the
halting number of $|\varphi\rangle$. Now, we assign code words $c_i$ to the halting
numbers $i$, that is, we construct a prefix code $\{c_i\}_{i=1}^N\subset\{0,1\}^*$.
We want the code words to be short; we claim that we can always choose the lengths as
\[
   \ell(c_i)=n+1-\lceil \log \dim \hr_M^{(n)}(t_i)\rceil\,\,.
\]
This can be verified by checking the Kraft inequality:
\begin{eqnarray*}
   \sum_{i=1}^N 2^{-\ell(c_i)}&=&2^{-n}\sum_{i=1}^N 2^{\lceil \log \dim \hr_M^{(n)}(t_i)\rceil
   -1}\\
   &\leq& 2^{-n} \sum_{i=1}^n \dim \hr_M^{(n)}(t_i)
   \leq 2^{-n} \dim \cn \\
   &\leq& 1,
\end{eqnarray*}
since the halting spaces are mutually orthogonal.

Putting classical and quantum part of $|\varphi\rangle$ together, we get
\[
   |\tilde\varphi\rangle:=c_i\otimes {\mathcal C} |\varphi\rangle\,\,,
\]
where $i$ is the halting number of $|\varphi\rangle$.
Thus, the length of $|\tilde\varphi\rangle$ is exactly $n+1$.

Let $s_M$ be a self-delimiting description of the QTM $M$. The idea is to construct
a QTM $\mathfrak U$ that, on input $s_M\otimes |\tilde\varphi\rangle$, proceeds as follows:
\begin{itemize}
\item By {\em classical} simulation of $M$, it computes descriptions of the halting spaces
$\hr_M^{(n)}(1), \hr_M^{(n)}(2), \hr_M^{(n)}(3),\ldots$ and the corresponding code words
$c_1,c_2,c_3,\ldots$ one after the other, until at step $\tau$, it finds the code word
$c_i$ that equals the code word in the input.
\item Afterwards, it applies a (quantum) decompression map to approximately reconstruct $|\varphi\rangle$
from ${\mathcal C}|\varphi\rangle$.
\item Finally, it simulates (quantum) for $\tau$ time steps the time evolution of $M$ on
input $|\varphi\rangle$ and then halts, whatever happens with the simulation.
\end{itemize}
Such a QTM $\mathfrak U$ will have the strong universality property as stated in
Theorem~\ref{MainTheorem1}. Unfortunately, there are many difficulties that
have to be overcome by the proof in Section~\ref{SecConstruction}:
\begin{itemize}
\item Also classically, QTMs can only be simulated approximately. Thus, it is for example impossible
for $\mathfrak U$ to decide by classical simulation whether the QTM $M$ halts on some input
$|\varphi\rangle$ perfectly or only approximately at some time $t$. Thus, we have to define
certain $\delta$-approximate halting spaces $\hr_M^{(n,\delta)}(t)$ and prove a lot of lemmas
with nasty inequalities.
\item According to the statement of Theorem~\ref{MainTheorem1}, we have to consider mixed
inputs and outputs, too.
\item The aforementioned prefix code must have the property that one code word can be constructed
after the other (since the sequence of all halting times is not computable), see Lemma~\ref{LemBlindCoding}.
\end{itemize}
We show that all these difficulties (and some more) can be overcome, and the idea outlined
above can be converted to a formal proof of Theorem~\ref{MainTheorem1}
which we give in full detail in Section~\ref{SecConstruction}.

\section{Construction of a Strongly Universal QTM}
\label{SecConstruction}
The aim of this section is to give a full proof of Theorem~\ref{MainTheorem1}. This will be done in
several steps: In Subsection~\ref{SubsecHaltingSubspaces}, we show that the domain of definition of
a QTM is given by mutually orthogonal halting spaces. Afterwards, we show in Subsection~\ref{SubsecApproxHalting}
that these subspaces have computable approximations, and we prove several properties of the corresponding
``approximate halting spaces''. In Subsection~\ref{SubsecCompDecomp}, we explain how the classical and
quantum part of some input can be coded and decoded by the UQTM $\mathfrak U$. Finally, in
Subsection~\ref{SubsecProofMainTheorems}, we put all these partial results together to construct
the strongly universal QTM $\mathfrak U$ mentioned in Theorem~\ref{MainTheorem1}.

\subsection{Halting Subspaces and their Orthogonality}
\label{SubsecHaltingSubspaces}
As already explained at the beginning of Section~\ref{SecHaltingUniv}, restricting
to pure input qubit strings $|\psi\rangle\in\hr_n$ of some fixed length $\ell(|\psi\rangle)=n$,
the vectors with equal halting time $t$ form a linear subspace of $\hr_n$. Moreover, inputs
with different halting times are mutually orthogonal, as depicted in Figure~\ref{AbbHaltingSpaces}.
We will now use the formalism for QTMs introduced in Subsection~\ref{SubsecQTMs} to give a formal proof
of these statements. We use the subscripts $\mathbf C$, $\mathbf I$, $\mathbf O$ and $\mathbf H$
to indicate to what part of the tensor product Hilbert space a vector belongs.

\begin{definition}[Halting Qubit Strings]
\label{DefHaltingQubitStrings}
\lineclear
Let $\sigma\in\mathcal{T}_1^+(\hr_\s)$ be a qubit string and $M$ a quantum Turing machine.
Then, $\sigma$ is called {\em $t$-halting (for $M$)}, if
$M$ halts on input $\sigma$ at time $t\in\N$. We define the {\em halting sets} and {\em halting subspaces}
\begin{eqnarray*}
   H_M(t)&:=&\{|\psi\rangle\in\hr_\s\,\, |\,\, |\psi\rangle\langle\psi|\mbox{ is }
   t\mbox{-halting for }M\},\\
   \hr_M(t)&:=&\{\alpha |\psi\rangle\enspace|\enspace |\psi\rangle\in H_M(t),
   \alpha\in\R\}, \\
   H_M^{(n)}(t)&:=&H_M(t)\cap \hr_n,\qquad \hr_M^{(n)}(t):=\hr_M(t)\cap \hr_n.
\end{eqnarray*}
\end{definition}
Note that the only difference between $H_M^{(n)}(t)$ and $\hr_M^{(n)}(t)$ is that the latter
set contains non-normalized vectors. It will be shown below that $\hr_M^{(n)}(t)$ is indeed
a linear subspace.

\begin{ftheorem}[Halting Subspaces]
\lineclear
For every QTM $M$, $n\in\N_0$ and $t\in\N$, the sets $\hr_M(t)$ and $\hr_M^{(n)}(t)$ are linear subspaces of
$\hr_\s$ resp. $\hr_n$, and
\[
   \hr_M^{(n)}(t) \perp \hr_M^{(n)}(t')\quad\mbox{and}\quad\hr_M(t) \perp \hr_M(t') \quad\mbox{for every }t\neq t'.
\]
\end{ftheorem}
{\bf Proof.} Let $|\varphi\rangle,|\psi\rangle\in H_M(t)$. The property that $|\varphi\rangle$
is $t$-halting is equivalent to the statement that there are states $|\Phi_q^{t'}\rangle
\in\hr_{\mathbf I}\otimes \hr_{\mathbf O}\otimes\hr_{\mathbf H}$ and coefficients $c_q^{t'}\in\C$
for every $t'\leq t$ and $q\in Q$ such that
\begin{eqnarray}
   V_M^t\left(\strut|\varphi\rangle_{\mathbf I}\otimes |\Psi_0\rangle\right)
   &=&|q_f\rangle_{\mathbf C} \otimes |\Phi_{q_f}^t\rangle\,\,,
   \label{Condition1}\\
   V_M^{t'}\left(\strut|\varphi\rangle_{\mathbf I}\otimes |\Psi_0\rangle\right)
   &=&\sum_{q \neq q_f}c_q^{t'}
   |q\rangle_{\mathbf C} \otimes |\Phi_q^{t'}\rangle\quad\mbox{for every } t'<t,
   \label{Condition2}
\end{eqnarray}
where $V_M$ is the unitary time evolution operator for the QTM $M$ as a whole, and
$|\Psi_0\rangle=|q_0\rangle_{\mathbf C}\otimes |\#\rangle_{\mathbf O}
\otimes |0\rangle_{\mathbf H}$ denotes the initial state of the control, output track and head.
Note that $|\Psi_0\rangle$ does not depend on the input qubit string (in this case $|\varphi\rangle$).

An analogous equation holds for $|\psi\rangle$, since it is also $t$-halting by assumption.
Consider a normalized superposition $\alpha |\varphi\rangle+\beta |\psi\rangle\in\hr_\s$:
\begin{eqnarray*}
   V_M^t\left(\strut\right.\left(\alpha|\varphi\rangle_{\mathbf I}+\beta|\psi\rangle_{\mathbf I}\right)\otimes |\Psi_0\rangle
   \left.\strut\right)
   &=&\alpha V_M^t |\varphi\rangle_{\mathbf I}\otimes |\Psi_0\rangle 
    +\beta V_M^t |\psi\rangle_{\mathbf I}\otimes |\Psi_0\rangle\\
   &=& \alpha |q_f\rangle_{\mathbf C} \otimes |\Phi_{q_f}^t\rangle
   +\beta |q_f\rangle_{\mathbf C}\otimes |\tilde\Phi_{q_f}^t\rangle\\
   &=&|q_f\rangle_{\mathbf C} \otimes \left(
      \alpha |\Phi_{q_f}^t\rangle+\beta |\tilde\Phi_{q_f}^t\rangle
   \right).
\end{eqnarray*}
Thus, the superposition also satisfies condition (\ref{Condition1}), and, by a similar calculation,
condition (\ref{Condition2}). It follows that
$\alpha |\varphi\rangle+\beta |\psi\rangle$ must also be $t$-halting.
Hence, $\hr_M(t)$ is a linear subspace of $\hr_\s$. As the intersection of linear
subspaces is again a linear subspace, so must be $\hr_M^{(n)}(t)$.

Let now $|\varphi\rangle\in H_M(t)$ and $|\psi\rangle\in H_M(t')$ such that
$t<t'$. Again by Equations~(\ref{Condition1}) and (\ref{Condition2}), it holds
\begin{eqnarray*}
   \langle \varphi |\psi\rangle &=& 
   \left(\strut\,_{\mathbf I}\langle \varphi| \otimes \langle \Psi_0|\right)
    \left(V_M^t\right)^* V_M^t \left(\strut| \psi\rangle_{\mathbf I}\otimes
    |\Psi_0\rangle\right)\\
   &=&\sum_{Q\ni q \neq q_f} c_q^t
   \underbrace{\enspace_{\mathbf{C}}\langle q_f|q\rangle_{\mathbf C}}_0
   \cdot  \langle \Phi_{q_f}^t | \tilde \Phi_q^t \rangle=0\,\,.
\end{eqnarray*}
It follows that $\hr_M(t)\perp \hr_M(t')$, and similarly for $\hr_M^{(n)}(\cdot)\subset \hr_M(\cdot)$.
\qed
The physical interpretation of the preceding theorem is straightforward: by linearity
of the time evolution, superpositions of $t$-halting strings are again $t$-halting,
and strings with different halting times can be perfectly distinguished by observing
their halting time.

It is now clear what the domain of definition of a QTM looks like:
\begin{flemma}[Domain of Definition of a QTM]
\lineclear
If $M$ is a QTM, then its domain of definition is given by
\[
   {\rm dom}\,M=\bigcup_{t\in\N} \mathcal{T}_1^+\left(\strut\hr_M(t)\right),
\]
i.e. the set of density operators on the linear subspaces of pure $t$-halting qubit strings.
\end{flemma}
\proof Let $\sigma\in{\rm dom}\,M$ have spectral decomposition $\sigma=\sum_i \lambda_i |\psi_i\rangle\langle\psi_i|$,
with $\lambda_i>0$. Let $t$ be the halting time that corresponds to $\sigma$. Then,
\[
   \sum_i \lambda_i \langle q_f|M_{\mathbf{C}}^{t'}(|\psi_i\rangle\langle\psi_i|)|q_f\rangle=\left\{
      \begin{array}{cl}
         0 & \mbox{if }t'<t,\\
         1 & \mbox{if }t'=t.
      \end{array}
   \right.
\]
It follows that each element of this convex combination must itself satisfy this equation.
Thus, $|\psi_i\rangle\in\hr_M(t)$, and $\sigma$ is a density operator on $\hr_M(t)$.
\qed

In general, different inputs $\sigma$ have different halting times
$t$ and the corresponding outputs are essentially results of
different unitary transformations given by $U_M^t$, where $U_M$ denotes $M$'s time evolution operator.
However, the action of the partial map
$M$ on ${\rm dom}\,M$  may be extended to a valid quantum operation on $\mathcal{T}(\hr_\s)$: 
\begin{flemma}[QTMs are Quantum Operations]
\label{LemmaQTMsAreOperations}
\lineclear
For every QTM $M$  there is a quantum operation
$\mathcal{M}:\mathcal{T}(\hr_\Fock)\to\mathcal{T}(\hr_\Fock)$, 
such that for every $\sigma\in {\rm dom}\,M$
\[
   M(\sigma)=\mathcal{M}(\sigma).
\]
\end{flemma}
{\bf Proof. } Let $\mathcal{B}_t$ and
$\mathcal{B}_{\perp}$ be an orthonormal basis of $\hr_M(t)$, $t \in \N$, and the
orthogonal complement of $\bigoplus_{t \in \N}\hr_M(t)$
within $\hr_\Fock$, respectively. 
We add an ancilla Hilbert space $\hr_{\mathbf{A}}:=\ell^2(\N_0)$ to the QTM,
and define a linear operator $V_M:\hr_\Fock\to\hr_{QTM}\otimes \hr_{\mathbf{A}}$
by specifying its action on the orthonormal basis vectors $\cup_{t \in \N}\mathcal{B}_t\cup
\mathcal{B}_{\perp}$:
\begin{eqnarray}
V_M |b \rangle :=\left\{
  \begin{array}{rl}
\left(\strut U_M^t | b \rangle\right) \otimes |t\rangle&
 \textrm{ if } 
|b\rangle \in \mathcal{B}_t,
\\ 
|b\rangle \otimes |0\rangle&  \textrm{ if } |b\rangle \in \mathcal{B}_{\perp}.
\label{eqONB}
  \end{array}
\right. 
\end{eqnarray}
Since the right hand side of (\ref{eqONB}) is a set of orthonormal vectors in
$\hr_{QTM}\otimes \hr_{\mathbf{A}}$, the map $V_M$ is an isometry (i.e. $V_M^* V_M=\idn$).
Thus, the map $\sigma\mapsto V_M \sigma V_M^*$ is trace-preserving, completely
positive (see \cite{holevo,Kraus,Paulsen}). Its composition with the partial trace, given
by $\mathcal{M}(\sigma):= {\rm Tr}_{\mathbf{CHIA}}(V_M \sigma V_M^{*})$, is a quantum operation.
\qed

In the following, it will turn out that it is interesting to study prefix QTMs, i.e. QTMs which are in a certain
sense quantum generalizations of classical prefix Turing machines. A classical TM is called prefix if its domain
of definition is a prefix-free set. We can define a natural quantum generalization by calling a QTM {\em prefix}
if its domain of definition in the qubit strings is in a certain sense prefix-free, too. Following the lines of
Schumacher and Westmoreland~\cite{SchumacherWestmoreland}, who have defined
{\em prefix-free quantum codes}, leads us to Definition~\ref{DefPrefixQTM} below.

To state the definition, we fix some notation. If a classical string $s\in\s$ has length $\ell(s)>n$, then
the string $s_1^n$ is defined to consist of the first $n$ bits of $s$. Thus, $s_1^n$ is the prefix of $s$ of
length $\ell(s_1^n)=s$.

Similarly, we can define the prefix $\sigma_1^n$ of a qubit string $\sigma\in\mathcal{T}_1^+(\hr_\s)$ in a
simple way. First, we identify the qubit string $\sigma$ with the corresponding density operator on the QTM's output
tape Hilbert space $\sigma'\in\mathcal{T}_1^+( \hr_{\mathbf O})$, such that the string is ``written'' onto the blank tape,
starting in cell $0$, and ending in cell $\ell(\sigma)-1$, as the input for a QTM has been defined in
Subsection~\ref{SubsecQTMs}. Then, we define the prefix $(\sigma_1^n)'$ by the partial trace
\[
   (\sigma_1^n)':={\rm Tr}_{(-\infty,-1]\cup[n,\infty)}\sigma'\in\mathcal{T}_1^+\left((\C^{\{0,1,\#\}})^{\otimes n}\right).
\]
Let $t\in\{0,1,\#\}^n$ be any configuration which is {\em not} of the form $s\#\#\ldots\#$, where $s\in\s$ is
a binary string (for example, $t=0\#1$). Then it is easy to see that $\langle t|(\sigma_1^n)'|t\rangle=0$. Thus,
$(\sigma_1^n)'$ 
is a superposition and mixture of classical strings embedded on the tape, and can be identified with a corresponding
qubit string $\sigma_1^n\in\mathcal{T}_1^+(\hr_\s)$.

\begin{definition}[Prefix QTM]
\label{DefPrefixQTM}
\lineclear
A QTM $M$ is called {\em prefix} if for every pair of pure qubit strings $|\varphi\rangle\langle\varphi|$, $|\psi\rangle\langle\psi|\in{\rm dom}\,M$
with $\ell(|\varphi\rangle)>\ell(|\psi\rangle)=:n$, it holds
\[
   \langle\psi|\left(|\varphi\rangle\langle\varphi|_1^n\right)|\psi\rangle=0,
\]
where $|\varphi\rangle\langle\varphi|_1^n$ is the qubit string consisting of the first $n$ qubits
of $|\varphi\rangle\langle\varphi|$ as defined above.
\end{definition}
The following lemma shows that the prefix property of QTMs resembles the prefix property of classical TMs:
\begin{emptyflemma}
\label{LemPrefixQTM}
If $M$ is a prefix QTM, then
\[
   \sigma\in{\rm dom}\, M \Rightarrow \sigma_1^n \not\in{\rm dom}\, M\mbox{ for every }n<\ell(\sigma).
\]
\end{emptyflemma}
\proof Let $M$ be a prefix QTM, and let $\sigma\in{\rm dom}\, M$ with $\ell(\sigma)>n\in\N_0$.
If $\sigma=\sum_j \lambda_j |\varphi_j\rangle\langle\varphi_j|$ is the spectral decomposition of $\sigma$
with $\lambda_j>0$ for every $j$, then there must be some $j$ such that $\ell(|\varphi_j\rangle)=\ell(\sigma)>n$;
fix this $j$ until the end of the proof.

Suppose $|\psi\rangle\langle\psi|\in{\rm dom}\, M$ with $\ell(|\psi\rangle)\leq n$. As $M$ is prefix, we get
\begin{eqnarray*}
   0 &=& \langle\psi|\left(|\varphi_j\rangle\langle\varphi_j|_1^{\ell(|\psi\rangle)}\right)|\psi\rangle\\
   &=&{\rm Tr}\left( |\psi\rangle\langle\psi|\otimes\idn_{[\ell(|\psi\rangle)+1,n]}
   |\varphi_j\rangle\langle\varphi_j|_1^n\right)\\
   &\geq& \langle\psi|\otimes\langle \#|\left(|\varphi_j\rangle\langle\varphi_j|_1^n\right)|\psi\rangle\otimes|\#\rangle\geq 0,
\end{eqnarray*}
identifying a qubit string $|\psi\rangle\in\hr_\s$ with the corresponding vector on the tape Hilbert space $\hr_{\mathbf O}$.
Thus, $|\psi\rangle\perp {\rm supp}\left(|\varphi_j\rangle\langle\varphi_j|_1^n\right)$, and since
$\ell(|\varphi_j\rangle\langle\varphi_j|_1^n)\leq n$, it follows that $|\varphi_j\rangle\langle\varphi_j|_1^n\not\in
{\rm dom}\, M$. But
\[
   \sigma_1^n=\sum_j \lambda_j |\varphi_j\rangle\langle\varphi_j|_1^n,
\]
so $\sigma_1^n$ as well cannot be halting for $M$, and so $\sigma_1^n\not\in{\rm dom}\, M$.
\qed

\subsection{Approximate Halting Spaces}
\label{SubsecApproxHalting}
We start by defining the notion of approximate halting.
\begin{definition}[$\eps$-$t$-halting Property]
\label{DefEpsTHalting}
A qubit string $\sigma\in\mathcal{T}_1^+(\hr_\s)$ will be called {\em $\eps$-$t$-halting for $M$} for
some $t\in \N$, $\eps\geq 0$ and $M$ a QTM, if and only if
\[
   \langle q_f | M_{\mathbf{C}}^{t'}(\sigma)|q_f\rangle \left\{
      \begin{array}{ll}
         \leq \eps & \mbox{for }t'<t\,\,,\\
         \geq 1-\eps & \mbox{for }t'=t\,\,.
      \end{array}
   \right.
\]
Let $S_n:=\left\{|\psi\rangle\in\cn\enspace | \enspace
\| |\psi\rangle\|=1\right\}$ be the unit sphere in $\cn$, and let
$U_\delta(|\varphi\rangle):=\left\{|\psi\rangle\in\cn\enspace |
\enspace \| |\psi\rangle-|\varphi\rangle\|< \delta\right\}$ be an open ball.
The ball $U_\delta(|\varphi\rangle)$ will be called
$\eps$-$t$-halting for $M$ if there is some $|\psi\rangle\in U_\delta(|\varphi\rangle)\cap S_n$
which is $\eps$-$t$-halting for $M$. Moreover, we use the following symbols:
\begin{itemize}
\item ${\rm dist}(S,|\varphi\rangle):=\inf_{s\in S} \|\,|s\rangle-|\varphi\rangle\|$
for any subset $S\subset \cn$ and $|\varphi\rangle\in\cn$,
\item $\cnq:=\left\{|\varphi\rangle\in\cn\,\,|\,\, \langle e_k | \varphi\rangle \in \mathbb{Q}+i\mathbb{Q}
\quad\forall k\right\}$,
where $\{|e_k\rangle\}_{k=1}^{2^n}$ denotes the computational basis vectors of $\cn$,
\item $|\varphi^0\rangle:=\frac{|\varphi\rangle}{\|\,|\varphi\rangle\|}$ for every vector
$|\varphi\rangle\in\cn\setminus\{0\}$.
\end{itemize}
\end{definition}
The set of vectors with rational coordinates, denoted $\cnq$, will in the following be used frequently
as inputs or outputs of algorithms. Such vectors can be symbolically added or multiplied with rational
scalars without any error. Also, given $|a\rangle,|b\rangle\in\cnq$, it is an easy task
to decide unambiguously which vector has larger norm than the other (one can compare the rational numbers
$\|\,|a\rangle\|^2$ and $\|\,|b\rangle\|^2$, for example).

\begin{flemma}[Algorithm for $\eps$-$t$-halting-Property of Balls]
\label{TheAlgBalls}
\lineclear
There exists a (classical) algorithm $B$ which, on input $|\varphi\rangle\in\cnq$,
$\delta,\eps\in\mathbb{Q}^+$, $t\in\N$ and a classical description $s_M\in\{0,1\}^*$ of
a fixed-length QTM $M$, always halts and returns either $0$ or $1$
under the following constraints:
\begin{itemize}
\item If $U_\delta(|\varphi\rangle)$ is not $\eps$-$t$-halting for $M$, then the output must be $0$.
\item If $U_\delta(|\varphi\rangle)$ is $\frac\eps 4$-$t$-halting for $M$, then the output must be $1$.
\end{itemize}
\end{flemma}
{\bf Proof.} The algorithm $B$ computes a set of vectors $\{|\varphi_k\rangle\}_{k=1}^N\subset
\cnq$ such that for every vector $|\psi\rangle\in U_\delta(|\varphi\rangle)
\cap S_n$ there is a $k\in\{1,\ldots,N\}$ such that $\|\,|\varphi_k\rangle-|\psi\rangle\|\leq
\frac 3 {64}\,\eps$, and also vice versa (i.e. ${\rm dist}\left(\strut U_\delta(|\varphi\rangle)\cap
S_n,|\varphi_k\rangle\right)\leq \frac 3 {64}\,\eps$ for every $k$).

For every $k\in\{1,\ldots,N\}$, the algorithm simulates the QTM $M$ on input $|\varphi_k\rangle$
classically for $t$ time steps and computes an approximation $a(t')$ of the quantity
$\langle q_f|M_{\mathbf{C}}^{t'}(|\varphi_k\rangle\langle\varphi_k|)|q_f\rangle$
for every $t'\leq t$, such that
\[
   \left| \strut a(t')-\langle q_f|M_{\mathbf{C}}^{t'}(|\varphi_k\rangle\langle\varphi_k|)|q_f\rangle\right|
   <\frac 3 {32}\,\eps \qquad\mbox{for every }t'\leq t\,\,.
\]
How can this be achieved? Since the number of time steps $t$ is finite, time evolution will
be restricted to a finite subspace $\tilde\hr_{\mathbf{T}}\subset
\hr_{\mathbf{T}}$ corresponding
to a finite number of tape cells, which also restricts the state space of the head
(that points on tape cells) to a finite subspace $\tilde\hr_{\mathbf{H}}$.
Thus, it is possible to give a matrix representation of
the time evolution operator $V_M$ on $\hr_{\mathbf{C}}\otimes \tilde\hr_{\mathbf{T}}
\otimes \tilde\hr_{\mathbf{H}}$, and the expression given above can be numerically calculated
just by matrix multiplication and subsequent numerical computation of the partial trace.

Every $|\varphi_k\rangle$ that satisfies \label{EinfachSo}$|a(t')-\delta_{t't}|\leq \frac 5 8 \, \eps$ for every $t'\leq t$
will be marked as ``approximately halting''. If there is at least one $|\varphi_k\rangle$ that
is approximately halting, $B$ shall halt and output $1$, otherwise it shall halt and output $0$.

To see that this algorithm works as claimed, suppose that $U_\delta(|\varphi\rangle)$ is not
$\eps$-$t$-halting for $M$, so for every $|\tilde \psi\rangle\in U_\delta(|\varphi\rangle)$ there
is some $t'\leq t$ such that $\left| \delta_{t't}-\langle q_f|M_{\mathbf{C}}^{t'}(
|\tilde\psi\rangle\langle\tilde\psi|)|q_f\rangle\right|>\eps$. Also, for every $k\in\{1,\ldots,N\}$,
there is some vector $|\psi\rangle\in U_\delta(|\varphi\rangle)\cap S_n$ with $\|\,|\varphi_k\rangle
-|\psi\rangle\|\leq\frac 3 {64}\,\eps$, so
\begin{eqnarray*}
   \Delta_k&:=&
   \left|\delta_{t't}-\langle q_f|M_{\mathbf{C}}^{t'}(|\varphi_k\rangle\langle\varphi_k|)|q_f\rangle
   \right|\\
   &\geq&
   \left|\delta_{t't}-\langle q_f|M_{\mathbf{C}}^{t'}(|\psi\rangle\langle\psi|)|q_f\rangle
   \right|\\
   &-&\left|\langle q_f|M_{\mathbf{C}}^{t'}(|\psi\rangle\langle\psi|)|q_f\rangle
   -\langle q_f|M_{\mathbf{C}}^{t'}(|\varphi_k^0\rangle\langle\varphi_k^0|)|q_f\rangle
   \right|\\
   &-&\left|\langle q_f|M_{\mathbf{C}}^{t'}(|\varphi_k\rangle\langle\varphi_k|)|q_f\rangle
   -\langle q_f|M_{\mathbf{C}}^{t'}(|\varphi_k^0\rangle\langle\varphi_k^0|)|q_f\rangle
   \right|\\
   &>&\eps-\|\,|\psi\rangle\langle\psi|-|\varphi_k^0\rangle\langle\varphi_k^0|\|_{\rm{Tr}}
   -2\cdot\left|1-\|\,|\varphi_k\rangle\|^2\right|\\
   &\geq& \eps-\|\,|\psi\rangle-|\varphi_k^0\rangle\|-2\left|\strut1-\|\,|\varphi_k\rangle\|\right|
   (1+\|\,|\varphi_k\rangle\|)\\
   &\geq& \eps-\frac 3 {64}\,\eps-\|\,|\varphi_k\rangle-|\varphi_k^0\rangle\|-4\cdot\frac 3 {64}\,\eps
   \geq \frac{23}{32}\eps\,\,,
\end{eqnarray*}
where we have used Lemma~\ref{LemNormInequalities} and Lemma~\ref{LemStability}. Thus, for every $k$ it holds
\begin{eqnarray*}
   \left|\strut a(t')-\delta_{t't}\right|&\geq&\Delta_k
   -\left|\strut \langle q_f| M_{\mathbf{C}}^{t'}(|\varphi_k\rangle\langle \varphi_k|)|q_f\rangle
   -a(t')\right|\\
   &>&\frac{23}{32}\eps-\frac 3 {32}\,\eps=\frac 5 8 \eps\,\,,
\end{eqnarray*}
which makes the algorithm halt and output $0$.

On the other hand, suppose that $U_\delta(|\varphi\rangle)$ is $\frac \eps 4$-$t$-halting for $M$,
i.e. there is some $|\psi\rangle\in U_\delta(|\varphi\rangle)\cap S_n$ which is
$\frac \eps 4$-$t$-halting for $M$. By construction, there is some $k$ such that
$\|\,|\varphi_k\rangle-|\psi\rangle\|\leq \frac 3 {64}\,\eps$. A similar calculation as above yields
$\left|\strut \delta_{t't}-\langle q_f|M_{\mathbf{C}}^{t'}(|\varphi_k\rangle\langle\varphi_k|)
|q_f\rangle\right|\leq \frac{17}{32}\eps$ for every $t'\leq t$, so $\left|\strut a(t')-\delta_{t't}\right|
\leq \frac{17}{32}\eps+\frac 3 {32}\,\eps=\frac 5 8\,\eps$, and
the algorithm outputs $1$.\qed
\pagebreak

\begin{flemma}[Algorithm $I$ for Interpolating Subspace]
\label{TheAlgInter}
\lineclear
There exists a (classical) algorithm $I$ which, on input $M,N\in\N$,
$|\tilde\varphi_1\rangle,\ldots,|\tilde\varphi_M\rangle$,
$|\varphi_1\rangle,\ldots,|\varphi_N\rangle\in\cnq$,
$d\in\N$, $\mathbb{Q}^+\ni\Delta>\delta$ and $\mathbb{Q}^+\ni\tilde\Delta>\tilde\delta$, always
halts and returns the description of a pair $(i,\tilde U)$ with $i\in\{0,1\}$ and $\tilde U\subset\cn$ a linear
subspace, under the following constraints:
\begin{itemize}
\item If the output is $(1,\tilde U)$, then $\tilde U\subset\cn$ must be a subspace
of dimension $\dim\tilde U=d$ such that ${\rm dist}(\tilde U,|\varphi_k\rangle)<\Delta$ for
every $k$ and ${\rm dist}(\tilde U,|\tilde\varphi_l\rangle)>\tilde\delta$ for every $l$.
\item If there exists a subspace $U\subset\cn$ of dimension $\dim U=d$ such that ${\rm dist}(U,
|\varphi_k\rangle)\leq\delta$ for every $k$ and ${\rm dist}(U,|\tilde\varphi_l\rangle)\geq\tilde\Delta$
for every $l$, then
the output must be of the\footnotemark form $(1,\tilde U)$.
\end{itemize}
The description of the subspace $\tilde U$ is a list of linearly independent
vectors $\{|\tilde u_i\rangle\}_{i=1}^d\subset\cnq
\cap \tilde U$.
\end{flemma}
\footnotetext{$\tilde U$ will then be an approximation of $U$.}
{\bf Proof.} Proving this lemma is a routine (but lengthy) exercise. The idea is to construct an
algorithm that looks
for such a subspace by brute force, that is, by discretizing the set of all
subspaces within some (good enough) accuracy. We omit the details.\qed

We proceed by defining the notion of an {\em approximate halting space}. Note that the definition
depends on the details of the previously defined algorithms in Lemma~\ref{TheAlgBalls}
and \ref{TheAlgInter}
(for example, there are always different possibilities to compute the necessary discretizations).
Thus, we fix a concrete instance of all those algorithms for the rest of the paper.

\begin{definition}[Approximate Halting Spaces]
\label{DefApproxHalt}
\lineclear
We define\footnote{From a formal point of view, the notation should rather read $\hr_{s_M}^{(n,\delta)}(t)$
instead of $\hr_M^{(n,\delta)}(t)$, since this space depends also on the choice of the classical
description $s_M$ of $M$.} the
$\delta$-approximate halting space $\hr_M^{(n,\delta)}(t)\subset\cn$ and the $\delta$-approximate
halting accuracy $\eps_M^{(n,\delta)}(t)\in\mathbb{Q}$
as the outputs
of the following classical algorithm on input $n,t\in\N$, $0<\delta\in\mathbb{Q}$ and
$s_M\in\{0,1\}^*$, where $s_M$ is a classical description of a fixed-length QTM $M$:
\begin{itemize}
\item[(1)] Let $\eps:=18\,\delta$.
\item[(2)] Compute a covering of $S_n$ of open balls of radius $\delta$, that is,
a set of vectors $\{|\psi_1\rangle,\ldots,|\psi_L\rangle\}\subset\cnq$
($L\in\N$) with $\|\,|\psi_k\rangle\|\in \left( 1-\frac\delta 2,1+\frac \delta 2\right)$
for every $k\in\{1,\ldots,L\}$
such that $S_n\subset\bigcup_{i=1}^L U_\delta(|\psi_i\rangle)$.
\item[(3)] For every $k\in\{1,\ldots,L\}$, compute $B(|\psi_k\rangle,\delta,\eps,t,s_M)$
and $B(|\psi_k\rangle,\delta,18\,\delta,t,s_M)$, where $B$ is
the algorithm for testing the $\eps$-$t$-halting property of balls of Lemma~\ref{TheAlgBalls}.
If the output is $0$ for every $k$, then output $\left(\{0\},\eps\right)$
and halt.
Otherwise set for $\N_0\ni N\leq L$ and $\N_0\ni K\leq L$
\begin{eqnarray*}
   \left\{|\varphi_i\rangle\right\}_{i=1}^N&:=&\left\{
      |\psi_k\rangle\enspace|\enspace
      B(|\psi_k\rangle,\delta,\eps,t,s_M)=1
   \right\},\\
   \left\{|\tilde\varphi_i\rangle\right\}_{i=1}^K&:=&\left\{
      |\psi_k\rangle\enspace|\enspace
      B(|\psi_k\rangle,\delta,18\,\delta,t,s_M)=0
   \right\}.
\end{eqnarray*}
If $N=0$, i.e. if the set $\{|\varphi_i\rangle\}_{i=1}^N$ is empty,
output $\left(\{0\},\eps\right)$ and halt.
\item[(4)] Set $d:=2^n$.
\item[(5)] Let $\Delta:=2\delta$, $\tilde\Delta:=\frac 7 4 \delta$ and $\tilde\delta:=\frac 3 2 \delta$.
Use the algorithm $I$ of Lemma~\ref{TheAlgInter} to search for an interpolating
subspace, i.e., compute $I(K,N,|\tilde\varphi_1\rangle,\ldots,|\tilde\varphi_K\rangle,
|\varphi_1\rangle,\ldots,|\varphi_N\rangle,d,\Delta,\delta,\tilde\Delta,\tilde\delta)$. If
the output of $I$ is $(1,\tilde U)$, output $\left(\tilde U,\eps\right)$ and halt.
\item[(6)] Set $d:=d-1$. If $d\geq 1$, then go back to step (5).
\item[(7)] Set $\eps:=\frac\eps 2$ and go back
to step (3).
\end{itemize}
Moreover, let $H_M^{(n,\delta)}(t):=\hr_M^{(n,\delta)}(t)\cap S_n$.
\end{definition}

The following theorem proves that this definition makes sense:

\begin{emptyftheorem}
The algorithm in Definition~\ref{DefApproxHalt} always terminates on any input; thus,
the approximate halting spaces $\hr_M^{(n,\delta)}(t)$ are well-defined.
\end{emptyftheorem}
{\bf Proof.} 
Define the function $\eps_{min}:S_n\to\R_0^+$
by $\eps_{min}(|\psi\rangle):=\inf\{{\eps>0}\enspace|\enspace |\psi\rangle \mbox{ is }\eps\mbox{-}
   t\mbox{-halting for }M\}$.
Lemma~\ref{LemNormInequalities}
and \ref{LemStability} yield
\begin{equation}
   \left|\strut \eps_{min}(|\psi_1\rangle)-\eps_{min}(|\psi_2\rangle)\right| \leq
   \|\,|\psi_1\rangle-|\psi_2\rangle\|\,\,,
   \label{eqEpsMin}
\end{equation}
so $\eps_{min}$ is continuous.
For the special case $H_M^{(n)}(t)=\emptyset$, it must thus hold that
$\eps_{min}(S_n):=\min_{|\psi\rangle\in S_n} \eps_{min}(|\psi\rangle)>0$. If the
algorithm has run long enough such that $\eps<\eps_{min}(S_n)$, it must then
be true that $B(|\psi_k\rangle,\delta,\eps,t,s_M)=0$ for every $k\in\{1,\ldots,L\}$,
since all the balls $U_\delta(|\psi_k\rangle)$ are
not $\eps$-$t$-halting. This makes the algorithm halt in step (3).

Now consider the case $H_M^{(n)}(t)\neq\emptyset$. The continuous function
$\eps_{min}$ attains a minimum
on every compact set $\bar U_\delta(|\psi_k\rangle)\cap S_n$, so let $\eps_k:=\min_{|\psi\rangle
\in\bar U_\delta(|\psi_k\rangle)\cap S_n} \eps_{min}(|\psi\rangle)$ ($1\leq k\leq N$).
If $\eps_k=0$ for every $k$, then for every $k$ and $\eps>0$, there is some vector $|\psi\rangle
\in U_\delta(|\psi_k\rangle)\cap S_n$ which is $\eps$-$t$-halting for $M$, so
$B(|\psi_k\rangle,\delta,\eps,t,s_M)=1$ for every $\eps>0$, and so $K=0$ in step (3).
Thus, the algorithm $I$ will by construction find the interpolating subspace $\tilde U=\left(
\C^2\right)^{\otimes n}$ and cause halting in step (5).

Otherwise,
let $\eps_0:=\min\{\eps_k\enspace|\enspace k\in\{1,\ldots,N\},\eps_k>0\}$. Suppose that
the algorithm has run long enough such that $\eps<\eps_0$. By construction of the algorithm
$B$, if $B(|\psi_k\rangle,\delta,\eps,t,s_M)=1$, it follows that $U_\delta(|\psi_k\rangle)$ is
$\eps$-$t$-halting for $M$, but then, $\eps_k\leq \eps < \eps_0$, so $\eps_k=0$, so there is some
$|\psi\rangle\in \bar U_\delta(|\psi_k\rangle)\cap S_n$ which is $0$-$t$-halting for $M$, so
${\rm dist}(\hr_M^{(n)}(t),|\psi_k\rangle)\leq\delta$.
On the other hand, if
$B(|\psi_k\rangle,\delta,18\,\delta,t,s_M)=0$, it follows that $U_\delta(|\psi_k\rangle)$ is not
$\left(\frac 9 2 \delta\right)$-$t$-halting for $M$.
Thus, ${\rm dist}\left(H_M^{(n)}(t),|\psi_k^0\rangle\right)\geq \frac 9 2 \delta$ according to
(\ref{eqEpsMin}),
so ${\rm dist}(\hr_M^{(n)}(t)\cap S_n,|\psi_k\rangle)>4\delta$, and
by elementary estimations ${\rm dist}(\hr_M^{(n)}(t),|\psi_k\rangle)>\frac 7 4 \delta$.
By definition of the
algorithm $I$, it follows that
$I(K,N,|\tilde\varphi_1\rangle,\ldots,|\tilde\varphi_K\rangle,
|\varphi_1\rangle,\ldots,|\varphi_N\rangle,d,\Delta,\delta,\tilde\Delta,\tilde\delta)=(1,\tilde U)$
for $d:=\dim \hr_M^{(n)}(t)\geq 1$ and some subspace $\tilde U\subset \cn$,
which makes the algorithm halt
in step (5).
\qed

We are now going to show some properties of the approximate halting spaces. These properties show
that these spaces are, in some sense, good approximation of a QTM's ``true'' halting spaces.

\begin{ftheorem}[Properties of Approximate Halting Spaces]
\label{TheProperties}
\lineclear
The approximate halting spaces $\hr_M^{(n,\delta)}(t)$ have the following properties:
\begin{itemize}
\item {\em Almost-Halting:} If $|\psi\rangle\in H_M^{(n,\delta)}(t)$, then $|\psi\rangle$
is $(20\,\delta)$-$t$-halting for $M$.
\item {\em Approximation:} For every $|\psi\rangle\in H_M^{(n)}(t)$, there is a vector $|\psi^{(\delta)}\rangle
\in H_M^{(n,\delta)}(t)$ which satisfies $\|\,|\psi\rangle-|\psi^{(\delta)}\rangle\|<\frac {11} 2 \delta$.
\item{\em Similarity:} If $\delta,\Delta\in\mathbb{Q}^+$ such that $\delta\leq \frac 1 {80}\,\eps_M^{(n,\Delta)}(t)$,
then for every $|\psi\rangle\in H_M^{(n,\delta)}(t)$ there is a vector $|\psi^{(\Delta)}\rangle
\in H_M^{(n,\Delta)}(t)$ which satisfies
$\|\,|\psi\rangle-|\psi^{(\Delta)}\rangle\|<\frac {11} 2 \Delta$.
\item {\em Almost-Orthogonality:} If $|\psi_t\rangle\in H_M^{(n,\delta)}(t)$ and
$|\psi_{t'}\rangle\in H_M^{(n,\delta)}(t')$ for $t\neq t'$, then it holds that $|\langle \psi_t | \psi_{t'}\rangle|
\leq 4\sqrt{5\delta}$.
\end{itemize}
\end{ftheorem}
{\bf Proof.} Assume that
$H_M^{(n,\delta)}(t)\neq\emptyset$.
Let $|\psi\rangle\in H_M^{(n,\delta)}(t)\subset S_n$, and let
$\{|\psi_1\rangle,\ldots,|\psi_L\rangle\}\subset\cn$
be the covering of $S_n$ from the algorithm in Definition~\ref{DefApproxHalt}.
By construction, there is some $k\in\{1,\ldots,L\}$ such that
$|\psi\rangle\in U_\delta(|\psi_k\rangle)$. The subspace $\hr_M^{(n,\delta)}(t)$
is computed in step (5) of the algorithm in Definition~\ref{DefApproxHalt}
via
$I(K,N,|\tilde\varphi_1\rangle,\ldots,|\tilde\varphi_K\rangle,
|\varphi_1\rangle,\ldots,|\varphi_N\rangle,d,\Delta,\delta,\tilde\Delta,\tilde\delta)
=(1,\hr_M^{(n,\delta)}(t))$, and since ${\rm dist}(\hr_M^{(n,\delta)}(t),|\psi_k\rangle)<\delta$,
it follows from the properties of the algorithm $I$ in Lemma~\ref{TheAlgInter} that
$|\psi_k\rangle\neq|\tilde\varphi_l\rangle$ for every $l\in\{1,\ldots,K\}$ in step (3)
of the algorithm. Thus, $B(|\psi_k\rangle,\delta,18\,\delta,t,s_M)=1$, and it follows
from the properties of the algorithm $B$ in Lemma~\ref{TheAlgBalls} that
$U_\delta(|\psi_k\rangle)$ is $(18\,\delta)$-$t$-halting for $M$, so there is some
$|\tilde\psi\rangle\in U_\delta(|\psi_k\rangle)\cap S_n$ which is $(18\,\delta)$-$t$-halting for $M$.
Since $\|\,|\tilde\psi\rangle-|\psi\rangle\|<2\delta$, the almost-halting property
follows from Equation~(\ref{eqEpsMin}).

To prove the approximation property, assume that
$H_M^{(n)}(t)\neq\emptyset$.
Let $|\psi\rangle\in H_M^{(n)}(t)\subset S_n$; again, there is some $j\in\{1,\ldots,L\}$ such that
$|\psi\rangle\in U_\delta(|\psi_j\rangle)$, so $U_\delta(|\psi_j\rangle)$ is $0$-$t$-halting for $M$,
and $B(|\psi_j\rangle,\delta,\eps,t,s_M)=1$ for every $\eps>0$ by definition of the
algorithm $B$. For step (3) of the algorithm in Definition~\ref{DefApproxHalt}, it thus
always holds that $|\psi_j\rangle\in\{|\varphi_i\rangle\}_{i=1}^N$. The output of the
algorithm is computed in step (5) via
$I(K,N,|\tilde\varphi_1\rangle,\ldots,|\tilde\varphi_K\rangle,
|\varphi_1\rangle,\ldots,|\varphi_N\rangle,d,\Delta,\delta,\tilde\Delta,\tilde\delta)
=(1,\hr_M^{(n,\delta)}(t))$.
By definition of $I$, it holds
${\rm dist}(\hr_M^{(n,\delta)}(t),|\psi_j\rangle)<\Delta$,
and by elementary estimations it follows that
${\rm dist}(\hr_M^{(n,\delta)}(t)\cap S_n,|\psi_j\rangle) < \frac \delta 2 + 2 \Delta$,
so there is some $|\psi^{(\delta)}\rangle\in H_M^{(n,\delta)}(t)$
such that $\|\,|\psi^{(\delta)}\rangle-|\psi_j\rangle\| < \frac \delta 2 +2\Delta$.
Since $\|\,|\psi\rangle-|\psi_j\rangle\|\leq\delta$ and $\Delta=2\delta$,
the approximation property follows.

Notice that under the assumptions
given in the statement of the similarity property, it follows from the almost-halting property
that if $|\psi\rangle\in H_M^{(n,\delta)}(t)$, then $|\psi\rangle$ must be
$\frac 1 4 \eps_M^{(n,\Delta)}(t)$-$t$-halting for $M$. Consider the computation
of $\hr_M^{(n,\Delta)}(t)$ by the algorithm in Definition~\ref{DefApproxHalt}.
By construction, it always holds that the parameter $\eps$ during the computation
satisfies $\eps\geq \eps_M^{(n,\Delta)}(t)$, so $|\psi\rangle$ is always
$\frac\eps 4$-$t$-halting for $M$, and if $|\psi\rangle\in U_\delta(|\psi_j\rangle)$,
it follows that $B(|\psi_j\rangle,\delta,\eps,t,s_M)=1$. The rest follows in complete
analogy to the proof of the approximation property.

For the almost-orthogonality property, suppose $|v\rangle\in H_M^{(n,\delta)}(t')$ and $|w\rangle\in
H_M^{(n,\delta)}(t)$
are two arbitrary qubit strings of length $n$ with different approximate halting times $t<t'\in\N$.
There is some $l\in\{1,\ldots,L\}$ such that $|w\rangle\in U_\delta(|\psi_l\rangle)$,
so ${\rm dist}(\hr_M^{(n,\delta)}(t),|\psi_l\rangle)<\delta<\tilde\delta$.
Since $I(K,N,|\tilde\varphi_1\rangle,\ldots,|\tilde\varphi_K\rangle,
|\varphi_1\rangle,\ldots,|\varphi_N\rangle,d,\Delta,\delta,\tilde\Delta,\tilde\delta)
=(1,\hr_M^{(n,\delta)}(t))$ at step (5) of the computation of $\hr_M^{(n,\delta)}(t)$,
it follows from the definition of $I$ that there is no $m\in\N$ such that
$|\psi_l\rangle=|\tilde \varphi_m\rangle$ for the sets defined in step (3) of the
algorithm above. Thus, $B(|\psi_l\rangle,\delta,18\,\delta,t,s_M)=1$, and by definition of $B$
it follows that $U_\delta(|\psi_l\rangle)$ must be $(18\,\delta)$-$t$-halting for $M$, so there
is some vector $|\tilde w\rangle\in U_\delta(\psi_l\rangle)\cap S_n$ which is $(18\,\delta)$-$t$-halting for $M$
and satisfies $\|\,|w\rangle-|\tilde w\rangle\|\leq \|\,|\tilde w\rangle-|\psi_l\rangle\|+
\|\,|\psi_l\rangle-|w\rangle\|<2\delta$. Analogously, there is some vector $|\tilde v\rangle\in S_n$
which is $(18\,\delta)$-$t'$-halting for $M$ and satisfies $\|\,|v\rangle-|\tilde v\rangle\|<2\delta$.

From the definition of the trace distance for pure states (see \cite[(9.99)]{NielsenChuang}
and of the $\eps$-$t$-halting property in Definition~\ref{DefEpsTHalting} together
with Lemma~\ref{LemNormInequalities} and Lemma~\ref{LemStability}, it follows that
\begin{eqnarray}
   \sqrt{1-\left| \langle w | v \rangle \right|^2}&=&\|\,|w\rangle\langle w | - |v\rangle\langle
   v|\,\|_{\rm Tr}\nonumber\\
   &\geq& \|\, |\tilde w\rangle\langle\tilde w|-|\tilde v\rangle\langle\tilde v|\,\|_{\rm Tr}
   -\|\,|w\rangle\langle w|-|\tilde w\rangle\langle\tilde w\|\,\|_{\rm Tr}\nonumber\\
   &&-\,\|\,|v\rangle\langle v|-|\tilde v\rangle\langle \tilde v|\,\|_{\rm Tr}\nonumber\\
   &\geq& \left| \langle q_f| M_{\mathbf{C}}^t (|\tilde w\rangle\langle \tilde w|)|q_f\rangle
   -\langle q_f| M_{\mathbf{C}}^t(|\tilde v\rangle\langle\tilde v|)|q_f\rangle\right|\nonumber\\
   &&-\|\,|w\rangle-|\tilde w\rangle\|-\|\,|v\rangle-|\tilde v\rangle\|\nonumber\\
   &\geq& 1-36\,\delta-2\delta-2\delta=1-40\,\delta.
   \label{eqInnerProdTr}
\end{eqnarray}
This proves the almost-orthogonality property.
\qed

The following corollary proves that the approximate halting spaces $\hr_M^{(n,\delta)}(t)$ are ``not too large''
if $\delta$ is small enough. Formally, we will need this property to prove the Kraft inequality for some code
in Subsection~\ref{SubsecProofMainTheorems}, as well as for some estimation in Section~\ref{SecQCofClass}
on the quantum complexity of classical strings.

\begin{fcorollary}[Dimension Bound for Halting Spaces]
\lineclear
\label{CorDimBound}
If $\delta<\frac 1 {80}\, 2^{-2n}$, then $\displaystyle\sum_{t\in\N} \dim \hr_M^{(n,\delta)}(t)\leq 2^n$.
\end{fcorollary}
{\bf Proof.} Suppose that $\sum_{t\in\N} \dim \hr_M^{(n,\delta)}(t)> 2^n$. Then, choose orthonormal bases
in each of the spaces $\hr_M^{(n,\delta)}(t)$, and let $\left\{
|\varphi_i\rangle\right\}_{i=1}^{2^n+1}$ be the union of the first $2^n+1$ of these basis vectors.
By construction and by the almost-orthogonality property of Theorem~\ref{TheProperties},
it follows that $|\langle \varphi_i|\varphi_j\rangle|\leq 4\sqrt{5\delta}<2^{-n}=\frac 1
{(2^n+1)-1}$ for every $i\neq j$. Lemma~\ref{LemInnerProduct} yields $\dim U\geq 2^n+1$ for
$U:={\rm span}\left\{|\varphi_i\rangle\right\}_{i=1}^{2^n+1}\subset \cn$, but
$\dim \cn=2^n$, which is a contradiction.\qed

\subsection{Compression, Decompression, and Coding}
\label{SubsecCompDecomp}
In this subsection, we define some compression and coding algorithms that will
be used in the construction of the strongly universal QTM.
\begin{definition}[Standard (De-)Compression]
\label{DefStandard}
\lineclear
Let $U\subset \hr_n$ be a linear subspace with $N:=\dim U$. Let
$P_U\in\mathcal{B}(\hr_n)$ be the orthogonal projector onto $U$, and
let $\left\{|e_i\rangle\right\}_{i=1}^{2^n}$ be the computational
basis of $\hr_n$.
The result of applying the Gram-Schmidt orthonormalization procedure to the vectors
$\left\{|\tilde u_i\rangle\right\}_{i=1}^{2^n}=\left\{P_U |e_i\rangle\right\}_{i=1}^{2^n}$ (dropping every null vector)
is called the {\em standard basis} $\{|u_1\rangle,\ldots,|u_N\rangle\}$ of $U$.
Let $|f_i\rangle$ be the $i$-th computational basis vector of $\hr_{\lceil \log N\rceil}$.
The {\em standard compression}
$\mathcal{C}_U:U\to\hr_{\lceil \log N\rceil}$ is then defined by linear extension of
$\mathcal{C}_U(|u_i\rangle):=|f_i\rangle$ for $1\leq i \leq N$,
that is, $\mathcal{C}_U$ isometrically embeds $U$ into $\hr_{\lceil \log N\rceil}$.
A linear isometric map $\mathcal{D}_U:\hr_{\lceil \log N\rceil}\to \hr_n$ will be
called a {\em standard decompression} if it holds that
\[
   \mathcal{D}_U\circ \mathcal{C}_U=\idn_U\,\,.
\]
\end{definition}
It is clear that there exists a classical algorithm that, given a description of $U$ (e.g. a list
of basis vectors $\{|u_i\rangle\}_{i=1}^{\dim U}\subset\cnq$), can effectively compute
(classically) an approximate description of the standard basis of $U$. Moreover,
a quantum Turing machine can effectively apply a standard decompression map to its input:

\begin{flemma}[Q-Standard Decompression Algorithm]
\label{LemStandardDecomp}
\lineclear
There is a QTM $\mathfrak{D}$ which, given a description\footnotemark of a
subspace $U\subset \hr_n$, the integer $n\in\N$, some $\delta\in\mathbb{Q}^+$, and a quantum state $|\psi\rangle\in
\hr_{\lceil \log\dim U\rceil}$, outputs some
state $|\varphi\rangle\in\hr_n$ with the property that $\|\,|\varphi\rangle-\mathcal{D}_U |\psi\rangle\|<\delta$,
where $\mathcal{D}_U$ is some standard decompression map.
\end{flemma}
\footnotetext{(a list of linearly independent vectors $\{|\tilde u_1\rangle,\ldots,|\tilde u_{\dim U}\rangle\}\subset U\cap \cnq$)}
{\bf Proof.} Consider the map $A:\hr_{\lceil \log\dim U\rceil}\to\hr_n$, given by
$A|v\rangle:=|0\rangle^{\otimes (n-\lceil \log\dim U\rceil)}
\otimes |v\rangle$. The map $A$ prepends zeroes to a vector; it maps the computational basis
vectors of $\hr_{\lceil \log\dim U\rceil}$ to the lexicographically first computational basis vectors
of $\hr_n$.
The QTM $\mathfrak{D}$ starts by applying this map $A$ to the input state $|\psi\rangle$ by
prepending zeroes on its tape,
creating a state $|\tilde\psi\rangle:=|0\rangle^{\otimes (n-\lceil \log\dim U\rceil)}
\otimes |\psi\rangle\in\hr_n$.

Afterwards, it applies (classically) the Gram-Schmidt orthonormalization procedure
to the list of vectors $\{|\tilde u_1\rangle,\ldots,|\tilde u_{\dim U}\rangle,
|e_1\rangle,\ldots,|e_{2^n}\rangle\}\subset \cnq$, where the vectors $\{|\tilde u_i\rangle\}_{i=1}^{\dim U}$
are the basis vectors of $U$ given in the input, and
the vectors $\{|e_i\rangle\}_{i=1}^{2^n}$
are the computational basis vectors of $\hr_n$. Since every vector has rational entries
(i.e. is an element of $\cnq$), the Gram-Schmidt procedure can be applied exactly,
resulting in a list
$\{|u_i\rangle\}_{i=1}^{2^n}$ of basis vectors of $\hr_n$ which have entries that are
square roots of rational numbers. Note that by construction,
the vectors $\{|u_i\rangle\}_{i=1}^{\dim U}$ are the standard basis vectors of $U$
that have been defined in Definition~\ref{DefStandard}.

Let $V$ be the unitary $2^n\times 2^n$-matrix that has the vectors $\{|u_i\rangle\}_{i=1}^{2^n}$
as its column vectors.
The algorithm continues by computing a rational approximation $\tilde V$ of $V$ such that the
entries satisfy
$|\tilde V_{ij}-V_{ij}|<\frac\delta {2^{n+1} (10\sqrt{2^n})^{2^n}}$, and thus,
in operator norm, it holds $\|\tilde V-V\|<\frac\delta{2(10\sqrt{2^n})^{2^n}}$.
Bernstein and Vazirani \cite[Sec. 6]{BernsteinVazirani} have shown that there are QTMs that
can carry out an $\eps$-approximation of a desired unitary
transformation $V$ on their tapes if given a matrix $\tilde V$ as input
that is within distance $\frac\eps {2(10\sqrt d)^d}$ of the $d\times d$-matrix $V$.
This is exactly the case here\footnote{Note that we consider $\hr_n$ as a subspace
of an $n$-cell tape segment Hilbert space
$\left(\C^{\{0,1,\#\}}\right)^{\otimes n}$, and we demand $V$ to leave blanks $|\#\rangle$
invariant.}, with $d=2^n$ and $\eps=\delta$, so let the $\mathfrak{D}$ apply $V$ within
$\delta$ on its tape to create the state $|\varphi\rangle\in\hr_n$ with
$\|\,|\varphi\rangle-V|\tilde\psi\rangle\|=\|\,|\varphi\rangle-V\circ A |\psi\rangle\|
<\delta$. Note that the map $V\circ A$
is a standard decompression map (as defined in Definition~\ref{DefStandard}), since
for every $i\in\{1,\ldots,\dim U\}$ it holds that
\[
   V\circ A\circ\mathcal{C}_U |u_i\rangle=V\circ A |f_i\rangle
   =V |e_i\rangle=|u_i\rangle\,\,,
\]
where the vectors $|f_i\rangle$ are the computational basis vectors of $\hr_{\lceil
\log\dim U\rceil}$.\qed

The next lemma will be useful for coding the ``classical part'' of a halting qubit string.
The ``which subspace'' information
will be coded into a classical string $c_i\in\s$ whose length $\ell_i\in\N_0$ depends on the dimension of the
corresponding halting space $\hr_M^{(n,\delta)}(t_i)$. The dimensions of the halting spaces
$\left(\dim \hr_M^{(n,\delta)}(t_1),\dim \hr_M^{(n,\delta)}(t_2),\ldots\right)$ can be computed one after the other,
but the complete list of the code word lengths $\ell_i$ is not computable due to the
undecidability of the halting problem.
Since most well-known prefix codes (like Huffman code, see \cite{CoverThomas}) start by initially sorting
the code word lengths in decreasing order, and thus require complete knowledge of the
whole list of code word lengths in advance, they are not suitable for our purpose. We thus give an easy algorithm
that constructs the code words one after the other, such that code word $c_i$ depends
only on the previously given lengths $\ell_1,\ell_2,\ldots,\ell_i$. We call this
``blind prefix coding'', because code words are assigned sequentially without looking at what is coming next.

\begin{flemma}[Blind Prefix Coding]
\label{LemBlindCoding}
\lineclear
Let $\{\ell_i\}_{i=1}^N\subset \N_0$ be a sequence of natural numbers (code word lengths)
that satisfies the Kraft inequality $\displaystyle\sum_{i=1}^N 2^{-\ell_i}\leq 1$.
Then the following (``blind prefix coding'') algorithm produces a list
of code words $\{c_i\}_{i=1}^N\subset \s$ with $\ell(c_i)=\ell_i$, such that
the $i$-th code word only depends on $\ell_i$ and the previously chosen codewords $c_1,\ldots,c_{i-1}$:
\begin{itemize}
\item Start with $c_1:=0^{\ell_1}$, i.e. $c_1$ is the string consisting of $\ell_1$ zeroes;
\item for $i=2,\ldots,N$ recursively, let $c_i$ be the first string in lexicographical order
of length $\ell(c_i)=\ell_i$ that is no prefix or extension of any of the previously
assigned code words $c_1,\ldots,c_{i-1}$.
\end{itemize}
\end{flemma}

{\bf Proof.} We omit the lengthy, but simple proof; it is based on identifying the binary
code words with subintervals of $[0,1)$ as explained in \cite{Vitanyibook}.
We also remark that the content of this lemma is given in \cite[Thm. 5.2.1]{CoverThomas}
without proof as an example for a prefix code.
\qed

\subsection{Proof of the Strong Universality Property}
\label{SubsecProofMainTheorems}
To simplify the proof of Main Theorem~\ref{MainTheorem1}, we show now that
it is sufficient to consider fixed-length QTMs only:

\begin{flemma}[Fixed-Length QTMs are Sufficient]
\label{LemFixedLengthEnough}
\lineclear
For every QTM $M$, there is a fixed-length QTM $\tilde M$ such that
for every $\rho\in\mathcal{T}_1^+(\hr_\s)$ there is a fixed-length qubit string
$\tilde\rho\in\bigcup_{n\in\N_0} \mathcal{T}_1^+(\hr_n)$ such that
$M(\rho)=\tilde M(\tilde\rho)$ and $\ell(\tilde\rho)\leq\ell(\rho)+1$.
\end{flemma}
{\bf Proof.} Since $\dim \hr_{\leq n}=2^{n+1}-1$, there is an isometric
embedding of $\hr_{\leq n}$ into $\hr_{n+1}$. One example is the map $V_n$,
which is defined as $V_n|e_i\rangle:=|f_i\rangle$ for $i\in\{1,\ldots,2^{n+1}-1\}$,
where $|e_i\rangle$ and $|f_i\rangle$ denote the computational basis vectors
(in lexicographical order) of $\hr_{\leq n}$ and $\hr_{n+1}$ respectively.
As $\hr_{n+1}\subset\hr_{\leq(n+1)}$ and $\hr_{\leq n}\subset \hr_{\leq(n+1)}$, we can extend $V_n$ to a unitary
transformation $U_n$ on $\hr_{\leq(n+1)}$, mapping computational basis vectors
to computational basis vectors.

The fixed-length QTM $\tilde M$ works as follows, given some fixed-length
qubit string $\tilde\rho\in\mathcal{T}_1^+(\hr_{n+1})$ on its input tape: first, it determines
$n+1=\ell(\tilde\rho)$
by detecting the first blank symbol $\#$. Afterwards, it computes a description
of the unitary transformation $U_n^*$ and applies it to the qubit string $\tilde\rho$
by permuting the computational basis vectors in the $(n+1)$-block of cells
corresponding to the Hilbert space $\left(\C^{\{0,1,\#\}}\right)^{\otimes(n+1)}$.
Finally, it calls the QTM $M$ to continue the computation on input 
$\rho:=U_n^*\,\tilde\rho\, U_n$. If $M$ halts, then the output will be $M(\rho$).
\qed

\vskip 0.5cm

{\bf Proof of Theorem~\ref{MainTheorem1}.} First, we show
how the input $\sigma_M$ for the strongly universal QTM $\mathfrak U$ is constructed
from the input $\sigma$ for $M$.
Fix some QTM $M$ and input length $n\in\N_0$, and let
$\eps_0:=\frac 1 {81}\, 2^{-2n}$.
Define the
halting time sequence $\{t_M^{(n)}(i)\}_{i=1}^N$ as the set of all integers $t\in\N$ such
that $\dim \hr_M^{(n,\eps_0)}(t)\geq 1$,
ordered such that $t_M^{(n)}(i)<t_M^{(n)}(i+1)$
for every $i$. The number $N$ is in general not computable, but must be somewhere between $0$
and $2^n$ due to Corollary~\ref{CorDimBound}.

For every $i\in\{1,\ldots,N\}$, define the code word length $\ell_i^{(M,n)}$ as
\[
   \ell_i^{(M,n)}:=n+1-\left\lceil \log\dim \hr_M^{(n,\eps_0)}\left(t_M^{(n)}(i)\right)\right\rceil\,\,.
\]
This sequence of code word lengths satisfies the Kraft inequality:
\begin{eqnarray*}
   \sum_{i=1}^N 2^{-\ell_i^{(M,n)}}&=&2^{-n}\sum_{i=1}^N 2^{\left\lceil \log\dim \hr_M^{(n,\eps_0)}\left(t_M^{(n)}(i)\right)\right\rceil
   -1}\\
   &\leq& 2^{-n}\sum_{i=1}^N \dim\hr_M^{(n,\eps_0)}\left(t_M^{(n)}(i)\right)\\
   &=&2^{-n}\sum_{t\in\N}\dim\hr_M^{(n,\eps_0)}(t)\leq 1\,\,,
\end{eqnarray*}
where in the last inequality, Corollary~\ref{CorDimBound} has been used. Let
$\left\{c_i^{(M,n)}\right\}_{i=1}^N
\subset\s$ be the blind prefix code corresponding to the sequence
$\left\{\ell_i^{(M,n)}\right\}_{i=1}^N$
which has been constructed in Lemma~\ref{LemBlindCoding}.

In the following, we use the space $\hr_M^{(n,\eps_0)}(t)$ as some kind of ``reference space''
i.e. we construct our QTM $\mathfrak U$ such that it expects the standard compression
of states $|\psi\rangle\in\hr_M^{(n,\eps_0)}(t)$ as part of the input. If the desired
accuracy parameter $\delta$ is smaller than $\eps_0$, then some ``fine-tuning'' must take place,
unitarily mapping the state $|\psi\rangle\in\hr_M^{(n,\eps_0)}(t)$ into halting spaces
of smaller accuracy parameter. In the next paragraph, these unitary transformations are constructed.

Recursively, for $k\in\N$, define $\eps_k:=\frac 1 {80} \eps_M^{(n,\eps_{k-1})}(t)$.
Since $\eps_M^{(n,\delta)}(t)\leq 18\delta$ by construction of the algorithm in Definition~\ref{DefApproxHalt},
we have $\eps_k\leq \left(\frac {18}{80}\right)^k\cdot \eps_0\stackrel{k\to\infty}\longrightarrow 0$.
It follows from the approximation property of Theorem~\ref{TheProperties} together with
Lemma~\ref{LemDimBoundSimilar} that $\dim\hr_M^{(n,\eps_k)}(t)\geq \dim \hr_M^{(n)}(t)$.
The similarity property and Lemma~\ref{LemDimBoundSimilar} tell us that
$\dim \hr_M^{(n,\eps_{k-1})}(t)\geq \dim \hr_M^{(n,\eps_{k})}(t)$ for every $k\in\N$,
and there exist isometries $U_k:\hr_M^{(n,\eps_{k})}(t)\to\hr_M^{(n,\eps_{k-1})}(t)$
that, for $k$ large enough, satisfy
\begin{equation}
\|U_k-\idn\|<\frac 8 3 \sqrt{\frac {11} 2 \eps_{k-1}} \left(\frac 5 2\right)^{2^n}
\leq {\rm const}_n \cdot \left(\frac{18}{80}\right)^{\frac k 2}.\label{EqConstn}
\end{equation}
Let now $d:=\lim_{k\to\infty} \dim \hr_M^{(n,\varepsilon_k)}(t)$ and
$c:=\min\left\{ k\in\N\,\,|\,\, \dim \hr_M^{(n,\varepsilon_k)}(t)=d\right\}$.
For any choice of the transformations $U_k$ (they are not unique), let
\[
   \tilde\hr_M^{(n,\eps_k)}(t):=\left\{
      \begin{array}{cl}
         U_{k+1}U_{k+2}\ldots U_c \hr_M^{(n,\eps_c)}(t) & \mbox{if } k<c\,\,,\\
         \hr_M^{(n,\eps_k)}(t) & \mbox{if }k\geq c\,\,.
      \end{array}
   \right.
\]
It follows that the spaces $\tilde\hr_M^{(n,\eps_k)}(t)$ all have the same dimension for every $k\in\N_0$,
and that $\tilde\hr_M^{(n,\eps_k)}(t)\subset\hr_M^{(n,\eps_k)}(t)$.
Define the unitary operators
$\tilde U_k:=U_k\upharpoonright \tilde\hr_M^{(n,\eps_k)}(t)$,
then $\|\tilde U_k^*-\idn\|\leq\|U_k-\idn\|$,
and so the sum $\sum_{k=1}^\infty \|\tilde U_k^*-\idn\|$ converges.
Due to Lemma~\ref{LemCompositionUnitary}, the product $U:=\prod_{k=1}^\infty \tilde U_k^*$ converges
to an isometry $U:\tilde\hr_M^{(n,\eps_0)}(t)\to\hr_n$.
It follows from the approximation property in Theorem~\ref{TheProperties} that\label{rmran}
$\hr_M^{(n)}(t)\subset{\rm ran}(U)$, so we can define a unitary map $U^{-1}:{\rm ran}(U)\to \tilde
\hr_M^{(n,\varepsilon_0)}(t)$ by $U^{-1}(Ux):=x$, and $\hr_M^{(n)}(t)\subset {\rm dom}(U^{-1})$.

Due to Lemma~\ref{LemFixedLengthEnough}, it is sufficient to consider
fixed-length QTMs $M$ only, so we can assume that our input $\sigma$ is a fixed-length
qubit string. Suppose $M(\sigma)$ is defined, and let $\tau\in\N$ be the corresponding
halting time for $M$. Assume for the moment that $\sigma=|\psi\rangle\langle\psi|$
is a pure state, so $|\psi\rangle\in H_M^{(n)}(\tau)$.
Recall the definition of the halting time sequence; it follows that there is some
$i\in\N$ such that $\tau=t_M^{(n)}(i)$. Let
\[
   |\psi^{(M,n)}\rangle:=|c_i^{(M,n)}\rangle\otimes
   \mathcal{C}_{\hr_M^{(n,\eps_0)}(\tau)}U^{-1}|\psi\rangle\,\,,
\]
that is, the blind prefix code of the halting number $i$, followed by
the standard compression (as constructed in Definition~\ref{DefStandard})
of some approximation $U^{-1}|\psi\rangle$ of $|\psi\rangle$ that
is in the subspace $\hr_M^{(n,\eps_0)}(\tau)$.
Note that
\begin{eqnarray*}
   \ell\left(|\psi^{(M,n)}\rangle\right)&=&\ell\left(
   c_i^{(M,n)}\right)+\ell\left(\mathcal{C}_{\hr_M^{(n,\eps_0)}(\tau)}U^{-1}|\psi\rangle\right)\\
   &=&\ell_i^{(M,n)}+\left\lceil \log\dim\hr_M^{(n,\eps_0)}(\tau)\right\rceil=n+1\,\,.
\end{eqnarray*}
If $\sigma=\sum_k \lambda_k |\psi_k\rangle\langle\psi_k|$ is a mixed fixed-length qubit string
which is $\tau$-halting for $M$, every convex component $|\psi_k\rangle$ must also be
$\tau$-halting for $M$, and it makes sense to define
$\sigma^{(M,n)}:=\sum_k\lambda_k |\psi_k^{(M,n)}\rangle
\langle\psi_k^{(M,n)}|$, where every $|\psi_k^{(M,n)}\rangle$
(and thus $\sigma^{(M,n)}$) starts with the same classical code word $c_i^{(M,n)}$,
and still $\sigma^{(M,n)}\in\mathcal{T}_1^+(\hr_{n+1})$.

The strongly universal QTM $\mathfrak U$ expects input of the form
\begin{equation}
   \left(s_M\otimes \sigma^{(M,n)},\delta\right)=:\left(\sigma_M,\delta\right)\,\,,
   \label{EqExpected}
\end{equation}
where $s_M\in\s$ is a self-delimiting description of the QTM $M$.
We will now give a description of how $\mathfrak U$ works; meanwhile, we will
always assume that the input is of the expected form (\ref{EqExpected}) and
also that the input $\sigma$ is a {\em pure} qubit string $|\psi\rangle\langle\psi|$ (we discuss the case
of mixed input qubit strings $\sigma$ afterwards):
\begin{itemize}
\item Read the parameter $\delta$ and the description $s_M$.
\item Look for the first blank symbol $\#$ on the tape to determine the length
$\ell(\sigma^{(M,n)})=n+1$.
\item Compute the halting time $\tau$. This is achieved as follows:
\begin{itemize}
\item[(1)] Set $t:=1$ and $i:=0$.
\item[(2)] Compute a description of $\hr_M^{(n,\eps_0)}(t)$. If $\dim \hr_M^{(n,\eps_0)}(t)=0$,
then go to step (5).
\item[(3)] Set $i:=i+1$ and set
$\ell_i^{(M,n)}:=n+1-\left\lceil \log\dim \hr_M^{(n,\eps_0)}\left(t\right)\right\rceil$.
From the previously computed code word lengths $\ell_j^{(M,n)}$ ($1\leq j\leq i$),
compute the corresponding blind prefix code word $c_i^{(M,n)}$. Bit by bit, compare
the code word $c_i^{(M,n)}$ with the prefix of $\sigma^{(M,n)}$. As soon as any difference is
detected, go to step (5).
\item[(4)] The halting time is $\tau:=t$. Exit.
\item[(5)] Set $t:=t+1$ and go back to step (2).
\end{itemize}
\item Let $|\tilde\psi\rangle$ be the rest of the input, i.e. $\sigma^{(M,n)}=:|c_i^{(M,n)}\rangle
\langle c_i^{(M,n)}|
\otimes |\tilde\psi\rangle\langle\tilde\psi|$ (up to a phase, this means that
$|\tilde\psi\rangle=\mathcal{C}_{\hr_M^{(n,\eps_0)}(\tau)}U^{-1}|\psi\rangle$). Apply
the quantum standard decompression algorithm $\mathfrak D$ given in Lemma~\ref{LemStandardDecomp},
i.e. compute $|\tilde\varphi\rangle:=\mathfrak{D}\left(\hr_M^{(n,\eps_0)}(\tau),
n,\frac\delta 3,|\tilde\psi\rangle\right)$.
Then,
\[
   \left\| \,|\tilde\varphi\rangle-\mathcal{D}_{\hr_M^{(n,\eps_0)}(\tau)}|\tilde\psi\rangle\right\|
   =\left\|\,|\tilde\varphi\rangle-U^{-1}|\psi\rangle\right\|<\frac\delta 3\,\,.
\]
\item Compute an approximation $V:\hr_n\to\hr_n$ of a unitary extension of $U$
with $\left\|U-V\upharpoonright\tilde\hr_M^{(n,\eps_0)}(\tau)\right\|<
\frac{\delta/3}{2(10\sqrt{2^n})^{2^n}}=:\eps$, where $U$ is some ``fine-tuning map'' as
constructed above.
This can be achieved as follows:
\begin{itemize}
\item Choose $N\in\N$ large enough such that $\sum_{k=N+1}^\infty {\rm const}_n\cdot
\left(\frac {18}{80}\right)^{\frac k 2}<\frac \varepsilon 2$, where ${\rm const}_n\in\R$
is the constant defined in Equation~(\ref{EqConstn}).
\item For every $k\in\{1,\ldots,N\}$, find matrices $V_k:\hr_n\to\hr_n$ that
approximate the forementioned\footnote{The isometries $U_k$ are not unique, so they can be chosen 
arbitrarily, except for the requirement that Equation~(\ref{EqConstn}) is satisfied,
and that every $U_k$ depends only on $\hr_M^{(n,\eps_k)}(t)$ and $\hr_M^{(n,\eps_{k-1})}(t)$
and not on other parameters.
} isometries $U_k:\hr_M^{(n,\eps_k)}(t)\to\hr_M^{(n,\eps_{k-1})}(t)$ such that
\[
   \left\| \prod_{k=1}^N \tilde U_k^* - \prod_{k=1}^N V_k^* \upharpoonright
   \tilde\hr_M^{(n,\eps_0)}(t)\right\| < \frac \eps 2\,\,.
\]
\end{itemize}
Setting $V:=\prod_{k=1}^N V_k^*$ will work as desired, since
\begin{eqnarray*}
   \left\| \prod_{k=1}^N \tilde U_k^*-U\right\| &\leq& \sum_{k=N+1}^\infty \| U_k-\idn \| \\
   &\leq& \sum_{k=N+1}^\infty {\rm const}_n\cdot \left(\frac{18}{80}\right)^{\frac k 2} < \frac \eps 2
\end{eqnarray*}
due to Equation~(\ref{EqConstn}) and the proof of Lemma~\ref{LemCompositionUnitary}.
\item Use $V$ to carry out a $\frac\delta 3$-approximation of a unitary extension $\tilde U$ of $U$ on the
state $|\tilde\varphi\rangle$ on the tape (the reason why this is possible
is explained in the proof of Lemma~\ref{LemStandardDecomp}). This results in a vector
$|\varphi\rangle$ with the property that $\|\,|\varphi\rangle-\tilde U|\tilde\varphi\rangle\|<\frac\delta 3$.
\item Simulate $M$ on input $|\varphi\rangle\langle\varphi|$ for $\tau$ time steps
within an accuracy of $\frac\delta 3$, that is, compute an output track state $\rho_{\mathbf O}
\in\mathcal{T}_1^+(\hr_{\mathbf O})$ with $\left\|\rho_{\mathbf O}-M_{\mathbf O}^{\tau}(|\varphi\rangle\langle\varphi|)
\right\|_{\rm Tr}<\frac\delta 3$, move this state to the own output track and halt.
(It has been shown by Bernstein and Vazirani in \cite{BernsteinVazirani} that
there are QTMs that can do a simulation in this way.)
\end{itemize}
Let $\sigma_M:=s_M\otimes \sigma^{(M,n)}$. Using the contractivity of the trace distance with respect to
quantum operations and Lemma~\ref{LemNormInequalities}, we get
\begin{eqnarray*}
   \left\| \mathfrak{U}\left(\sigma_M,\delta\right) \right.-\left. M(|\psi\rangle\langle\psi|)\right\|_{\rm Tr}&=&
   \left\|\mathcal{R}(\rho_{\mathbf O})-\mathcal{R}\left(M_{\mathbf O}^\tau(|\psi\rangle
   \langle\psi|)\right)\right\|_{\rm Tr}\\
   &\leq&\left\|\rho_{\mathbf O}-M_{\mathbf{O}}^\tau(|\varphi\rangle\langle\varphi|)\right\|_{\rm Tr}\\
   &&+\left\|M_{\mathbf{O}}^\tau(|\varphi\rangle\langle\varphi|)
   -M_{\mathbf{O}}^\tau(|\psi\rangle\langle\psi|)\right\|_{\rm Tr}\\
&<&\frac\delta 3 +\left\| |\varphi\rangle\langle\varphi|-|\psi\rangle\langle\psi|\right\|_{\rm Tr}\\
   &\leq& \frac\delta 3 +\|\,|\varphi\rangle-|\psi\rangle\|\\
   &\leq& \frac\delta 3 +\|\,|\varphi\rangle-
   \tilde U|\tilde\varphi\rangle\|+\|\tilde U|\tilde\varphi\rangle-|\psi\rangle\|\\
   &<& \frac 2 3 \delta +\left\|\,|\tilde\varphi\rangle-\tilde U^* |\psi\rangle\right\|<\delta\,\,.
\end{eqnarray*}
This proves the claim for pure inputs $\sigma=|\psi\rangle\langle\psi|$. If
$\sigma=\sum_k \lambda_k |\psi_k\rangle\langle \psi_k|$
is a mixed qubit string as explained right before Equation~(\ref{EqExpected}), the result just
proved holds for every convex component of $\sigma$ by the linearity of $M$, i.e.
$\left\| \rho_k-M(|\psi_k\rangle\langle\psi_k|)\right\|_{\rm Tr}<\delta$, and the assertion
of the theorem follows from the joint convexity of the trace distance and the observation
that $\mathfrak U$ takes the same number of time steps for every convex component
$|\psi_k\rangle\langle\psi_k|$.\qed

This proof relies on the existence of a universal QTM $\mathcal U$ in the sense of Bernstein
and Vazirani as given in Equation~(\ref{EqWeakUniversality}). Nevertheless, the proof does not imply
that every QTM that satisfies (\ref{EqWeakUniversality}) is automatically strongly universal
in the sense of Theorem~\ref{MainTheorem1}; for example, we can construct a QTM $\mathcal U$
that always halts after $T$ simulated steps of computation on input $(s_M,T,\delta,|\psi\rangle)$
and that does not halt at all if the input is not of this form. So formally,
\[
   \{{\mathcal U}\mbox{ QTM universal by~(\ref{EqWeakUniversality})}\}
   \supsetneq
   \{{\mathfrak U}\mbox{ QTM strongly universal}\}.
\]

We are now going to sketch the proof of Proposition~\ref{PropTwoParameter} and
the proof idea of Conjecture~\ref{TheAvLengthUQTM}. The reason why we do not give the full proof is
that this full proof would consist by a large part only of certain analytic
estimates that show to what accuracy the universal QTM $\mathfrak U$ should
do its calculations. This would be a very long proof, consisting of many
routine calculations which are not very helpful for a reader.

Remember the proof of Theorem~\ref{MainTheorem1}. The proof idea was to let the
universal QTM $\mathfrak U$ compute approximations of the halting spaces of
the other QTM $M$ and use this information to ``uncompress'' some cleverly chosen
input and simulate $M$ in a classically controlled manner. The subsequent lengthy
proof showed that the UQTM $\mathfrak U$ was really able to approximate these
halting spaces well enough to make the proof idea work. This had to be worked out in detail
at least once for this special situation, to be sure that there are no subtle difficulties inherent to
the computable approximations. Nevertheless, since every map and structure
that we encountered was continuous and finite-dimensional, it is not so surprising
that everything worked fine.

Consequently, we will now only sketch the proof of Proposition~\ref{PropTwoParameter}
and the proof idea of Theorem~\ref{TheAvLengthUQTM}, by only specifying what kind of structures (analogues
of the halting spaces) $\mathfrak U$ is supposed to approximate, but without
specifying in detail to what accuracy $\mathfrak U$ should do its approximations.

Both proof sketches that follow are based on the idea that a QTM which is universal in the
sense of Bernstein and Vazirani (i.e. as in Equation~(\ref{EqWeakUniversality})) has
a dense set of unitaries that it can apply exactly. We can call such unitaries on $\hr_n$
for $n\in\N$ {\em $\mathfrak U$-exact unitaries}.

This follows from the result by Bernstein and Vazirani that the corresponding UQTM $\mathcal U$
can apply a unitary map $U$ on its tapes within any desired accuracy, if it is
given a description of $U$ as input. It does so by decomposing $U$ into simple (``near-trivial'') unitaries
that it can apply directly (and thus exactly).

We can also call an $n$-block projector $P\in\mathcal{B}(\hr_n)$ $\mathfrak U$-exact if it has
some spectral decomposition $P=\sum_i |\psi_i\rangle\langle\psi_i|$ such that there is a $\mathfrak U$-exact
unitary that maps each $|\psi_i\rangle$ to some computational basis vector of $\hr_n$. If $P$
and $\idn-P$ are $\mathfrak U$-exact projectors on $\hr_n$, then $\mathfrak U$ can do something
like a ``yes-no-measurement'' according to $P$ and $\idn-P$: it can decide whether some vector $|\psi\rangle\in\hr_n$
on its tape is an element of ${\rm ran}\,P$ or of $({\rm ran}\,P)^\perp$ with certainty (if either one
of the two cases is true), just by applying the corresponding $\mathfrak U$-exact unitary, and then
by deciding whether the result is some computational basis vector or another.\\

{\bf Proof Sketch of Proposition~\ref{PropTwoParameter}.}
In analogy to Definition~\ref{DefHaltingQubitStrings}, we can define halting spaces
$\hr_M^{(n)}(t_1,t_2,\ldots,t_j)$ as the linear span of
\begin{eqnarray*}
   H_M^{(n)}(t_1,t_2,\ldots,t_j):=\{|\psi\rangle\in\hr_n\,\,|\,\,
   (|\psi\rangle\langle\psi|,i)\mbox{ is }t_i\mbox{-halting for }
    M\,\,(1\leq i\leq j)\}.
\end{eqnarray*}
Again, we have
$\hr_M^{(n)}\left((t_i)_{i=1}^j\right)\perp \hr_M^{(n)}\left((t'_i)_{i=1}^j\right)$ if $t\neq t'$,
and now it also holds that $\hr_M^{(n)}(t_1,\ldots,t_{j},t_{j+1})\subset
\hr_M^{(n)}(t_1,\ldots,t_j)$ for every $j\in\N$.
Moreover, we can define certain $\delta$-approximations $\hr_M^{(n,\delta)}(t_1,\ldots,t_j)$.
We will not get into detail; we will just claim that such a definition can be found in a way
such that these $\delta$-approximations
share enough properties with their counterparts from Definition~\ref{DefApproxHalt}
to make the algorithm given below work.

We are now going to describe how a machine $\mathfrak U$ with the properties given
in the assertion of the proposition works. It expects input of the form
$\left(k,f\otimes s_M\otimes \sigma^{(M,n)}\right)$, where $f\in\{0,1\}$ is a single bit,
$s_M\in\s$ is a self-delimiting description of the QTM $M$, $\sigma^{(M,n)}\in\mathcal{T}_1^+(\hr_\s)$
is a qubit string, and $k\in\N$ an arbitrary integer. For the same reasons as in the proof of
Theorem~\ref{MainTheorem1}, we may without loss of generality assume that the input
is a pure qubit string, so $\sigma^{(M,n)}=|\psi^{(M,n)}\rangle\langle\psi^{(M,n)}|$.
Moreover, due to Lemma~\ref{LemFixedLengthEnough}, we may also assume that $M$ is a fixed-length QTM,
and so $\sigma^{(M,n)}\in\mathcal{T}_1^+(\hr_n)$ is a fixed-length qubit string.

These are the steps that $\mathfrak U$ performs:
\begin{itemize}
\item[(1)] Read the first bit $f$ of the input. If it is a $0$, then proceed with the rest of
the input the same way as the QTM that is given in Theorem~\ref{MainTheorem1}.
If it is a $1$, then proceed with the next step.
\item[(2)] Read $s_M$, read $k$, and look for the first blank symbol $\#$ to determine
the length $n:=\ell(\sigma^{(M,n)})$.
\item[(3)] Set $j:=1$ and $\delta_0\in\mathbb{Q}^+$ (depending on $n$) small enough.
\item[(4)] Set $t:=1$.
\item[(5)] Compute $\hr_M^{(n,\delta_0)}(\tau_1,\ldots,\tau_{j-1},t)$. Find a $\mathfrak U$-exact
projector $P_M^{(n)}(\tau_1,\ldots,\tau_{j-1},t)$ with the following properties:
\begin{itemize}
\item[$\bullet$] $P_M^{(n)}(\tau_1,\ldots,\tau_{j-1},t')\cdot P_M^{(n)}(\tau_1,\ldots,\tau_{j-1},t)=0$
for every $1\leq t'<t$,
\item[$\bullet$] $P_M^{(n)}(\tau_1,\ldots,\tau_{j-1},t)\leq P_M^{(n)}(\tau_1,\ldots,\tau_{j-1})$,
\item[$\bullet$] the support of $P_M^{(n)}(\tau_1,\ldots,\tau_{j-1},t)$ is close enough to
$\hr_M^{(n,\delta_0)}(\tau_1,\ldots,\tau_{j-1},t)$.
\end{itemize}
\item[(6)] Make a measurement\footnote{It is not really a measurement, but rather some unitary branching:
if $|\psi^{(M,n)}\rangle$ is some superposition in between both subspaces $W:={\rm supp}\left(P_M^{(n)}(\tau_1,\ldots,\tau_{j-1},t)\right)$
and $W^\perp$, then the QTM will do both possible steps in superposition.
} described by $P_M^{(n)}(\tau_1,\ldots,\tau_{j-1},t)$.
If $|\psi^{(M,n)}\rangle$ is an element of the support of $P_M^{(n)}(\tau_1,\ldots,\tau_{j-1},t)$,
then set $\tau_j:=t$ and go to step (7). Otherwise, if $|\psi^{(M,n)}\rangle$ is an element of
the orthogonal complement of the support, set $t:=t+1$ and go back to step (5).
\item[(7)] If $j<2k$, then set $j:=j+1$ and go back to step (4).
\item[(8)] Use a unitary transformation $V$ (similar to the transformation $V$ from the proof
of Theorem~\ref{MainTheorem1}) to do some ``fine-tuning'' on $|\psi^{(M,n)}\rangle$, i.e.
to transform it closer (depending on the parameter $k$) to some space $\tilde\hr_M^{(n)}(\tau_1,\ldots,\tau_j)\supset
\hr_M^{(n)}(\tau_1,\ldots,\tau_j)$ containing the exactly halting vectors.
Call the resulting vector $|\tilde\psi^{(M,n)}\rangle:=V|\psi^{(M,n)}\rangle$.
\item[(9)] Simulate $M$ on input $\left(2k,|\tilde\psi^{(M,n)}\rangle\langle\tilde\psi^{(M,n)}|\right)$ for $\tau_{2k}$
time steps within some accuracy that is good enough, depending on $k$.
\end{itemize}

Let $\tilde\hr_M^{(n,\delta_0)}(t_1,\ldots,t_j)$ be the support of $P_M^{(n)}(t_1,\ldots,t_j)$.
These spaces (which are computed by the algorithm) have the properties
\begin{eqnarray*}
   \tilde\hr_M^{(n,\delta_0)}\left((t_i)_{i=1}^j\right)&\perp& \tilde\hr_M^{(n,\delta_0)}\left((t'_i)_{i=1}^j\right)
   \mbox{ if }t\neq t',\\
   \tilde\hr_M^{(n,\delta_0)}(t_1,\ldots,t_j,t_{j+1})&\subset& \tilde\hr_M^{(n,\delta_0)}(t_1,\ldots,t_j)
   \enspace \forall j\in\N,
\end{eqnarray*}
which are the same as those of the exact halting spaces $\hr_M^{(n)}(t_1,\ldots,t_j)$.
If all the approximations are good enough, then for every $|\psi\rangle\in H_M^{(n)}(t_1,\ldots,t_j)$
there will be a vector $|\psi^{(M,n)}\rangle\in \tilde\hr_M^{(n,\delta_0)}(t_1,\ldots,t_j)$ such
that $\|\,|\psi\rangle-V|\psi^{(M,n)}\rangle\|$ is small. If this $|\psi^{(M,n)}\rangle$ is given
to $\mathfrak U$ as input together with all the additional information explained above, then this algorithm will unambiguously find out
by measurement with respect to the $\mathfrak U$-exact projectors that it computes in step (5) what the
halting time of $|\psi\rangle$ is, and the simulation of $M$ will halt after the correct
number of time steps with probability one and an output which is close to the true output
$M(2k,\sigma)$.\qed\\

{\bf Proof Idea for Conjecture~\ref{TheAvLengthUQTM}.}
The first difficulty that arises in considering average length $\bar\ell$ instead of base length $\ell$
is that it is no more sufficient to consider fixed-length QTMs. Moreover, while the pure qubit strings $|\psi\rangle$
with base length $\ell(|\psi\rangle)\leq n$ are all elements of some (small) subspace $\hr_{\leq n}\subset\hr_\s$,
this is no more true for the qubit strings with average length $\bar\ell(|\psi\rangle)\leq n$.
But to do numerical approximations, we should be able to restrict to some finite-dimensional subspace.

To resolve this difficulty, note that if $\sigma\in\mathcal{T}_1^+(\hr_\s)$ is any input qubit string which makes
a QTM $M$ halt after $t$ time steps, then $M$ cannot have read more than $t$ cells of its tape. Thus, it
follows that also the restriction of $\sigma$ to the first $t$ cells (called $\sigma_1^t$ and defined
on page \pageref{DefPrefixQTM})
makes the QTM behave completely equivalently:
\[
   \sigma\mbox{ is }t\mbox{-halting for }M\Rightarrow M(\sigma)=M(\sigma_1^t).
\]
But if $M$ is a prefix QTM, as in the statement of the theorem that we are about to prove,
then it must hold that $\ell(\sigma)\leq t$, or equivalently, $\sigma=\sigma_1^t$, because otherwise,
Lemma~\ref{LemPrefixQTM} would be violated.
Thus,
\[
   \hr_M(t)\subset\hr_{\leq t}\qquad\mbox{since }M\mbox{ is a prefix QTM}.
\]
Again, we assume that we can define certain computable approximations $\hr_M^{(\eps)}(t)$,
where $\eps>0$ is some approximation parameter, that approximate the true halting spaces $\hr_M(t)$ good
enough to make the algorithm that follows work. We also assume that the approximate halting spaces $\hr_M^{(\eps)}(t)$
share the property $\hr_M^{(\eps)}(t)\perp\hr_M^{(\eps)}(t')$
for $t\neq t'$ with the true halting spaces $\hr_M(t)$ that they approximate.

Moreover, we want to use the prefix property of $M$, and demand that the approximate halting
spaces $\hr_M^{(\eps)}(t)$ have the prefix property of Definition~\ref{DefPrefixQTM}, i.e.
if $|\psi\rangle\in\hr_M^{(\eps)}(t)$ and $|\varphi\rangle\in\hr_M^{(\eps)}(t')$ for some $t,t'\in\N$
such that $\ell(|\varphi\rangle)>\ell(|\psi\rangle)=:n$, then it holds
\begin{equation}
   \langle\psi|\left(|\varphi\rangle\langle\varphi|_1^n\right)|\psi\rangle=0.
   \label{EqPrefix}
\end{equation}

For the same reason as in the proof of Theorem~\ref{MainTheorem1} (i.e. the number of steps that the following
algorithm takes depends only on the running time of the calculation that it simulates), we may restrict
to pure input qubit strings. The algorithm that $\mathfrak U$ performs expects input of the form
$(\delta,s_M\otimes |\psi^{(M)}\rangle\langle\psi^{(M)}|)$, where $\delta\in\mathbb{Q}^+$ is the parameter
from the statement of the theorem, $s_M\in\s$ is some description of a QTM $M$,
and $|\psi^{(M)}\rangle\in\hr_\s$ is an arbitrary indeterminate-length qubit string. It proceeds as follows:
\begin{itemize}
\item[(1)] Read $\delta$, read $s_M$, and let $t:=1$.
\item[(2)] Compute a description of the space $\hr_M^{(\eps)}(t)$. Find a $\mathfrak U$-exact projector $P_M^{(\eps)}(t)\in
\mathcal{B}(\hr_{\leq t})$ with the following properties:
\begin{itemize}
\item[$\bullet$] the support $\tilde\hr_M^{(\eps)}(t)$ of $P_M^{(\eps)}(t)$ is a good approximation of $\hr_M^{(\eps)}(t)$,
\item[$\bullet$] $P_M^{(\eps)}(t)\cdot P_M^{(\eps)}(t')=0$ for every $t'<t$, i.e. for all previously
computed $\mathfrak U$-exact projectors,
\item[$\bullet$] the collection of support subspaces $\bigcup_{t'=1}^t \tilde\hr_M^{(\eps)}(t')$ satisfies
Equation~(\ref{EqPrefix}), i.e. is prefix-free. It is not clear
if this is easy to achieve; this is exactly the point why the statement is just a conjecture, not a theorem.
\end{itemize}
\item[(3)] Make a measurement\footnote{Again, this is not really a measurement, but rather some unitary branching.} described
by the projectors $P_M^{(\eps)}(t)$ and $\idn_{\hr_{\leq t}}-P_M^{(\eps)}(t)$, i.e. decide whether $|\psi^{(M)}\rangle$ is an element
of $\tilde\hr_M^{(\eps)}(t)$ or of its orthogonal complement. In the first case, go to step (4). In the second case, let
$t:=t+1$ and go to step (2).
\item[(4)] Use a unitary transformation $V$ (similar to the transformation $V$ from the proof
of Theorem~\ref{MainTheorem1}) to do some ``fine-tuning'' on $|\psi^{(M)}\rangle$, i.e.
to transform it closer (depending on the parameter $\delta$) to some space $\tilde\hr_M(t)\supset
\hr_M(t)$ containing the exactly halting vectors.
Call the resulting vector $|\tilde\psi^{(M)}\rangle:=V|\psi^{(M)}\rangle$.
\item[(5)] Simulate the QTM $M$ for $t$ time steps on input $|\tilde\psi^{(M)}\rangle\langle\tilde\psi^{(M)}|$,
move the corresponding output to the output track and halt.
\end{itemize}
If all the approximations are good enough, then for every $|\psi\rangle\in H_M(t)$ there should be a vector $|\psi^{(M)}\rangle
\in\tilde\hr_M^{(\eps)}(t)$ such that $\|\,|\psi\rangle-V|\psi^{(M)}\rangle\|$ is small. If $|\psi^{(M)}\rangle$
is given to the QTM $\mathfrak V$ as input together with $s_M$ and $\delta$ as shown above, then this
algorithm will find out by measurement with respect to the $\mathfrak V$-exact projectors given above in step (2)
what the corresponding halting time is, and the simulation of $M$ will halt after the correct number of time
steps with probability one.

Note that the ``measurement'' in step (3) only works because $M$ is a prefix QTM: in the case that $|\psi^{(M)}\rangle\in
\tilde\hr_M^{(\eps)}(t')$ for some $t'>t$ and $\ell(|\psi^{(M)}\rangle)>t$, this fact guarantees that the measurement
result will always be that $|\psi^{(M)}\rangle$ is in the orthogonal complement of $\tilde\hr_M^{(\eps)}(t)$,
even though the measurement cannot access the state $|\psi^{(M)}\rangle$ completely.

It also seems that if $s_M$ and $\delta$ are encoded into the input in a clever way, then $\mathfrak V$ inherits
the property of being prefix-free from the QTMs that it simulates. But again, this has to be checked in more detail
once this proof idea will be turned into a complete proof.
\qed

\section{Halting Stability}
\label{SecHaltingStability}
In this thesis, we have defined that a QTM halts at some time $t$ according to Equation~(\ref{EqHalting})
if and only if its control is exactly in the halting state $|q_f\rangle$ at time $t$, and exactly orthogonal
to the halting state before. We have argued in Section~\ref{SecHaltingUniv} why this halting definition
is useful and natural, at least for our purpose to study quantum Kolmogorov complexity.

Yet, it may first seem that this halting definition is too restrictive, since it dismisses
every input which halts only approximately, but not perfectly, even if it is very close to halting.
In this section, we show that this definition of halting has some built-in error tolerance that
was not expected at the beginning: for every input which makes a QTM {\em almost} halt, there is
another input which is at most a constant number of qubits longer, and which makes the
universal QTM halt {\em perfectly}.

Thus, the definition of halting that we use in this thesis (and that was first considered by Bernstein
and Vazirani) is not as ``unphysical'' as it first seems, but makes perfect sense.

We start by showing that superpositions of almost halting input qubit strings are again almost
halting. To establish this result, we need some estimation of a matrix element
appearing in the superposition's density matrix.

\begin{flemma}[Halting Matrix Element]\lineclear
\label{LemHaltingMatrixElement}
Let $M$ be a QTM, let $|\varphi\rangle\in\hr_\s$ be $\varepsilon$-$t$-halting for $M$,
and let $|\psi\rangle\in\hr_\s$ be $\delta$-$t$-halting for $M$.
Then, the operator $|\varphi\rangle\langle\psi|$ satisfies
\begin{eqnarray*}
\left|\langle q_f| M_{\mathbf{C}}^{t'}(|\varphi\rangle\langle\psi|)|q_f\rangle\right|
&\leq&\sqrt{\varepsilon\delta}\qquad\mbox{for every }t'<t,\mbox{ and}\\
\left|\sum_{q\in Q:q\neq q_f}\langle q| M_{\mathbf{C}}^t (|\varphi\rangle\langle\psi|)
|q\rangle\right|&\leq&\sqrt{\varepsilon\delta}.
\end{eqnarray*}
\end{flemma}
\proof Let $V_M\in\mathcal{B}(\hr_{QTM})$ be the unitary time evolution operator of $M$.
Identifying $|\varphi\rangle\in\hr_\s$ with the initial state of the QTM $M$ on input $|\varphi\rangle$,
we write
\begin{equation}
   V_M^{t'}|\varphi\rangle=\sum_{q\in Q,b\in B} \alpha_{qb}^{t'}|q\rangle\otimes |b\rangle
   \label{eqDevelopPhi}
\end{equation}
for every $t'\in\N_0$, where $B$ is an arbitrary orthonormal basis of $\hr_{\mathbf I}
\otimes \hr_{\mathbf O}\otimes \hr_{\mathbf H}$. Multiplying and computing the partial trace,
we get
\[
   {\rm Tr}_{\mathbf{IOH}} V_M^{t'}|\varphi\rangle\langle\varphi|(V_M^{t'})^*=
   \sum_{q\in Q,q'\in Q,b\in B} \alpha_{qb}^{t'}\bar\alpha_{q'b}^{t'}|q\rangle\langle q'|.
\]
By the assumptions of the theorem, it follows
\[
   \langle q_f|{\rm Tr}_{\mathbf{IOH}} V_M^{t'}|\varphi\rangle\langle\varphi|(V_M^{t'})^*|q_f\rangle=
   \sum_{b\in B}|\alpha_{q_f b}^{t'}|^2\left\{
      \begin{array}{ll}
         \leq \varepsilon & \mbox{if }t'<t, \\
         \geq 1-\varepsilon & \mbox{if }t'=t.
      \end{array}
   \right.
\]
Similarly, for $|\psi\rangle$, we get the inequality
\[
   \langle q_f|{\rm Tr}_{\mathbf{IOH}} V_M^{t'}|\psi\rangle\langle\psi|(V_M^{t'})^*|q_f\rangle=
   \sum_{b\in B}|\beta_{q_f b}^{t'}|^2\left\{
      \begin{array}{ll}
         \leq \delta & \mbox{if }t'<t, \\
         \geq 1-\delta & \mbox{if }t'=t,
      \end{array}
   \right.
\]
where the coefficients $\beta_{qb}^{t'}$ are defined analogously as in Equation~(\ref{eqDevelopPhi}).
Now suppose $t'<t$. Then, we get by the Cauchy-Schwarz inequality
\begin{eqnarray*}
   \left|\langle q_f|M_{\mathbf{C}}^{t'}(|\varphi\rangle\langle\psi|)|q_f\rangle\right|
   &=&\left|\sum_{b\in B} \alpha_{q_f b}^{t'}\bar\beta_{q_f b}^{t'}\right|\leq \sqrt{\sum_{b\in B} |\alpha_{q_f b}^{t'}|^2}
   \cdot \sqrt{\sum_{b\in B} |\beta_{q_f b}^{t'}|^2}\\
   &\leq&\sqrt{\varepsilon\delta}.
\end{eqnarray*}
Using the Cauchy-Schwarz inequality again, we get for $t'=t$
\begin{eqnarray*}
   \left|\sum_{q\in Q:q\neq q_f}\langle q|
      M_{\mathbf{C}}^{t}(|\varphi\rangle\langle\psi|)|q\rangle\right|
   &=&\left|\sum_{q\in Q:q\neq q_f} \sum_{b\in B} \alpha_{qb}^t\bar\beta_{qb}^t\right|\\
   &\leq&\sqrt{\sum_{q\in Q:q\neq q_f} \sum_{b\in B} |\alpha_{qb}^t|^2}
   \cdot \sqrt{\sum_{q\in Q:q\neq q_f} \sum_{b\in B} |\beta_{qb}^t|^2}\\
   &=&\sqrt{1-\underbrace{\sum_{b\in B}|\alpha_{q_f b}^t|^2}_{\geq 1-\varepsilon}}
   \cdot \sqrt{1-\underbrace{\sum_{b\in B}|\beta_{q_f b}^t|^2}_{\geq 1-\delta}}\\
   &\leq& \sqrt{\varepsilon}\cdot \sqrt{\delta}.
\end{eqnarray*}
The claim follows.
\qed

\begin{flemma}[Approximate Halting of Superpositions]
\label{LemApproxHaltingSuperpositions}
\lineclear
Let $M$ be a QTM, $t\in\N$, and $\{\varepsilon_i\}_{i=1}^N\subset \R^+$ be a set of positive numbers.
Moreover, let $\{|\varphi_i\rangle\}_{i=1}^N\subset \hr_\s$ be a set of normalized vectors,
i.e. pure qubit strings, such that every $|\varphi_i\rangle$ is $\varepsilon_i$-$t$-halting for $M$.\\
If $|\varphi\rangle=\sum_{i=1}^N \alpha_i |\varphi_i\rangle$ is normalized,
then $|\varphi\rangle$ is $\left(\sum_{i=1}^N |\alpha_i| \sqrt{\varepsilon_i}\right)^2$-$t$-halting for $M$.
\end{flemma}

\proof Let $\rho:=|\varphi\rangle\langle\varphi|=\sum_{i,j=1}^N \alpha_i \bar\alpha_j |\varphi_i\rangle
\langle\varphi_j|$. Using Lemma~\ref{LemHaltingMatrixElement}, we get for $t'<t$
\begin{eqnarray*}
\langle q_f| M_{\mathbf C}^{t'}(\rho)|q_f\rangle &\leq& \sum_{i,j=1}^N |\alpha_i|
|\alpha_j| \left| \langle q_f| M_{\mathbf C}^{t'}(|\varphi_i\rangle\langle \varphi_j|)|q_f\rangle\right|\\
&\leq& \sum_{i,j=1}^N |\alpha_i| |\alpha_j| \sqrt{\varepsilon_i}\sqrt{\varepsilon_j}
=\left(\sum_{i=1}^N |\alpha_i|\sqrt{\varepsilon_i}\right)^2.
\end{eqnarray*}
Moreover, for $t'=t$, we have
\begin{eqnarray*}
\langle q_f|M_{\mathbf C}^t(\rho)|q_f\rangle &=& 1-\sum_{q\in Q:q\neq q_f}
\langle q|M_{\mathbf C}^t(\rho)|q\rangle\\
&\geq&1-\sum_{i,j=1}^N |\alpha_i| |\alpha_j| \left| \sum_{q\in Q: q\neq q_f}
\langle q| M_{\mathbf C}^t(|\varphi_i\rangle\langle \varphi_j|)|q\rangle\right|\\
&\geq& 1-\sum_{i,j=1}^N |\alpha_i| |\alpha_j| \sqrt{\varepsilon_i \varepsilon_j}
=1-\left(\sum_{i=1}^N |\alpha_i|\sqrt{\varepsilon_i}\right)^2.\mbox{\qed}
\end{eqnarray*}

To prove the result about halting stability, we need another lemma which states
that almost halting qubit strings with different halting times are almost
orthogonal to each other.

\begin{flemma}[Almost-Orthogonality]
\label{LemAlmostOrtho2}
Let $M$ be a QTM, and let $|\varphi\rangle,|\psi\rangle\in\hr_\s$ be two normalized
pure qubit strings. If $|\varphi\rangle$ is $\varepsilon$-$t$-halting for $M$, and
$|\psi\rangle$ is $\delta$-$t'$-halting for $M$ with $t\neq t'$, and if $\varepsilon+\delta\leq 1$, then
\[
   |\langle\psi|\varphi\rangle|\leq\sqrt{1-(1-\varepsilon-\delta)^2}.
\]
\end{flemma}
\proof We may assume that $t<t'$. Then we have
\begin{eqnarray*}
\langle q_f|M_{\mathbf C}^t(|\varphi\rangle\langle\varphi|)|q_f\rangle\geq 1-\varepsilon \quad\mbox{ and }\quad
\langle q_f|M_{\mathbf C}^t(|\psi\rangle\langle\psi|)|q_f\rangle\leq \delta.
\end{eqnarray*}
By the monotonicity of the trace distance with respect to quantum operations and
the definition of the trace distance for pure states together with Lemma~\ref{LemNormInequalities}, we get
\begin{eqnarray*}
   1-\varepsilon-\delta&\leq& \left|\langle q_f|M_{\mathbf C}^t(|\varphi\rangle\langle\varphi|)|q_f\rangle
   -\langle q_f|M_{\mathbf C}^t(|\psi\rangle\langle\psi|)|q_f\rangle\right|\\
   &\leq&\left\| M_{\mathbf C}^t(|\psi\rangle\langle\psi|)-M_{\mathbf C}^t(|\varphi\rangle
   \langle\varphi|)\right\|\\
   &\leq&\left\| M_{\mathbf C}^t(|\psi\rangle\langle\psi|)-M_{\mathbf C}^t(|\varphi\rangle
   \langle\varphi|)\right\|_{\rm Tr}\\
   &\leq&\|\,|\psi\rangle\langle\psi|-|\varphi\rangle\langle\varphi|\,\|_{\rm Tr}
   =\sqrt{1-|\langle\psi|\varphi\rangle|^2}.
\end{eqnarray*}
The claim follows by rearranging.\qed

We are now ready to prove the promised result about halting stability. The idea is to show in the first part of the proof
that every pure qubit string of fixed length $n$ which makes a QTM $M$ almost halt at time $t$
is close to some ``approximation subspace'' $L_M^{(n)}(t)\subset\hr_n$. Under certain assumptions
on the halting accuracy,
the dimensions of the spaces $L_M^{(n)}(t)$ for different $t$ add up to at most $2^n=\dim\hr_n$.

Then, as the second part of the proof, we can repeat the construction from Section~\ref{SecConstruction}, where the halting
spaces are replaced by these approxi\-mation spaces: we split every vector from $L_M^{(n)}(t)$ into
some classical and quantum part, and we can write a computer program for the UQTM $\mathfrak{U}$
that extracts the approximate halting time from the classical part, then simulates the QTM $M$
for the corresponding number of time steps, and finally halts with probability one.

Note that it is not trivial that such subspaces $L_M^{(n)}(t)$ with the aforementioned properties exist;
in particular, the halting spaces $\hr_M^{(n)}(t)$ themselves do not have this approximation property.
It is also does not seem that the approximate halting spaces $\hr_M^{(n,\delta)}(t)$
from Definition~\ref{DefApproxHalt} can be used instead.

\begin{ftheorem}[Halting Stability]
For every $\delta>0$, there is a sequence $\{a_n(\delta)\}_{n\in\N}\subset\R^+$ such that
every qubit string of length $n$ which is $a_n(\delta)$-halting can be enhanced to another
qubit string which is only a constant number of qubits longer, but which halts perfectly
and gives the same output up to trace distance $\delta$.

Moreover, the sequence $\{a_n(\delta)\}_{n\in\N}$ is computable.
\end{ftheorem}
{\bf Remark.} Here is the exact formal statement of the theorem: For every $\delta>0$,
there exists a sequence of positive real numbers $\{a_n(\delta)\}_{n\in\N}$ such that
for every QTM $M$, one can find a constant\footnote{Note that $c_{M,\delta}$ does not
depend on $n$.} $c_{M,\delta}\in\N$ such that for every qubit
string $\sigma\in\mathcal{T}_1^+(\hr_\s)$ which is $a_n(\delta)$-$t$-halting for $M$ for some $t\in\N$
and $\ell(\sigma)\leq n$, there is some qubit string $\sigma'\in\mathcal{T}_1^+(\hr_\s)$
with $\ell(\sigma')\leq n+c_{M,\delta}$ such that
\[
   \left\| \mathfrak{U}(\sigma')-\mathcal{R}\left(M_{\mathbf O}^t(\sigma)\right)\right\|_{\rm Tr}<\delta,
\]
where $\mathfrak U$ is some strongly universal QTM. Furthermore, $\mathfrak U$ halts perfectly on input $\sigma'$,
and the map $(n,\delta)\mapsto a_n(\delta)$ is computable.

\proof Assume that $\delta\in(0,1)\cap\mathbb{Q}$. We introduce two
different norms that will be useful in the proof.
For every $\Psi=\{|\psi_1\rangle,\ldots,|\psi_{2^n}\rangle\}\subset \cn$
which is a basis of $\cn$ consisting of normalized vectors,
and for every $|\varphi\rangle\in\cn$,
we define
\[
   \|\,|\varphi\rangle\|_\Psi:=\sum_{i=1}^{2^n} |\alpha_i|\quad\mbox{if }
   |\varphi\rangle=\sum_{i=1}^{2^n}\alpha_i|\psi_i\rangle.
\]
It is easily checked that $\|\cdot\|_\Psi$ is a norm on $\cn$ for every basis $\Psi$.
Suppose we have a set of vectors $\Psi=\{|\psi_1\rangle,\ldots,|\psi_{2^n}\rangle\}\subset S_n$ with the property
\begin{equation}
   \|\,|\psi_i\rangle-|\psi\rangle\|\geq \frac\delta 2\mbox{ for every }|\psi\rangle\in
   {\rm span}\{|\psi_1\rangle,\ldots,|\psi_{i-1}\rangle\},
   \label{EqChainProperty}
\end{equation}
then it is easily checked that the vectors of this set must be linearly independent. Since $\#\Psi=2^n$,
$\Psi$ must be a basis of $\cn$. Thus, the expression
\[
   \| \,|\varphi\rangle\|_{(\delta)}:=\sup\{\|\,|\varphi\rangle\|_\Psi\,\,|\,\,\Psi=\{|\psi_1\rangle,
   \ldots,|\psi_{2^n}\rangle\}\subset S_n,(\ref{EqChainProperty})\mbox{ holds for }\Psi
   \}
\]
is well-defined for every $|\varphi\rangle\in\cn$. Yet, it might be infinite for some $|\varphi\rangle$.
To see that it is finite for every $|\varphi\rangle\in\cn$, note that the set
\[
   \left\{ \Psi\in \underbrace{S_n\times S_n\times\ldots\times S_n}_{2^n\mbox{ factors}}\,\,|
   \,\, (\ref{EqChainProperty})\mbox{ holds for }\Psi\right\}
\]
is compact in $(\hr_n)^{2^n}$, and the map $\Psi\mapsto \|\,|\varphi\rangle\|_\Psi$ is continuous\footnote{To see
that this map is continuous, note that $\Psi$ can be interpreted as an invertible $2^n\times 2^n$-matrix. Thus,
$\|\,|\varphi\rangle\|_\Psi=\left\| \Psi^{-1}|\varphi\rangle\right\|_1$, and the map $\Psi\mapsto\Psi^{-1}$ is
continuous.}on this set and must thus have a maximum.

One easily checks that $\|\cdot\|_{(\delta)}$ is also a norm on $\cn$. Since all norms
on finite-dimensional linear spaces are equivalent, it follows that
\[
   m_n(\delta):=\sup_{|\varphi\rangle\in S_n} \|\,|\varphi\rangle\|_{(\delta)}
\]
is finite, and $m_n(\delta)\in\R^+$ for every $n$. Now we set
\[
   a_n(\delta):=\min\left\{\frac{1-\sqrt{1-2^{-2n}}}{6\left(m_n(\delta)\right)^2}\,\,,\,\,\frac\delta 3\right\}.
\]
It is clear that the map $(n,\delta)\mapsto a_n(\delta)$ is computable, although we do not
have an explicit formula for it.

According to Lemma~\ref{LemFixedLengthEnough}, we may assume that $M$ is a fixed-length QTM.
Fix some algorithm that on input $n\in\N$ and $\delta\in\mathbb{Q}^+$ computes some discretization
\[
   d^{(n)}(\delta):=\{|\varphi_1^{(n)}(\delta)\rangle,|\varphi_2^{(n)}(\delta)\rangle,
   \ldots,|\varphi_N^{(n)}(\delta)\rangle\}\subset S_n
\]
of the unit sphere $S_n\subset\cn$, with $N=\#d^{(n)}(\delta)<\infty$.
The discretization shall be $a_n(\delta)$-dense in the unit sphere $S_n\subset\cn$, i.e. for every
$|\varphi\rangle\in S_n$, there shall be some vector $|\varphi'\rangle\in d^{(n)}(\delta)$ such that
$\|\,|\varphi\rangle-|\varphi'\rangle\|<a_n(\delta)$. Moreover, we demand that ${\rm span}\, d^{(n)}(\delta)=\cn$.
For every $\eps>0$, let
\[
   d_M^{(n)}(\delta,\varepsilon,t):=\{|\varphi\rangle\in d^{(n)}(\delta)\,\,|\,\,|\varphi\rangle
   \mbox{ is }\varepsilon\mbox{-}t\mbox{-halting for }M\}.
\]
Now we construct some coarsening $D_M^{(n)}(\delta,\eps,t)\subset d_M^{(n)}(\delta,\eps,t)$ in the
following way: First, we choose an arbitrary vector $|\psi_1\rangle\in d_M^{(n)}(\delta,\eps,t)$.
Then, one after the other, we choose vectors $|\psi_i\rangle\in d_M^{(n)}(\delta,\eps,t)$ such that
no vector is $\frac\delta 2$-close to the span of the previously chosen vectors. We stop as soon as
there is no more such vector.

This way, we get a finite set $D_M^{(n)}(\delta,\eps,t)=\{|\psi_1\rangle,\ldots,|\psi_m
\rangle\}\subset S_n$ with the following properties:
\begin{itemize}
\item For every vector $|\varphi\rangle\in d_M^{(n)}(\delta,\eps,t)$, there is a vector
$|\varphi'\rangle\in {\rm span}\,D_M^{(n)}(\delta,\eps,t)$ such that $\|\,|\varphi\rangle-|\varphi'\rangle\|<\frac
\delta 2$.
\item Equation~(\ref{EqChainProperty}) is valid for every $i$.
\end{itemize}
Now we define the linear subspaces
\[
   L_M^{(n)}(\delta,\varepsilon,t):={\rm span}\,D_M^{(n)}(\delta,\eps,t).
\]
Suppose that $|\varphi\rangle\in L_M^{(n)}(\delta,\varepsilon,t)$ is a normalized vector.
In this case, $|\varphi\rangle$ can be written as $|\varphi\rangle=\sum_i \alpha_i |\varphi_i\rangle$,
where $\{|\varphi_i\rangle\}\subset D_M^{(n)}(\delta,\eps,t)\subset d_M^{(n)}(\delta,\eps,t)$
is a basis of $L_M^{(n)}(\delta,\varepsilon,t)$, and every $|\varphi_i\rangle$
is $\varepsilon$-$t$-halting for $M$. Choose some orthonormal basis of $L_M^{(n)}(\delta,\eps,t)^\perp$,
and append those vectors to $\{|\varphi_i\rangle\}$ to get a basis $\Psi$ of $\cn$.
It follows that $\sum_i |\alpha_i|=\|\,|\varphi\rangle\|_\Psi\leq \|\,|\varphi\rangle\|_{(\delta)}\leq m_n(\delta)$,
and Lemma~\ref{LemApproxHaltingSuperpositions}
implies:
\[
   \mbox{Every normalized vector }|\varphi\rangle\in L_M^{(n)}(\delta,\varepsilon,t)\mbox{ is }
   \left(m_n(\delta)^2\cdot\varepsilon\right)\mbox{-}t\mbox{-halting for }M.
\]
Now suppose that $\eps$ is any real number satisfying
\begin{equation}
   0<\eps<\frac{1-\sqrt{1-2^{-2n}}}{2 \left(m_n(\delta)\right)^2}.
   \label{EqBoundEpsStability}
\end{equation}
It follows that if $|\varphi\rangle\in L_M^{(n)}(\delta,\eps,t)$ is normalized,
then $|\varphi\rangle$ is better than $\frac {1-\sqrt{1-2^{-2n}}}2$-$t$-halting for $M$.
If $|\psi\rangle\in L_M^{(n)}(\delta,\eps,t')$ is another normalized vector with different
approximate halting time $t'\neq t$, then it follows from Lemma~\ref{LemAlmostOrtho2} that
$|\langle\psi|\varphi\rangle|<2^{-n}$.

Suppose now that $\sum_{t\in\N} \dim L_M^{(n)}(\delta,\eps,t)>2^n$. Then, by choosing
orthonormal bases in all spaces $L_M^{(n)}(\delta,\eps,t)$, we could choose $2^n+1$
vectors $\{|v_i\rangle\}_{i=1}^{2^n+1}$, such that their inner product satisfies
$|\langle v_i|v_j\rangle|<2^{-n}=\frac 1 {2^n + 1 -1}$ for every $i\neq j$.
Lemma~\ref{LemInnerProduct}
would then imply that the vectors were all linearly independent, which is impossible.
Thus,
\[
   \sum_{t\in\N} \dim L_M^{(n)}(\delta,\eps,t)\leq 2^n\quad\mbox{if }\eps\mbox{ satisfies }(\ref{EqBoundEpsStability}),
   \mbox{e.g. for }\eps=2 a_n(\delta).
\]
On the other hand, suppose that $|\varphi\rangle\in S_n$ is $a_n(\delta)$-$t$-halting for $M$.
Then, there is some vector $|\tilde\varphi\rangle\in d^{(n)}(\delta)$ such that $\|\,|\varphi\rangle-|\tilde\varphi\rangle
\|<a_n(\delta)$. According to Equation~(\ref{eqEpsMin}), the vector $|\tilde\varphi\rangle$ is $2 a_n(\delta)$-$t$-halting for $M$,
so $|\tilde\varphi\rangle\in d_M^{(n)}(\delta, 2 a_n(\delta),t)$. By construction, it follows
that there is another vector $|\varphi'\rangle\in L_M^{(n)}(\delta,2 a_n(\delta),t)$ with $\|\,|\tilde\varphi\rangle
-|\varphi'\rangle\|<\frac\delta 2$, so $\|\,|\varphi\rangle-|\varphi'\rangle\|<
a_n(\delta)+\frac \delta 2\leq\frac 5 6\delta$. The approximate
outputs of $M$ on inputs $|\varphi\rangle$ and $|\varphi'\rangle$ are then also $\delta$-close:
\begin{eqnarray}
\|\mathcal{R}\circ M_{\mathbf O}^t(|\varphi\rangle\langle \varphi|)-\mathcal{R}\circ M_{\mathbf O}^t
(|\varphi'\rangle\langle\varphi'|)\|_{\rm Tr}&\leq& \|\,|\varphi\rangle\langle\varphi|-|\varphi'\rangle
\langle \varphi'|\,\|_{\rm Tr}\nonumber\\
&\leq& \|\,|\varphi\rangle-|\varphi'\rangle\|<\frac 5 6 \delta,\label{EqIntrinsicError}
\end{eqnarray}
where we have used Lemma~\ref{LemContractivity} and \ref{LemStability}.

From that point on, we have the same situation as in Section~\ref{SecConstruction} where we proved the existence
of a strongly universal QTM: we have a collection of subspaces $\{L_M^{(n)}(\delta,2 a_n(\delta),t)\}_{t\in\N}$
such that their dimensions add up to at most $2^n$.
We can now use a construction that is analogous to that in Subsection~\ref{SubsecStronglyUniversalQTMs}:
For every vector $|\varphi\rangle\in S_n$ that is $a_n(\delta)$-halting for $M$, we can find
some vector $|\varphi'\rangle\in\bigcup_{t\in\N} L_M^{(n)}(\delta,2 a_n(\delta),t)$
such that (\ref{EqIntrinsicError}) holds.
We can divide $|\varphi'\rangle$ into a classical part, consisting of a prefix code $c_t\in\s$ of
the number of the corresponding subspace that contains $|\varphi'\rangle$, and a quantum part
$\mathcal{C}|\varphi'\rangle$,
consisting of a compression of $|\varphi'\rangle$ down to $\lceil \log \dim L_M^{(n)}(\delta,2 a_n(\delta),t)\rceil$
qubits.

The idea now is that the universal QTM $\mathfrak U$ works as follows:
On input $(\delta,s_M,c_t\otimes\mathcal{C}|\varphi'\rangle)$, where
$s_M$ is a description of the QTM $M$, the universal QTM
$\mathfrak U$ shall compute the halting time $t$ from $c_t$, approximately decompress
$|\varphi'\rangle$ from $\mathcal{C}|\varphi'\rangle$, and then simulate $M$
for $t$ time steps on input $|\varphi'\rangle$ and halt.

Again, $\mathfrak U$ cannot apply these steps exactly, but has to work with
numerical approximations of the spaces $L_M^{(n)}(\delta,2 a_n(\delta),t)$.
These approximations have to be good enough such that the resulting error
is bounded from above by $\frac 1 6\delta$, such that the resulting total
error (by adding (\ref{EqIntrinsicError})) is less than $\delta$.

This construction is completely analogous to the construction of the
strongly universal QTM $\mathfrak U$ in Section~\ref{SecConstruction};
it is even slightly simpler, since we do not need any ``fine tuning map'' $V$
as in the proof of Theorem~\ref{MainTheorem1}.
\qed

As $a_n(\delta)$ turns to zero exponentially fast
for $n\to\infty$, this theorem only applies to almost halting inputs that are extremely close to perfect halting.
Maybe it is possible to prove more general or less restrictive versions of this theorem
by allowing a larger blow-up of the program length (e.g. a factor larger than one, instead of an additive constant).
Another possibility might be to use a different definition of ``$\eps$-halting'': Instead of Definition~\ref{DefEpsTHalting},
one might instead define an input as $\eps$-halting at time $t$, if an outside observer who is continuously
measuring the halting state of the control observes halting at time $t$ with probability larger than $1-\eps$.

Despite this restriction, the theorem proves that the definition of halting by Bernstein and Vazirani~\cite{BernsteinVazirani}
has some unexpected built-in error tolerance, which makes that halting scheme look quite reasonable.

\chapter{Quantum Kolmogorov Complexity}
\label{ChapterQuantumKolmogorovComplexity}

\section{Definition of Quantum Kolmogorov Complexity}
\label{SecDefQComplexity}
The notion of quantum Kolmogorov complexity that we study in this thesis has first been defined
by Berthiaume, van Dam, and Laplante~\cite{Berthiaume}. 
They define the complexity $QC(\rho)$ of
a qubit string $\rho$ as the length of the shortest qubit string that, given as input
into a QTM $M$, makes $M$ output $\rho$ and halt.

Since there are uncountably many
qubit strings, but a QTM can only apply a countable number of transformations (analogously
to the circuit model), it is necessary to introduce a certain error tolerance $\delta>0$.

This can be done in essentially two ways: First, one can just fix some tolerance $\delta>0$.
Second, one can demand that the QTM outputs the qubit string $\rho$ as accurately as
one wants, by supplying the machine with a second parameter as input that represents
the desired accuracy. This is analogous to a classical computer program that computes
the number $\pi=3.14\ldots$: A second parameter $k\in\N$ can make the program output
$\pi$ to $k$ digits of accuracy, for example. We consider both approaches at once, and
get two different notions of quantum Kolmogorov complexity, namely $QC^\delta$ and $QC$.

Moreover, while Berthiaume et al. only allow inputs that are length eigenstates, base length $\ell$
and average length $\bar\ell$ coincide for their approach. We want to be more general and allow
arbitrary superpositions and mixtures, i.e. qubit strings $\sigma\in\mathcal{T}_1^+(\hr_\s)$ as inputs.
Thus, the number of possible definitions doubles again, depending on the way we quantify the length
of the input qubit strings. 
We get on the one hand the complexities $QC$ and $QC^\delta$
for base length $\ell$, and on the other hand the complexities $\qka$ and $\qka^\delta$ for average length $\bar\ell$.

According to Conjecture~\ref{TheAvLengthUQTM}, we can only hope to prove the invariance property
(cf. Section~\ref{SecInvariance}) for average-length complexities $\qka$ and $\qka^\delta$ if
we restrict them to prefix QTMs, i.e. if we define them as quantum analogues of classical prefix complexity.
Since classical prefix complexity is often denoted by $K$, while plain Kolmogorov complexity
(with no restriction on the reference Turing machine) is denoted by $C$, this explains why we
chose the notation $\qka$ and $\qka^\delta$.

Another difference to the definition by Berthiaume et al. is that we use the trace distance
rather than the fidelity to quantify the similarity of two qubit strings.

\begin{definition}[Quantum Kolmogorov Complexity]
\label{DefQComplexity}
Let $M$ be a QTM and $\rho\in\mathcal{T}_1^+(\hr_\s)$ an indeterminate-length qubit string.
For every $\delta>0$, we define the {\em finite-error quantum Kolmogorov complexity}
$QC_M^\delta(\rho)$ as the minimal length of any qubit string $\sigma\in\mathcal{T}_1^+(\hr_\s)$
such that the corresponding output $M(\sigma)$ has trace distance from $\rho$ smaller
than $\delta$,
\[
   QC_M^\delta(\rho):=\min\left\{\ell(\sigma)\,\,|\,\,\|\rho-M(\sigma)\|_{\rm Tr}<\delta\right\}.
\]
Similarly, we define the {\em approximation-scheme quantum Kolmogorov complexity} $QC_M(\rho)$
as the minimal length of any qubit string $\sigma\in\mathcal{T}_1^+(\hr_\s)$ such that when
given $M$ as input together with any integer $k$, the output $M(k,\sigma)$ has trace distance
from $\rho$ smaller than $1/k$:
\[
   QC_M(\rho):=\min\left\{\ell(\sigma)\,\,\left\vert
      \|\rho-M(k,\sigma)\|_{\rm Tr}< \frac 1 k \mbox{ for every } k\in\N
   \right.\right\}.
\]
We define two analogous notions of complexity, where base length $\ell$ is replaced by
average length $\bar\ell$: if $M$ is any QTM, then
\begin{eqnarray*}
   \qka_M^\delta(\rho)&:=&\inf\left\{\bar\ell(\sigma)\,\,|\,\,\|\rho-M(\sigma)\|_{\rm Tr}<\delta\right\}, \\
   \qka_M(\rho)&:=&\inf\left\{\bar\ell(\sigma)\,\,|\,\,\|\rho-M(k,\sigma)\|_{\rm Tr}<\frac 1 k \mbox{ for every }k\in\N\right\}.
\end{eqnarray*}
\end{definition}
Note that the specific choice of $f(k):=1/k$ as accuracy required on input $k$ is not important;
any other computable and strictly decreasing function $f$ that tends to zero for $k\to\infty$
such that $f^{-1}$ is also computable will give the same result within an additive constant, as long
as $M$ is a strongly universal QTM and the quantum complexity notions all have the invariance property
(which we discuss in Section~\ref{SecInvariance}).

The idea to define some notion like $\qka$ is due to Rogers and Vedral~\cite{Rogers}. In Chapter~\ref{ChapterSummary},
we argue that the notion of complexity $\qka$ is more useful for applications in statistical mechanics than
$QC$, since the average length sometimes has a physical interpretation as the expected energy cost
of communication.

In this thesis, we will most of the time restrict to the complexity notions $QC$ and $QC^\delta$,
since they are much easier to handle. The main technical reason for this is that the pure qubit strings $|\varphi\rangle$ with
base length $\ell(|\varphi\rangle)\leq n$ are all elements of one Hilbert space $\hr_{\leq n}$, which is not true
for average length. Nevertheless, we study the complexities $\qka$ and $\qka^\delta$ to some extent
in Section~\ref{SecInvariance}.

For later use, we note a simple relation between the two quantum complexities $QC^\delta$ and 
$QC$:

\begin{flemma}[Relation between Quantum Complexities]
\label{LemRelation}
\lineclear
For every QTM $M$ and every $k\in\N$, 
we have the relation
\begin{equation}
QC_M^{\frac 1 k}(\rho)\leq QC_M (\rho)+2\lfloor\log
k\rfloor+2
\qquad \mbox{for every }\rho \in \mathcal{T}^+_1(\hr_{\Fock} ).
\label{eqRelation}
\end{equation}
\end{flemma}
{\bf Proof. } Suppose that $QC_M(\rho)=l$, so there is
a density matrix $\sigma \in \mathcal{T}^+_1(\hr_{\Fock} )$ with 
$\ell(\sigma)=l$, such that $\|M(k,\sigma)-\rho\|_{\rm Tr}< \frac 1 k$ for every
$k\in\N$. Then $\sigma':= \langle k,\sigma\rangle$, where $\langle\cdot,\cdot\rangle$ 
is given in Definition \ref{DefEncoding}, is an input for $M$ such
that $\|M(\sigma')-\rho \|_{\rm Tr}< \frac 1 k$. 
Thus  $QC_M^{1/k}(\rho)\leq \ell(\sigma') \leq 2\lfloor\log k\rfloor
+2+ \ell(\sigma)= 2\lfloor\log k\rfloor +2+ QC_M(\rho)$, 
where the second inequality is by (\ref{encoding_length}).
\qed

The term $2\lfloor\log k\rfloor+2$ in (\ref{eqRelation}) depends on
our encoding $\langle\cdot,\cdot\rangle$ given in Definition~\ref{DefEncoding},
but if $M$ is assumed to be universal
(which will be discussed below), then (\ref{eqRelation}) will hold
for {\em every} encoding, if we replace the term $2\lfloor\log
k\rfloor+2$ by $K(k)+c_M$, where 
$K(k)\leq 2\lfloor\log k\rfloor+\mathcal{O}(1)$
denotes the classical (self-delimiting)
algorithmic complexity of the integer $k$, and $c_M$ is some
constant depending only on $M$. For more details we refer the reader
to~\cite{Vitanyibook}.

\section{Incompressibility Theorems}
\label{SecQCountingArgument}
In the theory of classical Kolmogorov complexity and in its applications, a simple but powerful argument used
frequently in proofs is the so-called incompressibility theorem. It can be stated in the following way
\cite[Theorem 2.2.1]{Vitanyibook}:

If $c$ is a positive integer, then every finite set $A$ of cardinality $m$ has at least $m(1-2^{-c})+1$ elements
$x$ with $C(x)\geq\log m -c$.

In this section, we are going to prove three quantum analogues of this theorem. The first version
is a very general theorem on the number of mutually orthonormal vectors that can be close in
trace distance to the output of some quantum operation. We call it ``quantum counting argument'',
because it is a quantization of a classical counting argument, saying that there can be no more
than $2^n$ different bit strings that have programs of length less than $n$.

Nevertheless, the theorem that follows is not restricted to the study of quantum computers, but
is a general result about quantum operations. Its proof is based on Holevo's $\chi$-quantity associated to any ensemble $\mathbb{E}_\rho
:=\left\{\lambda_i,\rho_i\right\}_i$, consisting of probabilities $0\leq\lambda_i\leq 1$,
$\sum_i \lambda_i=1$, and of density matrices $\rho_i$ acting on a Hilbert space $\hr$.
Setting $\rho:=\sum_i\lambda_i\rho_i$, the $\chi$-quantity is defined as follows:
\[
   \chi(\mathbb{E}_\rho):=S(\rho)-\sum_i \lambda_i S(\rho_i)=\sum_i\lambda_i S(\rho_i,\rho),
\]
where $S(\cdot,\cdot)$ denotes the relative entropy.

\begin{ftheorem}[Quantum Counting Argument]
\label{TheQCounting}
\lineclear
Let $\hr$ and $\hr'$ be separable Hilbert spaces with $0<d:=\dim\hr<\infty$, and let $0\leq\delta<\frac 1 {2e}$.
If $\mathcal{E}:\mathcal{T}(\hr)\to\mathcal{T}(\hr')$ is a quantum operation, then define
\[
   A_\delta:=\left\{|\psi\rangle\in\hr'\,\,|\,\,\exists\sigma\in\mathcal{T}_1^+(\hr):
   \|\mathcal{E}(\sigma)-|\psi\rangle
   \langle\psi|\,\|_{\rm Tr}\leq \delta\right\}.
\]
If $N_\delta\subset A_\delta$ is an orthonormal system, then
\[
   \log\# N_\delta\leq\frac{\log d + 4\delta\log\frac 1 \delta}{1-4\delta}.
\]
\end{ftheorem}
\proof
For $\delta=0$, the assertion of the theorem is trivial (setting, as usual, $0\log\frac 1 0:=0$), so
assume $\delta>0$. We may also assume that $N_\delta\neq\emptyset$. Let
\[
   N_\delta=:\left\{|\varphi_1\rangle,\ldots,|\varphi_N\rangle\right\},
\]
then by definition, there exist $\sigma_i\in\mathcal{T}_1^+(\hr)$ such that
$\|\mathcal{E}(\sigma_i)-|\varphi_i\rangle\langle\varphi_i|\,\|_{\rm Tr}\leq\delta$.
For $1\leq i\leq N$, define the projectors $P_i:=|\varphi_i\rangle\langle\varphi_i|$,
and set $P_{N+1}:=\idn-\sum_{i=1}^N |\varphi_i\rangle\langle\varphi_i|$. Let $\{|k\rangle\}_{k=1}^{\dim \hr'}$
be an orthonormal basis of $\hr'$.
Now we define a quantum operation $\mathcal{Q}:\mathcal{T}(\hr')\to\mathcal{T}(\C^{N+1})$ via
\[
   \mathcal{Q}(a):=\sum_{i=1}^{N+1}\sum_{k=1}^{\dim\hr'} |e_i\rangle\langle k|P_i a P_i |k\rangle\langle e_i|,
\]
where $\left\{|e_i\rangle\right\}_{i=1}^{N+1}$ denotes an arbitrary orthonormal basis of $\C^{N+1}$.
It is clear that $\mathcal{Q}$ is completely positive (Kraus representation), and one easily
checks that $\mathcal{Q}$ is also trace-preserving. This is also true if $\dim\hr'=\infty$; then,
the corresponding infinite series is absolutely convergent in $\|\cdot\|_{\rm Tr}$-norm, and inherits
complete positivity from its partial sums.
Moreover, for $1\leq j\leq N$,
we have
\[
   \mathcal{Q}(P_j)=\sum_k |e_j\rangle\langle k|P_j|k\rangle\langle e_j|=|e_j\rangle\langle e_j|.
\]

Consider the equidistributed ensemble $\mathbb{E}_\sigma:=\left\{\frac 1 N,\sigma_i\right\}_{i=1}^N$, and
let $\sigma:=\frac 1 N \sum_{i=1}^N \sigma_i$. Due to the monotonicity of relative entropy with
respect to quantum operations, we have
\begin{eqnarray*}
   \chi\left(\strut\mathcal{Q}\circ \mathcal{E}(\mathbb{E}_\sigma)\right)&=&\frac 1 N \sum_{i=1}^N
   S\left(\strut \mathcal{Q}\circ\mathcal{E}(\sigma_i),\mathcal{Q}\circ\mathcal{E}(\sigma)\right)
   \leq \frac 1 N \sum_{i=1}^N S(\sigma_i,\sigma)\\
   &=& \chi(\mathbb{E}_\sigma)\leq \log d.
\end{eqnarray*}
The trace distance is also monotone with respect to quantum operations (cf. Lemma~\ref{LemContractivity}).
Thus, for every $1\leq i\leq N$,
\[
   \left\|\mathcal{Q}\circ\mathcal{E}(\sigma_i)-\mathcal{Q}(P_i)\right\|_{\rm Tr}
   \leq \|\mathcal{E}(\sigma_i)-P_i\|_{\rm Tr}=\|\mathcal{E}(\sigma_i)-|\varphi_i\rangle
   \langle\varphi_i|\|_{\rm Tr}\leq \delta.
\]
Let now $\Delta:=\frac 1 N \sum_{i=1}^N \mathcal{Q}(P_i)=\frac 1 N \sum_{i=1}^N |e_i\rangle\langle e_i|$,
then $S(\Delta)=\log N$, and
\[
   \left\|\mathcal{Q}\circ\mathcal{E}(\sigma)-\Delta\right\|_{\rm Tr}
   \leq \frac 1 N \sum_{i=1}^N \left\|\mathcal{Q}\circ \mathcal{E}(\sigma_i)-\mathcal{Q}(P_i)\right\|_{\rm Tr}
   \leq \delta.
\]
The Fannes inequality \cite[11.44]{NielsenChuang} yields\footnote{Note that the notation in \cite{NielsenChuang}
differs from the notation in this thesis: it holds $T(\rho,\sigma)={\rm Tr}|\rho-\sigma|=\|\rho-\sigma\|_1=2\cdot\|\rho-\sigma\|_{\rm Tr}$.
} for $1\leq i\leq N$
\begin{eqnarray*}
   S\left(\strut \mathcal{Q}\circ \mathcal{E}(\sigma_i)\right)=\left| S\left(\strut
   \mathcal{Q}\circ \mathcal{E}(\sigma_i)
   \right)-S\left(\strut\mathcal{Q}(P_i)\right)\right|
   \leq 2\delta\log(N+1)+\eta(2\delta),\\
   \left| S\left(\strut \mathcal{Q}\circ \mathcal{E}(\sigma)\right)-S(\Delta)\right|
   \leq 2\delta\log(N+1)+\eta(2\delta),
\end{eqnarray*}
where $\eta(\delta)=-\delta\log\delta\geq 0$. Altogether, we get
\begin{eqnarray*}
   \log d &\geq &\chi\left(\strut \mathcal{Q}\circ\mathcal{E}(\mathbb{E}_\sigma)\right)
   =S\left(\strut \mathcal{Q}\circ \mathcal{E}(\sigma)\right)-\frac 1 N \sum_{i=1}^N
   S\left(\strut \mathcal{Q}\circ\mathcal{E}(\sigma_i)\right)\\
   &\geq& S(\Delta)-2\delta\log(N+1)-\eta(2\delta)-\frac 1 N \sum_{i=1}^N \left(\strut
   2\delta\log(N+1)+\eta(2\delta)\right)\\
   &=&\log N - 4\delta\log(N+1)-2\eta(2\delta)\\
   &\geq& (1-4\delta)\log N - 4\delta\log 2+4\delta\log(2\delta),
\end{eqnarray*}
where we have used the inequality $\log(N+1)\leq \log N +\log 2$ for $N\geq 1$.
The claim follows by rearranging.\qed

We will use this ``quantum counting argument'' later in Section~\ref{SecQCofClass} and \ref{SecQBrudno};
it will be useful in several proofs. Specifying it to the case that the quantum operation corresponds
to the action of a QTM, we get the following incompressibility theorem for quantum Kolmogorov complexity
$QC^\delta$:

\begin{fcorollary}[Incompressibility for Orthonormal Systems]
\lineclear
Let $M$ be a QTM,
let $0<\delta<\frac 1 {2e}$, and let $|\psi_1\rangle,\ldots,|\psi_n\rangle\in\hr_\s$
be a set of mutually orthonormal pure qubit strings. Then, there is some $i\in\{1,\ldots,n\}$
such that
\[
   QC_M^\delta(|\psi_i\rangle)>(1-4\delta)\log n - 1 - 4 \delta\log\frac 1 \delta.
\]
\end{fcorollary}
\proof
Let $l\in\N$ be a natural number such that $QC_M^\delta(|\psi_i\rangle)\leq l$ for every $i\in\{1,\ldots,n\}$.
Then, there exist qubit strings $\sigma_i\in\mathcal{T}_1^+(\hr_{\leq l})$ such that
$\|\mathcal{M}(\sigma_i)-|\psi_i\rangle\langle\psi_i|\,\|_{\rm Tr}<\delta$, where $\mathcal{M}$
is the quantum operation that corresponds to the QTM $M$, cf. Lemma~\ref{LemmaQTMsAreOperations}.
Thus, Theorem~\ref{TheQCounting} yields
\[
   \log n \leq \frac{\log\dim\hr_{\leq l}+4\delta\log\frac 1 \delta}{1-4\delta}
   <\frac{l+1+4\delta\log\frac 1 \delta}{1-4\delta}.
\]
It follows that $l>(1-4\delta)\log n -1 - 4\delta\log\frac 1 \delta$.
\qed

In \cite[Theorem 6]{Berthiaume}, Berthiaume et al. prove the following incompressibility
result for the approximation-scheme complexity $QC$: if $\rho_1,\ldots,\rho_M$ is any set of qubit
strings, then there is some $i\in\{1,\ldots,M\}$ such that\footnote{The ``-1''-term is missing in their paper.}
\[
   QC(\rho_i)\geq S\left(\frac 1 M \sum_{i=1}^M \rho_i\right)-\frac 1 M \sum_{i=1}^M S(\rho_i)-1.
\]
Note that the quantity on the right-hand side is exactly Holevo's $\chi$-quantity associated with
the ensemble $\left\{\frac 1 M,\rho_i\right\}_{i=1}^M$. Here, we give a generalization of this
result to the complexity notion $QC^\delta$. The proof is very similar to the proof of the quantum
counting argument, Theorem~\ref{TheQCounting}; the only difference is that we need a different quantum
operation $\mathcal{Q}$.

\begin{ftheorem}[Incompressibility for Pure Qubit Strings]
\lineclear
Let $M$ be a QTM, and
let $|\psi_1\rangle,\ldots,|\psi_n\rangle\in\hr_\s$ be a set of pure normalized
qubit strings. Then, there is some $i\in\{1,\ldots,n\}$ such that
\[
   QC_M^\delta(|\psi_i\rangle)>S\left(\frac 1 n \sum_{j=1}^n |\psi_j\rangle\langle \psi_j|\right)-4\delta
   \log\frac{n+1}{2\delta}-1,
\]
where $S$ denotes von Neumann entropy.
\end{ftheorem}
\proof
Let $l\in\N$ be a natural number such that $QC_M^\delta(|\psi_i\rangle)\leq l$ for every $i\in\{1,\ldots,n\}$.
Then, there exist qubit strings $\sigma_i\in\mathcal{T}_1^+(\hr_{\leq l})$ such that
$\|\mathcal{M}(\sigma_i)-|\psi_i\rangle\langle\psi_i|\,\|_{\rm Tr}<\delta$, where $\mathcal{M}$
is the quantum operation that corresponds to the QTM $M$, cf. Lemma~\ref{LemmaQTMsAreOperations}.

Let $\hr:={\rm span}\{|\psi_i\rangle\}_{i=1}^n$, let $N:=\dim\hr$, and let $U:\hr\to\C^{N+1}$ be an arbitrary
isometry (i.e. a unitary map from $\hr$ to some $N$-dimensional subspace of $\C^{N+1}$). Let $|e\rangle\in\C^{N+1}$
be a normalized vector from $({\rm ran}\,U)^\perp$. Then, define a quantum operation
$\mathcal{Q}:\mathcal{T}(\hr_\s)\to\mathcal{T}(\C^{N+1})$ via
\[
   \mathcal{Q}(a):=U P_\hr a P_\hr U^*+\sum_{k=1}^\infty |e\rangle\langle k|(\idn-P_\hr)a(\idn-P_\hr)|k\rangle\langle e|,
\]
where $\{|k\rangle\}_{k=1}^\infty$ denotes an orthonormal basis of $\hr^\perp$ in $\hr_\s$, and $P_\hr$ denotes
the orthogonal projector onto $\hr$.
It is easily checked
that $\mathcal Q$ is linear and trace-preserving, and it is clear that $\mathcal Q$ is completely positive
(Kraus representation). Moreover,
\[
   \mathcal{Q}(|\psi_i\rangle\langle\psi_i|)=U|\psi_i\rangle\langle \psi_i| U^*\quad\mbox{for every }1\leq i \leq n.
\]
As the trace distance is monotone with respect to quantum operations (cf. Lemma~\ref{LemContractivity}), we get
\[
   \| \mathcal{Q}\circ \mathcal{M}(\sigma_i)-\mathcal{Q}(|\psi_i\rangle\langle\psi_i|)\|_{\rm Tr}
   \leq\|\mathcal{M}(\sigma_i)-|\psi_i\rangle\langle\psi_i|\,\|_{\rm Tr}\leq\delta.
\]
Let $\Delta:=\frac 1 n \sum_{i=1}^n \mathcal{Q}(|\psi_i\rangle\langle\psi_i|)$. Since the trace distance is
jointly convex (cf. \cite{NielsenChuang}), we also get
\[
   \left\|\mathcal{Q}\circ\mathcal{M}\left(\frac 1 n\sum_{i=1}^n \sigma_i\right)-\Delta\right\|_{\rm Tr}
   \leq \frac 1 n \sum_{i=1}^n \|\mathcal{Q}\circ\mathcal{M}(\sigma_i)-\mathcal{Q}(|\psi_i\rangle\langle\psi_i|)\|_{\rm Tr}
   \leq \delta.
\]
For $1\leq i\leq n$, the Fannes inequality \cite[11.44]{NielsenChuang} yields
\begin{eqnarray*}
   \left| S(\Delta)-S\left(\frac 1 n \sum_{i=1}^n \mathcal{Q}\circ\mathcal{M}(\sigma_i)\right)\right|
   &\leq& 2\delta\log(N+1)+\eta(2\delta),\\
   |S(\mathcal{Q}\circ\mathcal{M}(\sigma_i)-\underbrace{S(U|\psi_i\rangle\langle\psi_i|U^*)}_0|&\leq&
   2\delta\log(N+1)+\eta(2\delta),
\end{eqnarray*}
where $\eta(x)=-x\log x>0$. Now consider the equidistributed ensemble $\mathbb{E}_\sigma:=\left\{
\frac 1 n,\sigma_i\right\}_{i=1}^n$. The monotonicity property of Holevo's $\chi$ quantity gives
\begin{eqnarray*}
   l+1 &>& \log\dim\hr_{\leq l}\geq \chi(\mathbb{E}_\sigma)\geq \chi(\mathcal{M}(\mathbb{E}_\sigma))\geq\chi(\mathcal{Q}
   \circ\mathcal{M}(\mathbb{E}_\sigma))\\
   &=&S\left(\frac 1 n \sum_{i=1}^n\mathcal{Q}\circ\mathcal{M}(\sigma_i))\right)-\frac 1 n
   \sum_{i=1}^n S(\mathcal{Q}\circ\mathcal{M}(\sigma_i))\\
   &\geq& S(\Delta)-2\delta\log(N+1)-\eta(2\delta)-\frac 1 n \sum_{i=1}^n \left(
      2\delta\log(N+1)+\eta(2\delta)
   \right)\\
   &=&S\left(\frac 1 n \sum_{i=1}^n |\psi_i\rangle\langle\psi_i|)\right)-4\delta\log(N+1)-4\delta\log\frac 1 {2\delta}.
\end{eqnarray*}
Using that $N\leq n$, the claim follows.
\qed

\section{The Invariance Property}
\label{SecInvariance}
The most important theorem for classical Kolmogorov complexity is the invariance theorem.
Basically, it says that Kolmogorov complexity does not depend too much on the choice of the corresponding TM.
In more detail, there is a (``universal'') TM $U$ such that for every TM $M$,
there is some constant $c_M\in\N$ such that
\[
   C_U(s)\leq C_M(s)+c_M\mbox{ for every }s\in\s
\]
(cf. \cite{Vitanyibook}). Consequently, if $U$ and $V$ are both universal TMs, then
the difference of the corresponding complexities $|C_U(s)-C_V(s)|$ is uniformly bounded
by a constant. Since additive constants do not matter so much for many applications,
this means that we can define Kolmogorov complexity with respect to any universal
computer we want.

It follows from the results in Section~\ref{SecHaltingUniv} and \ref{SecConstruction} that
both quantum Kolmogorov complexities $QC$ and $QC^\delta$ are invariant as well:

\begin{ftheorem}[Invariance of Q-Kolmogorov Complexity]
\label{MainTheorem2}
\lineclear
There is a fixed-length quantum Turing machine $\mathfrak{U}$ such
that for every QTM $M$ there is a constant $c_M\in\N$ such that
\[
   QC_\mathfrak{U}(\rho)\leq QC_M(\rho)+c_M\qquad\mbox{for every qubit string }\rho.
\]
Moreover, for every QTM $M$ and every $\delta,\Delta\in\mathbb{Q}^+$ with $\delta<\Delta$,
there is a constant $c_{M,\delta,\Delta}\in\N$ such that
\[
   QC^\Delta_{\mathfrak{U}}(\rho)\leq QC^\delta_M(\rho)+c_{M,\delta,\Delta}
   \qquad\mbox{for every qubit string }\rho.
\]
\end{ftheorem}
As a consequence, we now fix an arbitrary QTM $\mathfrak U$ with the property of Theorem~\ref{MainTheorem2},
and define $QC(\rho):=QC_{\mathfrak U}(\rho)$
and $QC^\delta(\rho):=QC_{\mathfrak U}^\delta(\rho)$ for every qubit string $\rho\in\mathcal{T}_1^+(\hr_\s)$
and $\delta>0$.\\

{\bf Proof of Theorem~\ref{MainTheorem2}.} First, we use Theorem~\ref{MainTheorem1} to prove
the second part of Theorem~\ref{MainTheorem2}. Let $M$ be an arbitrary QTM,
let $\mathfrak U$ be the (``strongly universal'')
QTM and $c_M$ the corresponding constant from Theorem~\ref{MainTheorem1}.
Let $\ell:=QC_M^\delta(\rho)$, i.e. there
exists a qubit string $\sigma\in\mathcal{T}_1^+(\hr_\s)$ with $\ell(\sigma)=\ell$ such that $\| M(\sigma)-\rho\|_{\rm Tr}< \delta$.
According to Theorem~\ref{MainTheorem1}, there exists a qubit string $\sigma_M\in\mathcal{T}_1^+(\hr_\s)$
with $\ell(\sigma_M)\leq \ell(\sigma)+c_M=\ell+c_M$ such that
\[
   \|\mathfrak{U}(\Delta-\delta,\sigma_M)-M(\sigma)\|_{\rm Tr}<\Delta-\delta\,\,.
\]
Thus, $\|\mathfrak{U}(\Delta-\delta,\sigma_M)-\rho\|_{\rm Tr}<\Delta$, and
$\ell(\Delta-\delta,\sigma_M)=\ell(\sigma_M)+\ell(\Delta-\delta)\leq \ell+c_M+c_{\delta,\Delta}$,
where $c_{\delta,\Delta}\in\N$ is some constant that only depends on $\delta$ and $\Delta$.
So $QC_{\mathfrak U}^\Delta(\rho)\leq \ell+c_{M,\delta,\Delta}$.

The first part of Theorem~\ref{MainTheorem2} uses Proposition~\ref{PropTwoParameter}. Again, let $M$
be an arbitrary QTM, let $\mathfrak U$ be the strongly universal QTM and $c_M$ the corresponding
constant from Proposition~\ref{PropTwoParameter}. Let $\ell:=QC_M(\rho)$, i.e. there
exists a qubit string $\sigma\in\mathcal{T}_1^+(\hr_\s)$ with $\ell(\sigma)=\ell$ such that
\[
   \| M(k,\sigma)-\rho\|_{\rm Tr}<\frac 1 k\qquad\mbox{for every }k\in\N\,\,.
\]
According to Proposition~\ref{PropTwoParameter}, there exists a qubit string $\sigma_M\in\mathcal{T}_1^+(\hr_\s)$
with $\ell(\sigma_M)\leq\ell(\sigma)+c_M=\ell+c_M$ such that
\[
   \left\|\mathfrak{U}\left(k,\sigma_M\right)-M\left(
   2 k,\sigma\right)\right\|_{\rm Tr}<\frac 1 {2k}\qquad\mbox{for every }k\in\N\,\,.
\]
Thus, $\|\mathfrak{U}(k,\sigma_M)-\rho\|_{\rm Tr}\leq \|\mathfrak{U}(k,\sigma_M)-M(2k,\sigma)\|_{\rm Tr}
+\|M(2k,\sigma)-\rho\|_{\rm Tr}<\frac 1 {2k}+\frac 1 {2k}=\frac 1 k$ for every $k\in\N$. So
$QC_{\mathfrak U}(\rho)\leq \ell+c_M$. \qed

Does the invariance property also hold for the average length complexities $\qka$ and $\qka^\delta$?
If Conjecture~\ref{TheAvLengthUQTM} holds true, then we can repeat the proof of invariance of
$QC^\delta$ without changes for $\qka^\delta$. Thus, we conjecture that the following holds true:

\begin{conjecture}[Invariance of Average-Length Complexity]
\label{ConInvALC}
There is a prefix QTM $\mathfrak V$ such that for every prefix QTM $M$ and every $\delta,\Delta\in\mathbb{Q}^+$ with $\delta<\Delta$,
there is some constant $c_{M,\delta,\Delta}$ such that
\[
   \qka_{\mathfrak{V}}^\Delta(\rho)\leq \qka_M^\delta(\rho)+c_{M,\delta,\Delta}\qquad\mbox{for every qubit string }\rho.
\]
\end{conjecture}

What about the complexity notion $\qka$? The question whether $\qka$ is invariant depends on
the question whether Proposition~\ref{PropTwoParameter} can be generalized to average length $\bar\ell$.
We think that this could be possible, but have no idea how to prove it.

A simple consequence of the invariance property is that the quantum Kolmogorov complexity of some qubit string
is bounded from above by its base length:

\begin{emptyflemma}
\label{LemQCBoundL}
There is some constant $c\in\N$ such that
\begin{equation}
   QC(\rho)\leq \ell(\rho)+c\qquad\mbox{for every qubit string }\rho.
   \label{EqLengthBound1}
\end{equation}
Similarly, for every $\delta\in\mathbb{Q}^+$, there is some constant $c_\delta\in\N$ such that
\begin{equation}
   QC^\delta(\rho)\leq \ell(\rho)+c_\delta\qquad\mbox{for every qubit string }\rho.
   \label{EqLengthBound2}
\end{equation}
\end{emptyflemma}
\proof Recall the construction used in the proof of Lemma~\ref{LemFixedLengthEnough} to compress indeterminate-length
qubit strings into fixed-length qubit strings which are only one qubit longer. We are using the same idea
to construct a fixed-length QTM $M$ with $QC_M(\rho)\leq\ell(\rho)+1$. Then, Equation~(\ref{EqLengthBound1})
follows immediately from Theorem~\ref{MainTheorem2} (the invariance property), and Equation~(\ref{EqLengthBound2}) follows from
Lemma~\ref{LemRelation}.

Going back to the idea of Lemma~\ref{LemFixedLengthEnough}, as $2^{n+1}-1=\dim\hr_{\leq n}<\dim\hr_{n+1}=2^{n+1}$, we can
embed $\hr_{\leq n}$ isometrically in $\hr_{n+1}$ in a simple way, e.g. by mapping computational basis vectors
to computational basis vectors. This transformation can be extended to a unitary transformation $U_n$ on $\hr_{\leq(n+1)}$,
again simply by mapping computational basis vectors to computational basis vectors, such that there is
a QTM that can apply each $U_n$ for every $n$ (and its inverse $U_n^{-1}$) exactly, i.e.
without any error.

The fixed-length QTM $M$ works as follows on input $(k,\sigma)$, where $k\in\N$ is some integer,
and $\sigma\in\bigcup_{n\in\N_0} \mathcal{T}_1^+(\hr_n)$ is some fixed-length qubit string:
First, it reads and ignores $k$. Then, it determines $n+1=\ell(\sigma)$ by detecting the first
blank symbol $\#$ on its input track. Afterwards, it applies $U_n^{-1}$ on the corresponding $n+1$-block
of input track cells exactly, moves this block to the output track and halts. Then $M$ has $QC_M(\rho)=
\ell(\rho)+1$.\qed

If the complexity notion $\qka^\delta$ is really invariant as stated in Conjecture~\ref{ConInvALC},
then the following result might give an analogue of Lemma~\ref{LemQCBoundL}.

\begin{emptyflemma}
\label{LemLogEllAverage}
For every $\delta\in\mathbb{Q}^+$, there is a QTM $M$ such that
\[
   \qka_M^\delta(\rho)\leq \bar\ell(\rho)+\mathcal{O}(\log\bar\ell(\rho))\qquad\mbox{for every qubit string }\rho.
\]
\end{emptyflemma}
For simplicity of the proof, we let $M$ depend on $\delta$ here, which can be avoided.
Unfortunately, it is not clear whether $M$ can be constructed to be prefix.

\proof The QTM $M$ expects input of the form $(c_{\lceil\bar\ell(\sigma)\rceil}\otimes \sigma)$, where
$\sigma$ is an arbitrary indeterminate-length qubit string, and $\{c_n\}_{n\in\N_0}\subset\s$ is a classical prefix code that
encodes the natural numbers into binary strings; it is well-known that this can be done in a way such that
$\ell(c_n)=\mathcal{O}(\log n)$.

The QTM starts by reading $c_{\lceil\bar\ell(\sigma)\rceil}$, and decodes $\lceil\bar\ell(\sigma)\rceil$ from it.
Then, it determines some $k\in\N$ such that $\|\sigma-\sigma_1^k\|_{\rm Tr}<\delta$, where $\sigma_1^k$
is defined in Definition~\ref{DefPrefixQTM}. Finally, it moves the first $k$ qubits of $\sigma$ from the input
to the output track and halts.

The only remaining question is how the aforementioned integer $k$ can be determined. First suppose that
$\sigma=|\psi\rangle\langle\psi|$ is a pure qubit string.

Let $s_{ln}\in\s$ be the $n$-th classical string in lexicographical order of length $l$. Then, we can write
\[
   |\psi\rangle=\sum_{l=0}^\infty \sum_{n=1}^{2^l} \alpha_{ln}|s_{ln}\rangle.
\]
Let $|\psi(k)\rangle:=\sum_{l=0}^k \sum_{n=1}^{2^l} \alpha_{ln}|s_{ln}\rangle$, then
\[
   \|\,|\psi\rangle-|\psi(k)\rangle\|^2=\left\|\sum_{l=k+1}^\infty \sum_{n=1}^{2^l}\alpha_{ln}|s_{ln}\rangle\right\|^2
   =\sum_{l=k+1}^\infty \sum_{n=1}^{2^l} |\alpha_{ln}|^2.
\]
Thus, we get
\begin{eqnarray*}
   (k+1)\|\,|\psi\rangle-|\psi(k)\rangle\|^2&=&\sum_{l=k+1}^\infty (k+1)\sum_{n=1}^{2^l} |\alpha_{ln}|^2
   \leq \sum_{l=k+1}^\infty l\sum_{n=1}^{2^l} |\alpha_{ln}|^2\\
   &\leq&\sum_{l=0}^\infty l\sum_{n=1}^{2^l} |\alpha_{ln}|^2=\langle\psi|\Lambda|\psi\rangle=\bar\ell(|\psi\rangle).
\end{eqnarray*}
From Lemma~\ref{LemNormInequalities}, it follows that
\[
   \|\,|\psi\rangle\langle\psi|-|\psi(k)\rangle\langle\psi(k)|\,\|_{\rm Tr}
   \leq \|\,|\psi\rangle-|\psi(k)\rangle\|\leq\sqrt{\frac{\bar\ell(|\psi\rangle)}{k+1}}.
\]
As quantum operations are contractive (cf. Lemma~\ref{LemContractivity}), restricting both states to
the first $k$ qubits yields $\|\,|\psi\rangle\langle\psi|_1^k-|\psi(k)\rangle\langle\psi(k)|\,\|_{\rm Tr}
\leq\sqrt{\frac{\bar\ell(|\psi\rangle)}{k+1}}$, and by the triangle inequality
\[
   \|\,|\psi\rangle\langle\psi|-|\psi\rangle\langle\psi|_1^k\|_{\rm Tr}\leq 2 \sqrt{\frac{\bar\ell(|\psi\rangle)}{k+1}}.
\]
Now suppose that $\sigma$ is an arbitrary mixed qubit string. Let $\sigma=\sum_i\lambda_i|\psi_i\rangle\langle\psi_i|$
be its spectral decomposition. Using the joint convexity of the trace distance and the Cauchy-Schwarz inequality, we get
\begin{eqnarray*}
\|\sigma-\sigma_1^k\|_{\rm Tr}&=&\left\|\sum_i \lambda_i |\psi_i\rangle\langle\psi_i|-\sum_i\lambda_i |\psi_i\rangle
\langle\psi_i|_1^k\right\|_{\rm Tr}\\
&\leq& \sum_i\lambda_i \|\,|\psi_i\rangle\langle\psi_i|-|\psi_i\rangle\langle\psi_i|_1^k\|_{\rm Tr}\\
&\leq&\sum_i \sqrt{\lambda_i} 2 \sqrt{\frac{\lambda_i\bar\ell(|\psi_i\rangle)}{k+1}}
\leq \sqrt{\sum_i \lambda_i}\cdot\sqrt{\sum_i \frac 4 {k+1} \lambda_i\bar\ell(|\psi_i\rangle)}\\
&=& 2\sqrt{\frac{\bar\ell(\sigma)}{k+1}}.
\end{eqnarray*}
Thus, $k$ just has to be chosen large enough such that the right-hand side is less than $\delta$.\qed

\section{Quantum Complexity of Classical Strings}
\label{SecQCofClass}
Quantum Kolmogorov was meant to be a generalization of classical Kolmogorov complexity. In this section,
we show that this point of view is justified by proving that at the domain of classical strings,
quantum and classical Kolmogorov complexity basically coincide up to an additive constant. Thus,
quantum Kolmogorov complexity extends classical complexity in a similar way as von Neumann entropy
generalizes Shannon entropy.

We start with a lemma which says that classical complexity is bounded from above by quantum complexity.
This was formulated as an open problem in the first paper on this complexity notion by Berthiaume et al. \cite{Berthiaume}.
Later, G\'acs proved some prefix-free analogue of (\ref{EqLowerBound1}) indirectly in \cite{Gacs}.

\begin{flemma}[Classical Complexity $\leq$ Quantum Complexity]
\label{LemClassCompl}
\lineclear
For every QTM $M$, there is a constant $c_M\in\N$ such that
\begin{equation}
   C(s)\leq QC_M(|s\rangle)+c_{M}\qquad\mbox{ for every }s\in\s.
   \label{EqLowerBound1}
\end{equation}
Moreover, for every $\delta\in\left(0,\frac 1 {2e}\right)\cap\mathbb{Q}$, there is a constant $c_{\delta,M}\in\N$ such that
\begin{equation}
  C(s)\leq \frac {QC_M^\delta(|s\rangle)} {1-4\delta}+c_{\delta,M}
  \qquad\mbox{ for every }s\in\s.
  \label{EqLowerBound2}
\end{equation}
\end{flemma}
\proof
According to Lemma~\ref{LemFixedLengthEnough},
we may without loss of generality assume that $M$ is a fixed-length QTM.
We give a classical computer program $P$ that, on input $i,n\in\N$ and
$\delta\in(0,\frac 1 {2e})\cap\mathbb{Q}$, together with a description of a QTM $M$,
approximately outputs the $i$-th string that is
generated by the QTM $M$ on some input of length $n$.
The program $P$ works as follows:
\begin{itemize}
\item[(1)] Set the time $t:=1$ and the counter $c:=0$. Compute some number
$\eps\in\left(0,\frac 1 {80}2^{-2n}\right)\cap\mathbb{Q}$ such that $\eps<\frac 1 {23}\left(\frac 1 {2e}-\delta
\right)$.
\item[(2)] Compute a description of the approximate halting
space $\hr_M^{(n,\eps)}(t)$. If $\hr_M^{(n,\varepsilon)}(t)=\{0\}$, go to step (4).
\item[(3)] Compute a finite set of self-adjoint matrices $\mathcal{\tilde T}$
such that for every $\sigma\in \mathcal{T}_1^+(\hr_M^{(n,\eps)}(t))$ there
is a matrix $\tilde \sigma\in\mathcal{\tilde T}$ such that $\|\tilde \sigma-\sigma\|_{\rm Tr}<\eps$ and vice versa.
For every matrix $\tilde \sigma\in\mathcal{\tilde T}$,
\begin{itemize}
\item simulate the QTM $M$ on input $\tilde \sigma$ for $t$ time steps, that is,
compute an approximation $\rho_{\tilde \sigma}$
of the output of $M$ on input $\tilde \sigma$
such that $\left\|\mathcal{R}\left(M_{\mathbf{O}}^t(\tilde \sigma)\right)-\rho_{\tilde \sigma}\right\|_{\rm Tr}
<\eps$;
\item for every $w\in\s$ with $\ell(w)\leq t$, compute an approximation
$\Delta_w$ of $\|\rho_{\tilde \sigma}-|w\rangle\langle w|\,\|_{\rm Tr}$ such that
$\left|\Delta_w-\|\rho_{\tilde \sigma}-|w\rangle\langle w|\,\|_{\rm Tr}\right|<\eps$;
\item if $\Delta_w<\delta+\frac{17}2\eps$, then set $c:=c+1$. If $c=i$, then
output $w$ and halt.
\end{itemize}
\item[(4)] Set $t:=t+1$ and go back to step (2).
\end{itemize}
The proof will consist of two parts: In the first part, we show that the program $P$
finally generates every string $s$ with $QC_M^{\delta}(|s\rangle)= n$
for some appropriate input $i$. In the second part, we show
that the number $i$ is not too large, such that it can be specified by a short binary string.

For the first part, suppose that $s\in\s$ is a binary string such that
$QC_M^{\delta}(|s\rangle)= n$.
By definition, it follows that there is some $\sigma\in\mathcal{T}_1^+(\hr_n)$ such that
$\left\| M(\sigma)-|s\rangle\langle s|\,\right\|_{\rm Tr}\leq \delta$.
If $T$ is the corresponding halting time and $\sigma=\sum_{i=1}^N\lambda_i |\varphi_i\rangle
\langle\varphi_i|$ is the spectral decomposition of $\sigma$, it follows that
$|\varphi_i\rangle\in H_M^{(n)}(T)$
for every $i$.
According to Theorem~\ref{TheProperties}, there are vectors $|\tilde\varphi_i\rangle\in H_M^{(n,\eps)}(T)$
such that $\|\,|\varphi_i\rangle-|\tilde\varphi_i\rangle\|\leq\frac{11}2\eps$.
Let $\sigma':=\sum_{i=1}^N \lambda_i |\tilde\varphi_i\rangle\langle\tilde\varphi_i|\in
\mathcal{T}_1^+(\hr_M^{(n,\eps)}(T))$, then $\|\sigma-\sigma'\|_{\rm Tr}\leq \frac{11}2\eps$
according to Lemma~\ref{LemNormInequalities}.

If the program $P$ has run long enough that $t=T$,
there is by assumption some $\tilde\sigma\in\mathcal{\tilde T}$ such that
$\|\sigma'-\tilde\sigma\|_{\rm Tr}<\eps$.
According to Lemma~\ref{LemContractivity}, we have
\begin{eqnarray*}
\left\|\mathcal{R}\left(M_{\mathbf{O}}^T(\tilde\sigma)\right)-|s\rangle\langle s|\right\|_{\rm Tr}
&\leq& \left\| M(\sigma)-|s\rangle\langle s|\right\|_{\rm Tr}+\left\|\mathcal{R}\left(M_{\mathbf{O}}^T(\tilde
\sigma)\right)-M(\sigma)\right\|_{\rm Tr}\\
&\leq&\delta+\|\tilde\sigma-\sigma\|_{\rm Tr}<\delta+\frac{13}2\eps.
\end{eqnarray*}
Thus, $\left\|\rho_{\tilde\sigma}-|s\rangle\langle s|\,\right\|_{\rm Tr}<\delta+\frac {15}2\eps$.
In step (3) of the program $P$, if $w=s$, it will then
hold that $\Delta_w<\delta+\frac{17}2\eps$, and the program $P$ will output the string $s$
if the input $i$ has been appropriately chosen.
This is true for every string $s\in\s$ with $QC_M^\delta(|s\rangle)= n$.

\vskip 1cm

Now suppose some classical string $w\in\s$ is output by $P$ on some input $i$.
In this case, it will hold $\Delta_w<\delta+\frac{17}2\eps$ in step (3) of the program $P$,
and thus, $\|\rho_{\tilde \sigma}-|w\rangle\langle w|\,\|_{\rm Tr}<\delta+\frac{19}2\eps$.
Thus, if $t$ is the corresponding halting time, we have
\begin{eqnarray*}
\left\|\mathcal{R}\left(M_{\mathbf{O}}^t(\tilde \sigma)\right)-|w\rangle\langle w|\,\right\|_{\rm Tr}
&\leq& \left\|\mathcal{R}\left(M_{\mathbf{O}}^t(\tilde \sigma)\right)-\rho_{\tilde \sigma}\right\|_{\rm Tr}
+\left\|\rho_{\tilde \sigma}-|w\rangle\langle w|\,\right\|_{\rm Tr}\\
&<&\delta+\frac{21}2\eps.
\end{eqnarray*}
By definition, there exists some $\sigma\in \mathcal{T}_1^+(\hr_M^{(n,\eps)}(t))$
such that $\|\sigma-\tilde \sigma\|_{\rm Tr}<\eps$, so
\begin{eqnarray*}
\left\|\mathcal{R}\left(M_{\mathbf{O}}^t(\sigma)\right)-|w\rangle\langle w|\,\right\|_{\rm Tr}&\leq&
\left\|\mathcal{R}\left(M_{\mathbf{O}}^t(\sigma)\right)-\mathcal{R}\left(M_{\mathbf{O}}^t(\tilde \sigma)\right)
\right\|_{\rm Tr}\\
&&+\left\|\mathcal{R}\left(M_{\mathbf{O}}^t(\tilde \sigma)\right)-|w\rangle\langle w|\,\right\|_{\rm Tr}\\
&<&\|\sigma-\tilde \sigma\|_{\rm Tr}+\delta+\frac{21}2\eps<\delta+\frac{23}2\eps.
\end{eqnarray*}
Define $\mathcal{E}_t:=\mathcal{R}\circ M_{\mathbf{O}}^t$ and $\Delta:=\delta+\frac{23}2\eps<\frac 1 {2e}$,
and set
\[
   N_\Delta(t):=\left\{w\in\s\,\,|\,\,
   \exists \sigma\in\mathcal{T}_1^+(\hr_M^{(n,\eps)}(t)):\|
   \mathcal{E}_t(\sigma)-|w\rangle\langle w|\|_{\rm Tr}<\Delta\right\}
\]
if $\dim\hr_M^{(n,\varepsilon)}(t)\geq 1$, and $N_\Delta(t):=\emptyset$ otherwise.
It follows from the quantum counting argument (Theorem~\ref{TheQCounting}) that
\[
   \log\# N_\Delta(t)\leq \frac{\log\dim\hr_M^{(n,\eps)}(t)+4\Delta\log\frac 1 \Delta}{1-4\Delta}.
\]
Let $L$ be the set of strings that are generated by the program $P$ on any input $i$. (It will turn
out that $L$ is finite; if the input $i$ is too large, then $P$ will not halt.)
As the function $x\mapsto x^c$ is superadditive on $[1,\infty)$ for $c\geq 1$,
and as $\eps<\frac 1 {80}2^{-2n}$ (compare Corollary~\ref{CorDimBound}), we get
\begin{eqnarray*}
\# L &\leq& \sum_{t\in\N:N_\Delta(t)\neq\emptyset} \#N_\Delta(t)
\leq \sum_{t\in\N:\dim\hr_M^{(n,\eps)}(t)\neq 0} 2^{\frac{\log\dim\hr_M^{(n,\eps)}(t)+4\Delta\log\frac
1 \Delta}{1-4\Delta}}\\
&\leq&\left(\sum_{t\in\N:\dim\hr_M^{(n,\eps)}(t)\neq 0} 2^{\log\dim\hr_M^{(n,\eps)}(t)+4\Delta\log\frac 1 \Delta}\right)
^{\frac 1 {1-4\Delta}}\\
&=&\left( \sum_{t\in\N}\dim \hr_M^{(n,\eps)}(t)\cdot \left(\frac 1 \Delta\right)^{4\Delta}\right)
^{\frac 1 {1-4\Delta}}\leq 2^{\frac n {1-4\Delta}}\cdot\left(\frac 1 \Delta\right)^{\frac{4\Delta}{1-4\Delta}}.
\end{eqnarray*}
Thus, we get
\begin{equation}
   \log\# L\leq\frac{n}{1-4\Delta}+\frac{4\Delta}{1-4\Delta}\log\frac 1 \Delta.
   \label{eqClassLBound}
\end{equation}

\vskip 1cm
Now we join both parts of the proof together to show the assumption of the lemma.
Let $T$ be a classical Turing machine that expects input $x_s$ of the following form:
\[
   \s\ni x_s=\left(\underbrace{\mbox{description of }M,\mbox{ description of }\delta}
   _{\mbox{prefix coded}},
   \mbox{ classical string }s\in\s\right).
\]
The machine $T$ first determines the length $\ell(s)$ by detecting the first blank
symbol $\#$ on its tape. Then, it computes the number $\eps$ in the same way as given above
in step (1) of the computer program $P$, and $\Delta:=\delta+\frac{23}2\eps$.
Afterwards, it computes the number $n\in\N$ as the unique\footnote{If such an integer $n\in\N$ exists,
it is unique. Otherwise, we may define the program to continue in an arbitrary way, e.g. to halt immediately.
} integer satisfying
\[
   \ell(s)=\left\lceil \frac{n}{1-4\Delta}+\frac {4\Delta}{1-4\Delta}\log\frac 1 \Delta\right\rceil.
\]
Let $i$ be the number of the string $s$ in the set $\{0,1\}^{\ell(s)}$.
The machine $T$
computes the output $w$ of $P$ on input $i$, $n$, $M$ and $\delta$, outputs $w$ and halts.
We know from Equation~(\ref{eqClassLBound}) that every word $w$ on the list $L$ can be constructed in this way
by choosing $s$ appropriately.
We have
\[
   \ell(x_s)\leq \frac n {1-4\Delta}+{\rm const}_\Delta.
\]
Since $\left| \frac 1 {1-4\Delta}-\frac 1 {1-4\delta}\right| n\leq
\sup_{0<x<\frac 1 {2e}} \left(\frac 1 {1-4\bullet}\right)'(x)\cdot (\Delta-\delta) n
<{\rm const}\cdot 2^{-2n}n$
is bounded, we even have
\[
   \ell(x_s)\leq \frac n {1-4\delta}+{\rm const}_{\delta}.
\]
Thus, if $w\in L$ is any string on the list, then $C_T(w)\leq \frac n {1-4\delta}+{\rm const}_\delta$.
Equation~(\ref{EqLowerBound2}) now follows from the invariance of classical Kolmogorov complexity.

To prove Equation~(\ref{EqLowerBound1}), let $V$ be the classical Turing machine that
expects input of the form
\[
   \s\ni x_s=\left(\underbrace{\mbox{description of }M}_{\mbox{prefix coded}},\mbox{ classical string }s\in\s\right).
\]
The machine $V$ first determines the length $\ell(s)$ by detecting the first blank
symbol $\#$ on its tape. Then, it computes $n:=\ell(s)-4$ and $\delta:=\frac 1 {4(n+1)}
-\frac {23}2\cdot\frac 1 {80}2^{-2n}\in\left(0,\frac 1 {2e}\right)$
as well as $k:=\left\lceil \frac 1 {\delta}\right\rceil$.
Moreover, it computes a classical description of the QTM $M_k$, defined by
$M_k(\sigma):=M(k,\sigma)$ (compare Lemma~\ref{LemIsAQTM}).
Let $i$ be the number of the string $s$ in the set $\{0,1\}^{\ell(s)}$.
The machine $V$
computes the output $w$ of $P$ on input $i$, $n$, $M_k$ and $\delta$, outputs $w$ and halts.

Suppose that $w\in\s$ is a classical string with $QC_M(|w\rangle)= n$, then there
is a qubit string $\sigma\in\mathcal{T}_1^+(\hr_n)$ such that $\|M(k,\sigma)-|w\rangle\langle w|\,\|_{\rm Tr}
\leq \frac 1 k$ for every $k\in\N$, in particular for the $k$ given above. Thus,
\[
   \|M_k(\sigma)-|w\rangle\langle w|\,\|_{\rm Tr}\leq\delta.
\]
It follows that the string $w$ is an element of the set
$L$ corresponding to the input specified above.
Moreover, since $\Delta<\delta+\frac{23}2 \cdot\frac 1 {80} 2^{-2n}=\frac 1 {2(n+1)}$,
the length of the list is bounded by
\[
   \log\# L\leq \frac {n}{1-4\Delta}+\frac {4\Delta}{1-4\Delta}
   \log\frac 1 \Delta<\frac n {1-4\Delta}+3< n+4.
\]
Thus, the length $\ell(s)=n+4$ is enough to specify any element of the set $L$, and
$C_V(w)\leq \ell(x_s)\leq n+{\rm const}$. Equation~(\ref{EqLowerBound1}) now follows again from the
invariance of classical Kolmogorov complexity.
\qed

This was the most difficult part. Now, we use a few more arguments to prove the main result of this section.
\begin{ftheorem}[Quantum Complexity of Classical Strings]
\lineclear
For every classical string $s\in\s$, it holds
\[
   C(s)=QC(|s\rangle)+\mathcal{O}(1),
\]
i.e. the absolute value of the difference 	of $C$ and $QC$ is bounded by a constant on the
domain of classical strings. Moreover, for every rational $0<\delta<\frac 1 {2e}$,
there are constants $c_\delta,c_\delta'\in\N$ such that
\[
   QC^\delta(|s\rangle)\leq C(s)+c_\delta\leq\frac{QC^\delta(|s\rangle)}{1-4\delta}+c_\delta'.
\]
\end{ftheorem}
\proof If $k\in\N$ is large enough such that $\frac 1 k<\delta$, then we have
\[
   QC^\delta(|s\rangle)\leq QC^{\frac 1 k}(|s\rangle)\leq QC(|s\rangle)+k_\delta,
\]
where $k_\delta$ is a constant that depends only on $\delta$. This follows from the
obvious monotonicity property $\eps\leq\delta\Rightarrow QC^\delta\leq QC^\eps$ and Lemma~\ref{LemRelation}.

Also, we claim that there is some constant $c\in\N$ such that for every classical string $s\in\s$,
it holds
\[
   QC(|s\rangle)\leq C(s)+c.
\]
This can be seen as follows: According to Bennett~\cite{Bennett}, we can choose the classical TM
which is used in the definition of $C(s)$ to be reversible. But every reversible TM is also
a (special case of a) QTM. Thus, this equation follows from Theorem~\ref{MainTheorem2},
the invariance theorem for $QC$.

All the remaining inequalities are shown in Lemma~\ref{LemClassCompl}.
\qed

\section{Quantum Brudno's Theorem}
\label{SecQBrudno}
In this section, we prove a theorem that relates the von Neumann entropy rate and the
quantum Kolmogorov complexity rate of ergodic quantum information sources. This generalizes a classical
theorem that has first been conjectured by Zvonkin and Levin~\cite{Zvonkin}, and was later proved
by Brudno~\cite{Brudno}. The content of this section is joint work with F. Benatti, T. Kr\"uger,
Ra. Siegmund-Schultze and A. Szko\l a, and has already been published in \cite{QBrudno}.

The idea of the classical theorem is to compare two different notions of randomness:  Kolmogorov
complexity, which measures the randomness of {\em single} binary strings, and Shannon entropy,
which is a measure of randomness for information sources, i.e. probability distributions.

In more detail, if $p$ is a stationary classical information source, the most important parameter
is its {\em entropy rate} $h(p)=\lim_{n\to\infty}\frac 1 n H(p^{(n)})$, where $H(p^{(n)})$
denotes the Shannon entropy of the ensembles of strings of length $n$ that are emitted according
to the probability distribution $p^{(n)}$. According to the
Shannon-McMillan-Breiman theorem~\cite{Billingsley,CoverThomas},
$h(p)$ represents the optimal compression rate at which the
information provided by classical ergodic sources can be
compressed and then retrieved with negligible probability of error
(in the limit of longer and longer strings). Essentially, $n\cdot h(p)$
is the number of bits that are needed for reliable compression of
bit strings of length $n$. Thus, $h(p)$ can be interpreted as a measure
of randomness of the source $p$ and of the ensembles it emits.

On the other hand, one can look at the randomness of the single strings
that are emitted by the source. If $x$ is an infinite binary string and
$x^{(n)}$ denotes its first $n$ bits, then one can similarly define
its {\em complexity rate} as $c(x):=\lim_{n\to\infty}\frac 1 n C(x^{(n)})$
(if that limit exists),
where $C$ denotes classical Kolmogorov complexity.

Intuitively, one expects a connection between the randomness of
single strings and the average randomness of ensembles of strings.
In the classical case, this is exactly the content of a theorem by
Brudno~\cite{Brudno,Whi,Keller,Sow} which states that for
ergodic sources, the complexity rate of $p$-almost all infinite sequences
$x$ coincides with the entropy rate, i.e. $c(x)=h(p)$ holds $p$-almost surely.

In this section, we prove that a similar relation holds for the von Neumann
entropy rate and the quantum Kolmogorov complexity rate of quantum ergodic
information sources (we explain this notion below in Subsection~\ref{SecErgodicQuantumSources}).
This is an interesting result in its own right, and it also supports the point of view
that the quantum Kolmogorov complexity notions $QC$ and $QC^\delta$ are useful and natural.

\subsection{Ergodic Quantum Sources}
\label{SecErgodicQuantumSources}

In order to  formulate our main result rigorously, we start with a
brief introduction to the relevant concepts of the formalism of
quasi-local $C^*$-algebras, which is the most suitable formalism for dealing with
quantum information sources. At the same time, we fix some notation.

We would like to consider a spin chain of infinitely many qubits.
This chain is modelled by some $C^*$-algebra $\mathcal{A}^\infty$, the
{\em quasi-local algebra}, which is constructed as follows.

We consider the lattice $\mathbb{Z}$ and assign to each site
$x \in \mathbb{Z}$  a $C^*$-algebra $\mathcal{A}_x$ being
a copy of a fixed finite-dimensional algebra $\mathcal{A}$, in the sense
that there exists a $*$-isomorphism $i_x:\mathcal{A}\to\mathcal{A}_x$.
To simplify notations, we write $a\in\mathcal{A}_x$ for 
$i_x(a)\in\mathcal{A}_x$ and $a\in\mathcal{A}$.
The algebra of observables associated to a finite $\Lambda \subset
\mathbb{Z}$ is defined by $\mathcal{A}_{\Lambda}:= \bigotimes _{x
\in \Lambda} \mathcal{A}_x$. Observe that for $\Lambda \subset
\Lambda^{'}$ we have $
\mathcal{A}_{\Lambda^{'}}=\mathcal{A}_{\Lambda}\otimes
\mathcal{A}_{\Lambda^{'} \backslash \Lambda}$ and there is a
canonical embedding of $\mathcal{A}_{\Lambda}$ into
$\mathcal{A}_{\Lambda^{'}}$ given by $a \mapsto a \otimes
\mathbf{1}_{\Lambda^{'}\backslash  \Lambda}$,  where $a \in
\mathcal{A}_{\Lambda}$ and $\mathbf{1}_{\Lambda^{'} \backslash
\Lambda}$ denotes the identity of $\mathcal{A}_{\Lambda^{'}
\backslash \Lambda}$. The infinite-dimensional quasi-local
$C^*$-algebra $\mathcal{A}^{\infty}$ is the norm completion of the
normed algebra $\bigcup_{\Lambda \subset \mathbb{Z}}
\mathcal{A}_{\Lambda}$, where the union is taken over all finite
subsets $\Lambda$.

In this thesis, we only deal with qubits.
Thus, in the following, we restrict our considerations to the case where
$\mathcal{A}$  is the algebra of observables of a qubit, i.e. the
algebra $\mathcal{M}_2(\mathbb{C})$ of $2\times 2$ matrices acting on
$\mathbb{C}^2$.

Similarly, we think of  $\mathcal{A}_{\Lambda}$ as the algebra of
observables of qubit strings of length $|\Lambda|$, namely the algebra
$\mathcal{M}_{2^{|\Lambda|}}(\mathbb{C})
=\mathcal{M}_2(\mathbb{C})^{\otimes|\Lambda|}$ of 
$2^{|\Lambda|}\times 2^{|\Lambda|}$ matrices acting on the Hilbert
space $\mathcal{H}_\Lambda:=(\mathbb{C}^2)^{\otimes |\Lambda|}$. The quasi-local
algebra  $\mathcal{A}^{\infty}$ corresponds to the doubly-infinite
qubit strings.

The (right) shift $T$ is a $*$-automorphism on $\mathcal{A}^\infty$
uniquely defined by its action on local observables
\begin{eqnarray}
   T:a\in\mathcal{A}_{[m,n]}\mapsto a\in\mathcal{A}_{[m+1,n+1]}
\end{eqnarray}
where $[m,n] \subset \mathbb{Z}$ is an integer interval.

A state $\Psi$ on $\mathcal{A}^{\infty}$ is a normalized positive
linear functional on $\mathcal{A}^\infty$. Each local state
$\Psi_{\Lambda}:=\Psi\upharpoonright \mathcal{A}_{\Lambda}$,
$\Lambda \subset \mathbb{Z} $ finite, corresponds to a density
operator $\rho_\Lambda \in \mathcal{A}_{\Lambda}$ by the relation
$\Psi_{\Lambda}(a)= \hbox{Tr}\left(\rho_\Lambda a\right)$, for all
$a \in \mathcal{A}_\Lambda$, where $\rm{Tr}$ is the trace on $(\C^2)^{\otimes |\Lambda|}$.
The density operator $\rho_\Lambda$
is a positive matrix acting on the Hilbert space 
$\mathcal{H}_\Lambda$ associated with
$\mathcal{A}_\Lambda$ satisfying the normalization
condition $\hbox{Tr}\rho_\Lambda=1$. The simplest $\rho_\Lambda$ 
correspond to one-dimensional projectors 
$P:=\vert\psi_\Lambda\rangle\langle\psi_\Lambda\vert$ onto
vectors
$\vert\psi_\Lambda\rangle\in\mathcal{H}_\Lambda$
and are called pure states, while
general density operators are linear convex combinations of
one-dimensional projectors: $\rho_\Lambda=\sum_i\lambda_i
\vert\psi^i_\Lambda\rangle\langle\psi^i_\Lambda\vert$, $\lambda_i\geq 0$,
$\sum_j\lambda_j=1$.

A state $\Psi $ on
$\mathcal{A}^{\infty}$ corresponds one-to-one to a family of
density operators $\rho_\Lambda \in \mathcal{A}_{\Lambda}$, $\Lambda
\subset \mathbb{Z} $ finite, fulfilling the consistency condition
$\rho_{\Lambda}= \hbox{Tr}_{\Lambda ' \backslash
\Lambda}\left(\rho_{\Lambda '}\right)$ for $\Lambda\subset \Lambda'$,
where $\hbox{Tr}_{\Lambda}$
denotes the partial trace over the local algebra
$\mathcal{A}_{\Lambda}$ which is computed with respect to any
orthonormal basis in the associated Hilbert space 
$\mathcal{H}_\Lambda$. 
Notice that a state $\Psi$ with $\Psi \circ
T= \Psi$, i.e. a shift-invariant state, is uniquely determined by a
consistent sequence of density operators
$\rho^{(n)}:=\rho_{\Lambda(n)}$ in 
$\mathcal{A}^{(n)}:= \mathcal{A}_{\Lambda(n)}$
corresponding to the local states $\Psi^{(n)}:= \Psi_{\Lambda(n)}$,
where  $\Lambda(n)$ denotes the integer interval $[1,n]\subset
\mathbb{Z}$, for each  $n \in \mathbb{N}$.

As motivated in the introduction, in the information-theoretical
context, we interpret the tuple $(\mathcal{A}^{\infty}, \Psi)$
describing the quantum spin chain as a stationary quantum source.

The von Neumann entropy of a density matrix $\rho$ is 
$S(\rho ):=-\hbox{Tr}(\rho \log \rho)$. By the subadditivity of $S$ for a
shift-invariant state $\Psi$ on $\mathcal{A}^{\infty}$, the
following limit, the quantum entropy rate, exists $$
s(\Psi):=\lim_{n\to\infty}\frac{1}{n}S(\rho^{(n)})\ .
$$
The set of shift-invariant states  on $\mathcal{A}^{\infty}$ is
convex and compact in the  weak$*$-topology. The extremal points of
this set are called ergodic states: they are those states which cannot
be decomposed into linear convex combinations of other shift-invariant states.
Notice that in particular the
shift-invariant product states defined by a sequence of density
matrices $\rho^{(n)}= \rho^{\otimes n}$, $n \in \N$, where $\rho $ is a fixed
$2 \times 2$ density matrix, are ergodic. They are the quantum
counterparts of Bernoulli (i.i.d.) processes. Most of the results in
quantum information theory concern such sources, but more general ergodic quantum sources allowing 
correlations can be considered. This is often useful, since such sources
naturally appear, for example, in statistical mechanics.

\subsection{Proof of Quantum Brudno's Theorem}
It turns out that the rates of the quantum Kolmogorov complexities
$QC$ and $QC^\delta$
of the typical pure states (i.e. typical pure qubit strings) generated by an
ergodic quantum source $(\mathcal{A}^\infty, \Psi)$ are
asymptotically equal to the entropy rate $s(\Psi)$ of the
source. A precise formulation of this result is the content of the 
following theorem. It can be seen  as a quantum extension of Brudno's
theorem as a convergence in probability
statement, while the original formulation of Brudno's result is an
almost sure statement.

In the remainder of this section, we call a sequence of projectors 
$p_n \in \mathcal{A}^{(n)}$, $n \in \N$, satisfying 
$\lim_{n \to \infty} \Psi^{(n)}(p_n)=1$
a {\sl sequence of $\Psi$-typical projectors}.

\begin{ftheorem}[Quantum Brudno Theorem]
\label{TheQBrudno}
\lineclear
Let $(\mathcal{A}^\infty,\Psi)$ be an ergodic quantum source with 
entropy rate $s$.
For every $\delta>0$, there exists a sequence of
$\Psi$-typical projectors $q_n(\delta)\in\mathcal{A}^{(n)}$, 
$n \in \N$, i.e. $\lim_{n \to \infty} \Psi^{(n)}(q_n(\delta))=1$, 
such that for $n$ large enough every
one-dimensional projector $q\leq q_n(\delta)$ satisfies
\begin{eqnarray}
& & \frac 1 n QC(q) \in \left(s-\delta,s+\delta\right),\label{eq1}\\
& & \frac 1 n QC^\delta(q)\in\left( s-\delta(4+\delta)s,s+\delta\right).\label{eq2}
\end{eqnarray}
Moreover, $s$ is the optimal expected asymptotic complexity rate, in the sense that every sequence of projectors $q_n\in\mathcal{A}^{(n)}$, 
$n \in \N$, that for large $n$ may be represented as a sum of mutually orthogonal one-dimensional projectors that all violate the lower bounds in (\ref{eq1}) and (\ref{eq2}) for some $\delta>0$,  has an asymptotically vanishing expectation value with respect to $\Psi$.
\end{ftheorem}

\subsubsection{Proof of the Lower Bound}
A key argument in the proof of the lower bound is the following theorem \cite[Prop. 2.1]{QSMPaper}.
It is closely related to the quantum
Shannon-McMillan Theorem and concerns the minimal
dimension of the $\Psi-$typical subspaces.

\begin{ftheorem}[\cite{QSMPaper}]
\label{QAEP}
Let $(\mathcal{A}^\infty,\Psi)$ be an ergodic quantum source with
entropy rate $s$.
Then, for every $0<\varepsilon<1$,
\begin{equation}
   \lim_{n\to\infty}\frac 1 n \beta_{\eps,n}(\Psi)=s,
   \label{eqBoltzmann}
\end{equation}
where $
   \beta_{\eps,n}(\Psi):=\min\left\{
      \log {\rm Tr}_n(q)\enspace|\enspace q\in\mathcal{A}^{(n)}
\mbox{ projector },
      \Psi^{(n)}(q)\geq 1-\eps
   \right\}$.
\end{ftheorem}
Notice that the limit (\ref{eqBoltzmann}) is valid for every
$\eps\in(0,1)$. By means of this property, we will first prove the lower
bound for the complexity notion $QC^\delta$, and then use
Lemma~\ref{LemRelation} to extend it to $QC$.
\begin{fcorollary}[Lower Bound for $\frac 1 n QC^\delta$]
\label{Cor1}
\lineclear
Let $(\mathcal{A}^\infty,\Psi)$ be an
ergodic quantum source with entropy rate $s$. Moreover, let
$0<\delta< \frac 1 {2e}$, and let
$\left(p_n\right)_{n\in\N}$ be a sequence of $\Psi$-typical projectors.
Then, there is another sequence of $\Psi$-typical projectors
$q_n(\delta)\leq p_n$, such that for $n$ large enough
\[
\frac{1}{ n} QC^{\delta}(q)>s-\delta(4+\delta)s\]
is true for every one-dimensional projector $q\leq q_n(\delta)$.
\end{fcorollary}
{\bf Proof. } The case $s=0$ is trivial, so let $s>0$.
Fix $n \in \N$ and $0<\delta<\frac 1 {2e}$, and consider the set
\[
   A_n(\delta):=\left\{
      p\leq p_n\ |\ p \textrm{ one-dim. proj., }
      QC^\delta(p)\leq ns(1-\delta(4+\delta))\right\}.
\]
From the definition of $QC^\delta(p)$, for all $p\in A_n(\delta)$
there exist associated density matrices $\sigma_p$ with
$\ell(\sigma_p)\leq ns(1-\delta(4+\delta))$
such that
$\|\mathcal{U}(\sigma_p)-p\|_{\rm Tr}\leq \delta$, where $\mathcal{U}$ denotes
the quantum operation $\mathcal{U}:\mathcal{T}(\hr_\s)\to
\mathcal{T}(\hr_\s)$ of the corresponding strongly universal QTM $\mathfrak{U}$,
as explained in Lemma~\ref{LemmaQTMsAreOperations}.
Let $p_n(\delta)\leq p_n$ be a sum of
a maximal number of mutually orthogonal projectors from the set $A_n(\delta)$.
Lemma~\ref{TheQCounting} implies that
\[
   \log{\rm Tr} \, p_n(\delta)\leq \frac{\log\dim \hr_{\leq\lfloor ns(1-\delta(4+\delta))\rfloor}+4\delta\log\frac 1 \delta}{1-4\delta}
\]
and there are no one-dimensional projectors $p \leq p_n(\delta)^\perp:= p_n - p_n(\delta)$ such that 
$p \in A_n(\delta)$. Thus,
one-dimensional projectors $p \leq p_n(\delta)^{\perp}$ must satisfy
$\frac{1}{n}QC^{\delta}(p)> s- \delta(4+\delta)s$.
Since $\log\dim\hr_{\leq c}<c+1$ for every $c\in\N$, we conclude
\begin{eqnarray}
   \limsup_{n\to\infty}\frac 1 n \log {\rm Tr}\, p_n(\delta)
   \leq \frac{s(1-\delta(4+\delta))}{1-4\delta}=s-\frac{s\delta^2}{1-4\delta}<s.
   \label{eqSmallerS}
\end{eqnarray}
Using Theorem~\ref{QAEP}, we obtain that
$\lim_{n\to\infty}\Psi^{(n)}(p_n(\delta))=0$.
Finally, set $q_n(\delta):=p_n(\delta)^\perp$.
The claim follows. \qed

\begin{fcorollary}[Lower Bound for $\frac 1 n QC$]
\label{Cor2}
\lineclear
Let $(\mathcal{A}^\infty,\Psi)$ be an
ergodic quantum source with entropy rate $s$.
Let $\left(p_n\right)_{n\in\N}$ with $p_n\in\mathcal{A}^{(n)}$ be an
arbitrary sequence of $\Psi$-typical projectors. 
Then, for every $0<\delta <\frac 1 {2e}$,
there is a sequence of $\Psi$-typical projectors $q_n(\delta)\leq p_n$ such
that for $n$ large enough
\[
\frac 1 n QC(q)>s-\delta
\]
is satisfied for every one-dimensional projector
$q\leq q_n(\delta)$.
\end{fcorollary}
{\bf Proof. } According to Corollary~\ref{Cor1}, for
every $k \in \N$, there exists a sequence of $\Psi$-typical 
projectors $p_n(\frac 1 k)\leq p_n$
with $\frac 1 n QC^\frac 1 k (q)>s-\frac 1 k (4+\frac 1 k)s$ for
every one-dimensional projector $q\leq p_n(\frac 1 k )$ if $n$ is large enough.
We have
\begin{eqnarray*}
   \frac 1 n QC(q)&\geq&\frac 1 n QC^{1/k}(q)-
   \frac {2+2\lfloor\log k\rfloor}{n}\\
   &>& s-\frac 1 k \left(4+\frac 1 k\right)s-\frac{2(2+\log k)} n,
\end{eqnarray*}
where the first estimate is by Lemma~\ref{LemRelation}, and the second
one is true for one-dimensional projectors $q \leq p_n(\frac 1 k )$
and $n\in\N$ large enough. Fix some large $k$ satisfying 
$\frac 1 k (4 + \frac 1 k )s \leq \frac \delta 2 $. 
The result follows by setting $q_n(\delta)=p_n(\frac 1 k)$.
\qed

\subsubsection{Upper Bound}
In the previous paragraph, we have shown that with high probability
and for large $m$, the quantum complexity rate
$\frac 1 m QC^{\delta}$ is bounded from
below by $s(1-\delta(4+\delta))$, and the
quantum complexity rate $\frac 1 m QC$ by $s-\delta$.
We are now going to establish the upper bounds.
\begin{proposition}[Upper Bound]
\label{PropUpperBound}
\lineclear
Let $(\mathcal{A}^\infty,\Psi)$ be an ergodic quantum source with entropy rate
$s$.
Then, for every $0<\delta<1/e$, there is a sequence of 
$\Psi$-typical projectors $q_m(\delta)\in\mathcal{A}^{(m)}$ such that
for every one-dimensional
projector $q\leq q_m(\delta)$ and $m$ large enough
\begin{eqnarray}
   \frac 1 m QC(q)&<&s+\delta\qquad\mbox{and}\label{UpperZero}\\
   \frac 1 m QC^\delta(q)&<&s+\delta\,\,.\label{UpperDelta}
\end{eqnarray}
\end{proposition}
We prove the above proposition by explicitly providing a quantum
algorithm (with program length increasing like $m(s+\delta)$)
that computes $q$ within arbitrary accuracy. This will be done by
means of quantum universal typical subspaces constructed by Kaltchenko and Yang in \cite{KaltchenkoYang}. 
\begin{ftheorem}[Universal Typical Subspaces \cite{KaltchenkoYang}]
\label{Kaltc}
\lineclear
Let $s > 0$ and $\varepsilon>0$. There exists a sequence of
projectors $Q^{(n)}_{s,\varepsilon} \in \mathcal{A}^{(n)} $, $n\in\N$, 
such that for $n$ large enough
\begin{eqnarray}
\displaystyle
{\rm Tr}\Bigl(Q^{(n)}_{s,\varepsilon}\Bigr)\leq
2^{n(s+\varepsilon)}\label{eqTraceIsSmall}
\end{eqnarray}
and  for every ergodic quantum state $\Psi \in \mathcal{S}(\mathcal{A}^\infty)$
with entropy rate $s(\Psi) \leq s$ it holds that
\begin{eqnarray}
\lim_{n\to\infty}\Psi^{(n)}(Q^{(n)}_{s,\varepsilon})=1\,\,.
\end{eqnarray}
\end{ftheorem}
We call the orthogonal projectors $Q_{s,\varepsilon}^{(n)}$ in the
above theorem universal typical projectors at level $s$.
Suited for designing an appropriate quantum algorithm,
we slightly modify the proof given by Kaltchenko and
Yang in \cite{KaltchenkoYang}. 

{\bf Proof. } Let $l \in \mathbb{N}$ and $R > 0$. We consider an Abelian
quasi-local subalgebra $\mathcal{C}_{l}^\infty \subseteq
\mathcal{A}^\infty $ constructed from a maximal Abelian $l-$block
subalgebra $\mathcal{C}_{l}\subseteq \mathcal{A}^{(l)} $. The results in
\cite{Ziv, Kieffer} imply that there exists a universal sequence 
of projectors $p_{l,R}^{(n)} \in \mathcal{C}^{(n)}_l 
\subseteq \mathcal{A}^{(ln)}$ with $\frac{1}{n} \log \textrm{Tr }
p_{l,R}^{(n)}\leq R$ such that $\lim_{n \to \infty} \pi^{(n)}(p_{l,
R}^{(n)})=1$ 
for any ergodic state $\pi$ on the Abelian algebra $\mathcal{C}_l^{\infty}$
with entropy rate $s(\pi) < R$. Notice that ergodicity and entropy
rate of $\pi$ are defined with respect to
the shift on $\mathcal{C}_l^\infty$, which corresponds to
the $l$-shift on $\mathcal{A}^{\infty}$.

The first step in \cite{KaltchenkoYang} is to
apply unitary operators of the form $U ^{\otimes n}$,
$U\in \mathcal{A}^{(l)}$
unitary, to the $p_{l,R}^{(n)} $ and to introduce the projectors
\begin{eqnarray}
w_{l,R}^{(ln)}:=\bigvee_{U\in\mathcal{A}^{(l)}\mbox{ unitary}}
   U^{\otimes n} p_{l,R}^{(n)} U^{*\otimes n} \in \mathcal{A}^{(ln)}.
   \label{eqJoin1}
\end{eqnarray}
Let $p_{l,R}^{(n)}=\sum_{i\in I}|i_{l,R}^{(n)}\rangle\langle i_{l,R}^{(n)}|$ be
a spectral decomposition of $p_{l,R}^{(n)}$ (with $I\subset\N$ some
index set), and let $\mathbb{P}(V)$ denote the orthogonal projector
onto a given subspace $V$.
Then, $w_{l,R}^{(ln)}$ can also be written as
\[
   w_{l,R}^{(ln)}=\mathbb{P}\left(
      {\rm span}\{
         U^{\otimes n}|i_{l,R}^{(n)}\rangle:
         i\in I, U\in\mathcal{A}^{(l)}\mbox{ unitary}
      \}
   \right).
\]
It will be more convenient for the construction of our algorithm 
in \ref{subsubconstr} to consider the projector
\begin{equation}
   W_{l,R}^{(ln)}:=\mathbb{P}\left(
      {\rm span}\{
         A^{\otimes n}|i_{l,R}^{(n)}\rangle:
         i\in I, A\in\mathcal{A}^{(l)}
      \}
   \right).
   \label{eqJoin2}
\end{equation}
It holds that $w_{l,R}^{(ln)}\leq W_{l,R}^{(ln)}$.
For integers $m = nl + k$ with $n \in \mathbb{N}$ and $k \in \{0,\dots, l-1\}$
we introduce the projectors in $\mathcal{A}^{(m)}$
\begin{eqnarray}
   w_{l,R}^{(m)}:=w_{l,R}^{(ln)} \otimes \mathbf{1}^{\otimes k},\qquad
   W_{l,R}^{(m)}:=W_{l,R}^{(ln)} \otimes \mathbf{1}^{\otimes k}.
\end{eqnarray}
We now use an argument of \cite{JHHH} to estimate the trace of
$W_{l,R}^{(m)} \in \mathcal{A}^{(m)}$. The dimension of the symmetric subspace
$\textrm{SYM}^n(\mathcal{A}^{(l)}):={\rm span}\{A^{\otimes
n}:A\in\mathcal{A}^{(l)}\}$ is upper bounded by $(n+1)^{\dim
\mathcal{A}^{(l)}}$, thus
\begin{eqnarray}
   \hbox{Tr } W_{l,R}^{(m)} = \hbox{Tr } W_{l,R}^{(ln)}
   \cdot  \hbox{Tr }\mathbf{1}^{\otimes k}  &\leq& (n+1)^{2^{2l}}
   \hbox{Tr }p_{l,R}^{(n)} \cdot 2^{l} \nonumber\\&\leq&  (n+1)^{2^{2l}}
   \cdot 2^{Rn} \cdot 2^{l} \label{dim_estimate}.
\end{eqnarray}
Now we consider a stationary ergodic state $\Psi$  on the
quasi-local algebra $\mathcal{A}^\infty$ with entropy rate $s(\Psi)
\leq s$. Let $\eps, \delta > 0$. If  $l$ is chosen large enough then the
projectors $w_{l,R}^{(m)}$, where
$R:=l(s+ \frac{\eps}{2})$, are $\delta-$typical for $\Psi$, i.e. 
$\Psi^{(m)}(w_{l,R}^{(m)})\geq 1- \delta$, for $m \in \N$ sufficiently
large. This can be seen as
follows. Due to the result in \cite[Thm. 3.1]{QSMPaper} the ergodic state
$\Psi$ convexly decomposes into $k(l)\leq l$ states
\begin{eqnarray}\label{erg_decomp}
\Psi = \frac{1}{k(l)}\sum_{i=1}^{{k(l)}} \Psi_{i,l},
\end{eqnarray}
each $\Psi_{i,l}$ being ergodic with respect to the $l-$shift on
$\mathcal{A}^\infty$ and having an entropy rate (with respect to the
$l-$shift) equal to $s(\Psi)\cdot l$. 
We define for $\Delta >0$ the set of integers
\begin{eqnarray}
A_{l, \Delta}:= \{i \in \{1, \dots, k(l)\}: \ S(\Psi^{(l)}_{i,l})
\geq l(s(\Psi)+ \Delta) \}.
\end{eqnarray}
Then, according to a density lemma proven in \cite[Lemma 3.1]{QSMPaper} it holds
\begin{eqnarray}
\lim_{l \to \infty} \frac{|A_{l,\Delta}|}{k(l)}=0.
\end{eqnarray}
Let $\mathcal{C}_{i,l}$ be the maximal Abelian subalgebra of
$\mathcal{A}^{(l)}$ generated by the one-dimensional
eigenprojectors of $\Psi_{i,l}^{(l)}\in
\mathcal{S}(\mathcal{A}^{(l)} )$.
The restriction of a component $\Psi_{i,l}$ to the Abelian
quasi-local algebra $\mathcal{C}_{i,l}^\infty $ is again an ergodic
state. It holds in general
\begin{eqnarray}
l \cdot s(\Psi) = s(\Psi_{i,l})\leq s(\Psi_{i,l} \upharpoonright
\mathcal{C}_{i,l}^\infty)  \leq
S(\Psi_{i,l}^{(l)}\upharpoonright \mathcal{C}_{i,l})
=S(\Psi^{(l)}_{i,l}).
\end{eqnarray}
For $i \in A_{l,\Delta}^c$, where we set $\Delta:=\frac{R}{l}-s(\Psi)$,
we additionally have
the upper bound $S(\Psi^{(l)}_{i,l}) < R$.
Let $U_i \in \mathcal{A}^{(l)}$ be a unitary operator such that
$U_i^{\otimes n} p_{l,R}^{(n)}U_i^{*\otimes n}
\in \mathcal{C}_{i,l}^{(n)}$.
For every $i \in A_{l,\Delta}^c$, it holds that
 \begin{eqnarray}\label{erg_comp}
\Psi_{i,l}^{(ln)}(w_{l,R}^{(ln)}) \geq
\Psi_{i,l}^{(ln)}(U_i^{\otimes n} p_{l,R}^{(n)}U_i^{*\otimes n})
\longrightarrow 1\qquad
\textrm{as } n \to \infty.
\end{eqnarray}
We fix an $l \in \N$ large enough to fulfill
$\frac{|A_{l,\Delta}^c|}{k(l)}\geq 1-\frac{\delta}{2}$ and use the
ergodic decomposition  (\ref{erg_decomp})
 to obtain the lower bound
\begin{eqnarray}
\Psi^{(ln)}(w_{l,R}^{(ln)})\geq \frac{1}{k(l)}
\sum_{i \in A_{l,\Delta}^c}\Psi_{l,i}^{(nl)}(w_{l,R}^{(ln)})
\geq \left(1- \frac{\delta}{2}\right)\min_{i \in A_{l,\Delta}^c}
\Psi_{i,l}^{(nl)}(w_{l,R}^{(ln)}).
\end{eqnarray}
\,From (\ref{erg_comp}) we conclude that for $n$ large enough
\begin{eqnarray}
   \Psi^{(ln)}(W_{l,R}^{(ln)})\geq \Psi^{(ln)}(w_{l,R}^{(ln)}) \geq 1-\delta.
\end{eqnarray}
We proceed by following the lines of \cite{KaltchenkoYang} by
introducing  the sequence $l_m$, $m \in \N$, where each $l_m$ 
is a power of $2$ fulfilling the inequality
\begin{eqnarray}\label{l_m}
l_m 2^{3\cdot l_m} \leq m < 2 l_m 2^{3\cdot 2 l_m}.
\end{eqnarray}
Let the integer sequence $n_m$ and the real-valued sequence $R_m$ be defined
 by $n_m:=\lfloor \frac m {l_m}\rfloor$ and 
$R_m:=l_m \cdot \left(s+ \frac \eps 2\right)$. Then we set
\begin{equation}
   Q_{s,\eps}^{(m)}:=\left\{
      \begin{array}{ll}
          W_{l_m,R_m}^{(l_m n_m)} & \mbox{if }m=l_m 2^{3\cdot l_m}\,\,,\\
          W_{l_m,R_m}^{(l_m n_m)}\otimes \idn^{\otimes(m-l_m n_m)}
          & \mbox{otherwise}\,\,.
      \end{array}
   \right.
   \label{eqTypicalProjector}
\end{equation}
Observe that
\begin{eqnarray}
\frac{1}{m}\log \hbox{Tr }Q_{s,\eps}^{(m)} & \leq& \frac{1}{n_ml_m}
\log \hbox{Tr }Q_{s,\eps}^{(m)}
\nonumber \\ 
&\leq& \frac{4^{l_m}}{l_m}\frac{\log(n_m+1)}{n_m}
+\frac{R_m}{l_m}+\frac{1}{n_m}\\ 
&\leq & \frac{4^{l_m}}{l_m}\frac{6l_m +2}{2^{3l_m}-1}
+ s+\frac{\eps}{2} +\frac{1}{2^{3l_m}-1},
\end{eqnarray}
where the second inequality is by estimate (\ref{dim_estimate}) 
and the last one by the bounds on $n_m$
\begin{eqnarray*}
2^{3l_m}-1\leq \frac{m}{l_m}-1 \leq n_m \leq \frac{m}{l_m}\leq 2^{6l_m +1}.
\end{eqnarray*}
Thus, for large $m$, it holds
\begin{eqnarray}
   \frac{1}{m} \log \textrm{Tr }Q_{s,\eps}^{(m)} \leq s +\eps.
   \label{eqLogTrQm}
\end{eqnarray}
By the special choice (\ref{l_m}) of $l_m$ it is ensured that the sequence of
projectors $Q_{s,\eps}^{(m)} \in \mathcal{A}^{(m)}$ is indeed typical for
any quantum state $\Psi$ with entropy rate $s(\Psi) \leq s$, 
compare \cite{KaltchenkoYang}. This means that
$\{Q_{s,\eps}^{(m)}\}_{m \in \mathbf{N}}$ is a sequence of universal
typical projectors at level $s$. \qed
\subsubsection{Construction of the Decompression Algorithm}
\label{subsubconstr}
We proceed by applying the latter result to universal typical subspaces
for our proof of the upper bound. Let
$0<\eps<\delta/2$ be an arbitrary real number such that $r:=s+\eps$ is
rational, and let $q_m:=Q_{s,\eps}^{(m)}$ be the universal
projector sequence of Theorem~\ref{Kaltc}. Recall that
the projector sequence $q_m$ is {\em independent} of the choice of
the ergodic state $\Psi$, as long as $s(\Psi)\leq s$.

Because of (\ref{eqTraceIsSmall}), for $m$ large enough,
there exists some unitary transformation $U^*$ that transforms the projector
$q_m$ into a projector belonging to $\mathcal{T}_1^+(\hr_{\lceil mr\rceil})$,
thus transforming every one-dimensional projector
$q\leq q_m$ into a qubit string $\tilde q:=U^* q U$ of length 
$\ell(\tilde q)=\lceil mr\rceil$.

As shown in \cite{BernsteinVazirani}, a {\sl UQTM} can implement
every classical algorithm, and it can apply every unitary
transformation $U$ (when given an algorithm for the computation of $U$) on
its tapes within any desired accuracy.
We can thus feed $\tilde q$ (plus some classical instructions
including a subprogram for the computation of $U$) as
input into the {\sl UQTM} $\mathfrak U$. This {\sl UQTM} starts by
computing a classical description of the transformation $U$, and
subsequently applies $U$ to $\tilde q$, recovering the original
projector $q=U\tilde q U^*$ on the output tape.

Since $U=U(q_m)$ depends on $\Psi$ only through
its entropy rate $s(\Psi)$,
the subprogram that computes $U$ does not have to be
supplied with additional information on $\Psi$ and will thus
have fixed length.

We give a precise definition of a quantum decompression algorithm 
$\mathfrak{A}$, which is, formally, a mapping ($r$ is rational)
\begin{eqnarray*}
   \mathfrak{A}:\N\times\N
   \times{\mathbb Q}\times\hr_{\Fock}\to\hr_{\Fock}\,\,,\\
   (k,m,r,\tilde q)\mapsto      q=\mathfrak{A}(k,m,r,\tilde q)\,\,.
\end{eqnarray*}
We require that $\mathfrak{A}$ is a  ``short algorithm'' in the sense 
of ``short in description'', {\em not}
short (fast) in running time or resource consumption. Indeed, the algorithm
$\mathfrak{A}$ is very slow and memory consuming, but this does
not matter, since Kolmogorov complexity only cares about the
description length of the program.

The instructions defining the quantum algorithm $\mathfrak{A}$ are:
\begin{itemize}
\item[1.] Read the value of $m$, and find a solution $l\in\N$ for the 
inequality
\[
   l \cdot 2^{3 l}\leq m < 2 \cdot l\cdot 2^{3\cdot 2 l}
\]
such that $l$ is a power of two. (There is only one such $l$.)
\item[2.]Compute $n:=\lfloor\frac m l\rfloor$.
\item[3.] Read the value of $r$. Compute $R:=l\cdot r$.
\item[4.] Compute a list of codewords $\Omega_{l,R}^{(n)}$,
belonging to a classical universal block
code sequence of rate $R$. (For the construction of an appropriate algorithm,
see \cite[Thm. 2 and 1]{Kieffer}.) Since
\[
   \Omega_{l,R}^{(n)}\subset\left(\{0,1\}^l\right)^n\,\,,
\]
$\Omega_{l,R}^{(n)}=\{\omega_1,\omega_2,\ldots,\omega_M\}$ 
can be stored as a list
of binary strings. Every string has length $\ell(\omega_i)=nl$. 
(Note that the exact value
of the cardinality $M\approx 2^{nR}$ depends on the choice of 
$\Omega_{l,R}^{(n)}$.)
\end{itemize}
During the following steps, the quantum algorithm $\mathfrak{A}$
will have to deal with
\begin{itemize}
\item rational numbers,
\item square roots of rational numbers,
\item binary-digit-approximations (up to some specified accuracy) of
real numbers,
\item (large) vectors and matrices containing such numbers.
\end{itemize}
A classical {\sl TM} can of course deal with all such objects
(and so can a QTM):
For example, rational numbers can be stored
as a list of two integers (containing numerator and denominator),
square roots can be stored as such a list and an additional bit
denoting the square root, and binary-digit-approximations can be
stored as binary strings. Vectors and matrices are arrays containing
those objects. They are always assumed to be given in the
computational basis. Operations on those objects, like addition or
multiplication, are easily implemented.

The quantum algorithm $\mathfrak{A}$ continues as follows:
\begin{itemize}
\item[5.] Compute a basis $\left\{A_{\{i_1,\ldots,i_n\}}\right\}$
of the symmetric subspace
\[
   \textrm{SYM}^n(\mathcal{A}^{(l)}):={\rm span}\{A^{\otimes n}:A
   \in\mathcal{A}^{(l)}\}\,\,.
\]
This can be done as follows: For every $n$-tuple $\{i_1,\ldots,i_n\}$,
where $i_k\in\{1,\ldots,2^{2l}\}$,
there is one basis element $A_{\{i_1,\ldots,i_n\}}\in\mathcal{A}^{(ln)}$,
given by the formula
\begin{equation}
   A_{\{i_1,\ldots,i_n\}}=\sum_{\sigma}
   e^{(l,n)}_{\sigma(i_1,\ldots,i_n)}\,\,,
   \label{eqWillBeRational}
\end{equation}
where the summation runs over all $n$-permutations $\sigma$, and
\[
   e_{i_1,\ldots,i_n}^{(l,n)}:=e_{i_1}^{(l)}\otimes
   e_{i_2}^{(l)}\otimes\ldots
   \otimes e_{i_n}^{(l)}\,\,,
\]
with $\left\{e_k^{(l)}\right\}_{k=1}^{2^{2l}}$ a system of matrix
units\footnote{In the computational basis, all entries are zero,
except for one entry which is one.} in $\mathcal{A}^{(l)}$.

There is a number of $d=\binom{n+2^{2l}-1} {2^{2l}-1}
=\dim(\textrm{SYM}^n(\mathcal{A}^{(l)}))$
different matrices $A_{\{i_1,\ldots,i_n\}}$ which we can label by 
$\left\{A_k\right\}_{k=1}^d$.
It follows from (\ref{eqWillBeRational}) that these matrices
have integer entries.

They are stored as a list of $2^{ln}\times 2^{ln}$-tables of integers.
Thus, this step of the computation is exact, that is without approximations.
\item[6.] For every $i\in\{1,\ldots,M\}$ and $k\in\{1,\ldots,d\}$, let
\[
   |u_{k,i}\rangle:=A_k|\omega_i\rangle\,\,,
\]
where $|\omega_i\rangle$ denotes the computational basis vector which
is a tensor product of $|0\rangle$'s and $|1\rangle$'s according to
the bits of the string $\omega_i$.
Compute the vectors $|u_{k,i}\rangle$ one after the other. For every
vector that has been computed, check if it can be written as a
linear combination of already computed vectors.
(The corresponding system of linear equations can
be solved exactly, since every vector is given as an array
of integers.) If yes, then discard the new vector
$|u_{k,i}\rangle$, otherwise store it and give it a number.

This way, a set of vectors $\left\{|u_k\rangle\right\}_{k=1}^D$ is
computed. These vectors linearly span the support of the projector
$W_{l,R}^{(ln)}$ given in (\ref{eqJoin2}).
\item[7.] Denote by $\left\{|\phi_i\rangle\right\}_{i=1}^{2^{m-ln}}$
the computational basis vectors of $\hr_{m-ln}$. If $m= l\cdot
2^{3\cdot l}$, then let $\tilde D:=D$, and let
$|x_k\rangle:=|u_k\rangle$. Otherwise, compute
$|u_k\rangle\otimes|\phi_i\rangle$ for every $k\in\{1,\ldots,D\}$
and $i\in\{1,\ldots,2^{m-ln}\}$. The resulting set of vectors
$\left\{|x_k\rangle\right\}_{k=1}^{\tilde D}$ has cardinality
$\tilde D:=D\cdot 2^{m-ln}$.

In both cases, the resulting vectors $|x_k\rangle\in\hr_m$ will span
the support of the projector $Q_{s,\eps}^{(m)}=q_m$.
\item[8.] The set $\left\{|x_k\rangle\right\}_{k=1}^{\tilde D}$ is completed
to linearly span the whole space $\hr_m$. This will be
accomplished as follows:

Consider the sequence of vectors
\[
   (|\tilde x_1\rangle,|\tilde x_2\rangle,\ldots,|\tilde x_{\tilde D+2^m}
\rangle):=   (|x_1\rangle,|x_2\rangle,\ldots,|x_{\tilde D}\rangle,
   |\Phi_1\rangle,|\Phi_2\rangle,\ldots,|\Phi_{2^m}\rangle),
\]
where $\left\{\Phi_k\right\}_{k=1}^{2^m}$ denotes the computational
basis vectors of $\hr_m$. Find the smallest $i$ such that $|\tilde
x_i\rangle$ can be written as a linear combination of $|\tilde
x_1\rangle, |\tilde x_2\rangle, \ldots,|\tilde x_{i-1}\rangle$, and
discard it (this can still be decided exactly, since all the vectors
are given as tables of integers). Repeat this step $\tilde
D$ times until there remain only $2^m$ linearly independent vectors,
namely all the $|x_j\rangle$ and $2^m-\tilde D$ of the
$|\Phi_j\rangle$.
\item[9.] Apply the Gram-Schmidt orthonormalization
procedure to the resulting vectors, to get
an orthonormal basis $\left\{|y_k\rangle\right\}_{k=1}^{2^m}$ of $\hr_m$,
such that
the first $\tilde D$ vectors are a basis for the support of 
$Q_{s,\eps}^{(m)}=q_m$.

Since every vector $|x_j\rangle$ and $|\Phi_j\rangle$ has only
integer entries, all the resulting vectors $|y_k\rangle$ will have
only entries that are (plus or minus) the square root of some
rational number.
\end{itemize}

Up to this point, every calculation was {\em exact} without any
numerical error,
comparable to the way that well-known computer algebra systems work.
The goal of the next steps is to compute an approximate description of the
desired unitary decompression map $U$ and subsequently apply it
to the quantum state $\tilde q$.

According to Section 6 in \cite{BernsteinVazirani}, a {\sl UQTM} is
able to apply a unitary transformation $U$ on some segment of its
tape within an accuracy of $\delta$, if it is supplied with a
complex matrix $\tilde U$ as input which is within operator norm
distance $\frac \delta{2(10\sqrt d)^d}$ of $U$ (here, $d$ denotes
the size of the matrix). Thus, the next task is to compute the number
of digits $N$ that are necessary to guarantee that the output will
be within trace distance $\delta=\frac 1 k$ of $q$.
\begin{itemize}
\item[10.] Read the value of $k$ (which denotes an approximation parameter; the
larger $k$, the more accurate the output of the algorithm will be).
Due to the considerations above and the calculations below, the
necessary number of digits $N$ turns out to be
$N=1+\lceil \log(2k 2^m(10\sqrt {2^m})^{2^m})\rceil$.
Compute this number.

Afterwards, compute the components of all the vectors
$\left\{|y_k\rangle\right\}_{k=1}^{2^m}$ up to $N$ binary digits of
accuracy. (This involves only calculation of the square root of
rational numbers, which can easily be done to any desired accuracy.)

Call the resulting numerically approximated vectors $|\tilde
y_k\rangle$. Write them as columns into an array (a matrix) $\tilde
U:=\left(\tilde y_1,\tilde y_2,\ldots,\tilde y_{2^m}\right)$.

Let $U:=\left(y_1,y_2,\ldots,y_{2^m}\right)$ denote the unitary
matrix with the exact vectors $|y_k\rangle$ as columns. Since $N$
binary digits give an accuracy of $2^{-N}$, it follows that
\[
   \left\vert\tilde U_{i,j}-U_{i,j}\right\vert < 2^{-N} <
   \frac{1/k}{2\cdot 2^m(10\sqrt {2^m})^{2^m}}\,\,.
\]
If two $2^m\times 2^m$-matrices $U$ and $\tilde U$ are $\eps$-close
in their entries, they also must be $2^m\cdot\eps$-close in norm, so
we get
\[
   \|\tilde U-U\|<\frac{1/k}{2(10\sqrt {2^m})^{2^m}}\,\,.
\]
\end{itemize}
So far, every step was purely classical and could have been done on
a classical computer.
Now, the quantum part begins: $\tilde q$ will be touched for the first time.
\begin{itemize}
\item[11.] Compute $\lceil mr\rceil$, which gives the length $\ell(\tilde q)$.
Afterwards, move $\tilde q$ to some free space on the input tape,
and append zeroes, i.e. create the state
\[
   q'\equiv|\psi_0\rangle\langle\psi_0|:=\left(|0\rangle\langle
   0|\right)^{\otimes(m-\ell(\tilde q))}\otimes\tilde q
\]
on some segment of $m$ cells on the input tape.
\item[12.]
Approximately apply the unitary transformation $U$ on the tape segment
that contains the state $q'$.

The machine cannot apply $U$ exactly (since it only knows an
approximation $\tilde U$), and it also cannot apply $\tilde U$
directly (since $\tilde U$ is only approximately unitary, and the
machine can only do unitary transformations). Instead, it will
effectively apply another unitary transformation $V$ which is close
to $\tilde U$ and thus close to $U$, such that
\[
\| V-U \|<\frac 1 k\,\,.
\]

Let $|\psi\rangle:=U|\psi_0\rangle$ be the output that we want to have, and let
$|\phi\rangle:=V|\psi_0\rangle$ be the approximation that is really computed
by the machine. Then,
\[
\|\, |\phi\rangle-|\psi\rangle\| <\frac 1 k\,\,.
\]
A simple calculation proves that the trace distance must then also be small:
\[
\||\phi\rangle\langle\phi|-|\psi\rangle\langle\psi|\|_{\rm Tr}<\frac 1 k\,\,.
\]

\item[14.] Move $q:=|\phi\rangle\langle\phi|$ to the output tape and halt.
\end{itemize}
\subsubsection{Proof of Proposition~\ref{PropUpperBound}}
We have to give a precise definition how the parameters $(m,r,\tilde
q)$ are encoded into a single qubit string $\sigma$. (According to
the definition of $QC$, the parameter $k$ is not a part
of $\sigma$, but is given as a second parameter. See 
Definitions \ref{DefEncoding} and \ref{DefQComplexity} for details.)

We choose to encode $m$ by giving $\lfloor\log m\rfloor$ 1's, followed
by one 0, followed by the  $\lfloor\log m\rfloor +1$ binary digits of $m$.
Let $|M\rangle\langle M|$ denote the corresponding projector in the
computational basis.

The parameter $r$ can be encoded in any way, since it does not
depend on $m$. The only constraint is that the description must be
self-delimiting, i.e. it must be clear and decidable at what
position the description for $r$ starts and ends. The descriptions
will also be given by a computational basis vector (or rather the
corresponding projector) $|R\rangle\langle R|$.

The descriptions are then stuck together, and the input 
$\sigma(\tilde q)$ is given by
\[
   \sigma(\tilde q):=|M\rangle\langle M|\otimes|R\rangle\langle R|\otimes
   \tilde q\,\,.
\]
If $m$ is large enough such that (\ref{eqLogTrQm}) is fulfilled, it
follows that
$\ell(\sigma(\tilde q))=2\lfloor\log m\rfloor + 2 + c + \lceil mr\rceil$,
where $c\in\N$ is some constant which depends on $r$, but not on $m$.

It is clear that this qubit string can be fed into the reference
{\sl UQTM} $\mathfrak{U}$ together with a description of the
algorithm $\mathfrak{A}$ of fixed length $c'$ which depends on $r$, 
but not on $m$. This will give a
qubit string $\sigma_{\mathfrak U}(\tilde q)$ of length
\begin{eqnarray}
   \ell(\sigma_{\mathfrak U}(\tilde q))&=&2\lfloor\log m\rfloor
   +2+c+\lceil mr\rceil+c'\nonumber\\
   &\leq& 2 \log m + m\left(s+\frac 1 2 \delta\right)+c''\,\,,
   \label{eqLength}
\end{eqnarray}
where $c''$ is again a constant which depends on $r$, but not on $m$.
Recall the matrix $U$ constructed in step 11 of our algorithm
$\mathfrak A$, which rotates (decompresses) a compressed
(short) qubit string $\tilde q$ back into the typical subspace.
Conversely, for every one-dimensional projector $q\leq q_m$, where
$q_m=Q_{s,\eps}^{(m)}$ was defined in (\ref{eqTypicalProjector}), let
$\tilde q\in\hr_{\lceil mr\rceil}$ be the projector given by
$\left(|0\rangle\langle 0|\right)^{\otimes(m-\lceil
mr\rceil)}\otimes \tilde q =U^* q U$. Then, since $\mathfrak A$ has
been constructed such that
\[
   \|\mathfrak{U}(k,\sigma_{\mathfrak U}(\tilde q))-q\|_{\rm Tr}
   < \frac 1 k\qquad\mbox{ for every }k\in\N\,\,,
\]
it follows from (\ref{eqLength}) that
\[
   \frac 1 m QC(q)\leq 2 \frac{\log m} m + s+
   \frac 1 2 \delta +\frac{c''}m\,\,.
\]
If $m$ is large enough, Equation~(\ref{UpperZero}) follows.

Now we continue by proving Equation~(\ref{UpperDelta}).
Let $k:=\lceil\frac 1 {2\delta}\rceil$. Then, we have
for every one-dimensional projector $q\leq q_m$ and $m$ large enough
\begin{eqnarray}
   \frac 1 m QC^{2\delta}(q)&\leq& \frac 1 m QC^{1/k}(q)
   \leq \frac 1 m QC^(q)
   +\frac{2\lfloor\log k\rfloor+2} m\nonumber\\
   &<&s+\delta  +\frac{2\log k +2} m<s+2\delta\,\,,
   \label{eqOmegaEquation}
\end{eqnarray}
where the first inequality follows from the obvious monotonicity property
$\delta\geq \eps \Rightarrow QC^\delta\leq QC^\eps$, the second one is
by Lemma~\ref{LemRelation},
and the third estimate is due to Equation~(\ref{UpperZero}).
\qed
{\bf Proof of the Main Theorem~\ref{TheQBrudno}.} Let $\tilde
q_m(\delta)$ be the $\Psi$-typical projector sequence given in 
Proposition~\ref{PropUpperBound}, i.e.
the complexities $\frac 1 m QC$ and $\frac 1 m QC^\delta$ of every
one-dimensional projector $q\leq \tilde q_m(\delta)$ are upper bounded 
by $s+\delta$.
Due to Corollary~\ref{Cor1}, there exists another sequence of
$\Psi$-typical projectors
$p_m(\delta)\leq \tilde q_m(\delta)$ such that additionally, 
$\frac 1 m QC^\delta(q)>s-\delta(4+\delta)s$
is satisfied for $q\leq p_m(\delta)$. From Corollary~\ref{Cor2}, 
we can further deduce that there is another sequence of 
$\Psi$-typical projectors $q_m(\delta)\leq p_m(\delta)$
such that also $\frac 1 m QC(q)>s-\delta$ holds. Finally, the optimality assertion
is a direct consequence of Lemma~\ref{TheQCounting},
combined with Theorem~\ref{QAEP}. \qed

\chapter{Summary and Outlook}
\label{ChapterSummary}
\vskip -0.5cm
In this thesis, we have formally defined quantum Kolmogorov complexity, based on work by Berthiaume et al. \cite{Berthiaume},
and have given rigorous mathematical proofs of its basic properties. In particular, we have shown
that the quantum Kolmogorov complexity notions $QC$ and $QC^\delta$ are invariant, that they coincide with classical complexity for classical strings,
they have incompressi\-bility properties, and the corresponding quantum Kolmogorov complexity rates agree with the von Neumann
entropy rate for ergodic quantum information sources.

The most complicated step to achieve these results was to give a rigorous formal proof that there
exists a universal quantum Turing machine (QTM) $\mathfrak U$ in the following sense: that QTM $\mathfrak U$ can simulate
every other QTM for an arbitrary number of time steps, without knowing the running time in advance,
and then halt itself with probability one. The question whether this is possible has been ignored in
previous work on quantum Kolmogorov complexity, but it is necessary to prove the invariance property,
i.e. the feature that quantum Kolmogorov complexity depends on the choice of the universal quantum computer
only up to an additive constant.

We also discussed the question how the halting of a QTM can be defined. We argued that for the
purpose of studying quantum Kolmogorov complexi\-ty, the most useful and natural definition
is to demand perfect halting. To show that this definition is not as restrictive as one might first suppose,
we proved that every input that makes a QTM halt approximately can be enhanced by at most a constant
number of qubits to make the universal QTM halt entirely.

Furthermore, we have defined the average-length complexities $\qka$ and $\qka^\delta$ and studied them
to some extent. Because of
Lemma~\ref{LemLogEllAverage} and the proof idea of Conjecture~\ref{TheAvLengthUQTM}, we think that
these complexities are closely related to {\em prefix QTMs}, which we have defined in Definition~\ref{DefPrefixQTM}.
Studying prefix QTMs may also be interesting for another reason:
it may give an alternative approach to Tadaki's definition \cite{Tadaki} of the quantum halting probability,
and it may help to clarify the relation of the complexity notions $\qka$ or $\qka^\delta$ to the
universal density matrix approach by G\'acs. This speculation is supported by the fact that
classical prefix complexity is related to universal probability and Chaitin's halting probability by Levin's theorem \cite{Vitanyibook}.

Classical Kolmogorov complexity has a large variety of applications in different fields of mathematics
and computer science. Hence it may be worthwhile to look for applications of quantum Kolmogorov complexity.
A very promising field for application is quantum statistical mechanics, since classical
Kolmogorov complexity has already turned out to be useful in the classical version of that theory.

A concrete proposal for an application of quantum Kolmogorov complexity is to analyze a quantum
version of the thought experiment of Maxwell's demon. In one of the versions of this thought experiment, some microscopic
device tries to decrease the entropy of some gas in a box, without the expense of energy, by
intelligently opening or closing some little door that separates both halves of the box.

It is clear that a device like this cannot work as described, since its existence would violate
the second law of thermodynamics. But then, the question is what prevents such a little device
(or ``demon'') from operating. Roughly, the answer is that the demon has to make observations
to decide whether to close or open the door, and these observations accumulate information.
From time to time, the demon must erase this additional information, which is only possible
at the expense of energy, due to Landauer's principle.

In \cite{Vitanyibook}, this cost of energy is analyzed under very weak assumptions with the help
of Kolmogorov complexity. Basically, the energy that the demon can extract from the gas is
limited by the difference of the entropy of the gas, {\em plus} the difference of the Kolmogorov complexity of the
demon's memory before and after the demon's actions.
The power of this
analysis is that it even encloses the case that the demon has a computer to do clever calculations,
e.g. to compress the accumulated information before erasing it.

It seems that quantum Kolmogorov complexity might have all the properties needed to extend this
analysis to the quantum case. Yet, the average-length complexities $\qka$ or $\qka^\delta$
are probably more useful in this case than $QC$ or $QC^\delta$, since they resemble more closely the fact
that the {\em expectation value} of the amount of information that has to be erased is physically important,
not the maximal size of the system.

To conclude, we found that quantum Kolmogorov complexity is a beautiful concept with a promising
potential for new applications. Applications aside, quantum Kolmogorov complexity offers the opportunity
to deepen our understanding of the theoretical aspects of quantum computation and is
interesting as a subject in its own right.

\setcounter{section}{0}
\setcounter{theorem}{0}
\renewcommand{\thetheorem}{\thechapter.\arabic{theorem}}
\setcounter{equation}{0}
\renewcommand{\theequation}{\thechapter.\arabic{equation}}
\setcounter{figure}{0}
\setcounter{table}{0}

\appendix
\chapter{Appendix}
\renewcommand{\thechapter}{A}

The following lemma is due M. B. Ruskai (\cite{Ruskai}) and can also be found in
\cite{NielsenChuang} for the finite-dimensional case.

\begin{flemma}[Quantum Operations are Contractive]
\label{LemContractivity}
\lineclear
Let $\hr$ and $\hr'$ be Hilbert spaces, and let
$\mathcal{E}:\mathcal{T}(\hr)\to\mathcal{T}(\hr')$ be linear,
positive and trace-preserving.
If $A=A^*\in\mathcal{T}(\hr)$, then
\[
   \|\mathcal{E}(A)\|_{\rm Tr}\leq \|A\|_{\rm Tr}.
\]
\end{flemma}
\proof
If $P\geq 0$ is any positive trace-class operator on $\hr$, then
\[
   \|P\|_{\rm Tr}=\frac 1 2 {\rm Tr}|P|=\frac 1 2 {\rm Tr} P.
\]
Since every self-adjoint operator $A$ can be written as $A=A_+-A_-$, where $A_+$ and
$A_-$ are positive operators, we get
\begin{eqnarray*}
\|\mathcal{E}(A)\|_{\rm Tr}&=&\|\mathcal{E}(A_+-A_-)\|_{\rm Tr}
\leq \|\mathcal{E}(A_+)\|_{\rm Tr}+\|\mathcal{E}(A_-)\|_{\rm Tr}\\
&=&\frac 1 2 {\rm Tr}\mathcal{E}(A_+)+\frac 1 2 {\rm Tr}\mathcal{E}(A_-)
=\frac 1 2 {\rm Tr} A_+ +\frac 1 2 {\rm Tr}A_-\\
&=&\frac 1 2{\rm Tr}\left(A_+ + A_-\right)=\frac 1 2 {\rm Tr}|A|=\|A\|_{\rm Tr}.\qquad\qquad\qquad\qquad\qquad\mbox{\qed}
\end{eqnarray*}

\begin{flemma}[Inner Product and Dimension Bound]
\label{LemInnerProduct}
\lineclear
Let $\hr$ be a Hilbert space, and let $|\psi_1\rangle,\ldots,|\psi_N\rangle\in\hr$ with
$\|\,|\psi_i\rangle\|=1$ for every $i\in\{1,\ldots,N\}$, where $2\leq N\in\N$. Suppose that
\[
   \left\vert\strut\langle \psi_i|\psi_j\rangle\right\vert<\frac 1 {N-1}\qquad\mbox{for every }i\neq j\,\,.
\]
Then, $\displaystyle \dim\hr\geq N$. In particular, the vectors $\{|\psi_i\rangle\}_{i=1}^N$
are linearly independent.
\end{flemma}
{\bf Proof.} We prove the statement by induction in $N\in\N$. For $N=2$, the statement
of the theorem is trivial. Suppose the claim holds for some $N\geq 2$, then consider
$N+1$ vectors $|\psi_1\rangle,\ldots,|\psi_{N+1}\rangle\in\hr$, where $\hr$ is an arbitrary
Hilbert space.
Suppose that $|\langle \psi_i|\psi_j\rangle|<\frac 1 N$ for every $i\neq j$.
Let $P:=\idn-|\psi_{N+1}\rangle\langle\psi_{N+1}|$, then $P|\psi_i\rangle\neq 0$ for every
$i\in\{1,\ldots,N\}$, and let
\[
   |\varphi_i'\rangle:=P |\psi_i\rangle\,\,,\qquad
   |\varphi_i\rangle:=\frac{|\varphi_i'\rangle}{\|\,|\varphi_i'\rangle\|}\,\,.
\]
The $|\varphi_i\rangle$ are normalized vectors in the Hilbert
subspace $\tilde\hr:={\rm ran}(P)$ of $\hr$. Since $\|\,|\varphi_i'\rangle\|^2
=\langle \psi_i|\psi_i\rangle-|\langle\psi_i|\psi_{N+1}\rangle|^2
   >1-\frac 1 {N^2}$,
it follows that the vectors $|\varphi_i\rangle$ have small inner product: Let $i\neq j$, then
\begin{eqnarray*}
   |\langle \varphi_i | \varphi_j\rangle|&=&\frac 1 {\|\,|\varphi_i'\rangle\|\cdot\|\,|\varphi_j'\rangle\|}
   |\langle \varphi_i'|\varphi_j'\rangle|\\
   &<& \frac 1 {\sqrt{1-\frac 1 {N^2}}\sqrt{1-\frac 1 {N^2}}}\left\vert
      \langle \psi_i|\psi_j\rangle- \langle\psi_{N+1}|\psi_j\rangle\langle\psi_i|\psi_{N+1}\rangle
   \right\vert\\
   &<& \frac 1 {1-\frac 1 {N^2}} \left(\frac 1 N + \frac 1 {N^2}\right) = \frac 1 {N-1}\,\,.
\end{eqnarray*}
Thus, $\dim\tilde\hr\geq N$, and so $\dim\hr\geq N+1$.\qed

\begin{flemma}[Composition of Unitary Operations]
\label{LemCompositionUnitary}
\lineclear
Let $\hr$ be a finite-dimensional Hilbert space, let $(V_i)_{i\in\N}$ be a sequence
of linear subspaces of $\hr$ (which have all the same dimension), and let $U_i:V_i\to V_{i+1}$
be a sequence
of unitary operators on $\hr$ such that $\sum_{k=1}^\infty \| U_k-\idn\|$ exists.
Then, the product $\prod_{k=1}^\infty U_k=\ldots\cdot U_3\cdot U_2\cdot U_1$ converges
in operator norm to an isometry $U: V_1\to\hr$.
\end{flemma}
{\bf Proof.} We first show by induction that $\left\| \prod_{k=1}^N U_k - \idn\right\|
\leq \sum_{k=1}^N \| U_k-\idn\|$. This is trivially true for $N=1$; suppose it is true
for $N$ factors, then
\begin{eqnarray*}
   \left\| \prod_{k=1}^{N+1} U_k - \idn\right\|&\leq&\left\|
    \prod_{k=1}^{N+1} U_k-\prod_{k=1}^N U_k\right\| +
    \left\| \prod_{k=1}^N U_k-\idn\right\|\\
    &\leq& \left\|(U_{N+1}-\idn)\prod_{k=1}^N U_k\right\|
    +\sum_{k=1}^N \| U_k-\idn\|
    \leq\sum_{k=1}^{N+1} \| U_k-\idn\|\,\,.
\end{eqnarray*}

By assumption, the sequence $a_n:=\sum_{k=1}^n\| U_k-\idn\|$ is a Cauchy sequence; hence, for every
$\eps>0$ there is an $N_\eps\in\N$ such that for every $L,N\geq N_\eps$ it holds that
$\sum_{k=L+1}^N \| U_k-\idn\|<\eps$. Consider now the sequence $V_n:=\prod_{k=1}^n U_k$.
If $N\geq L\geq N_\eps$, then
\begin{eqnarray*}
   \| V_N-V_L\| &=& \left\| \prod_{k=L+1}^N U_k \cdot \prod_{k=1}^L U_k - \prod_{k=1}^L U_k\right\|
   \leq\left\| \prod_{k=L+1}^N U_k - \idn\right\| \cdot \left\| \prod_{k=1}^L U_k\right\|\\
   &\leq& \sum_{k=L+1}^N \| U_k-\idn\| <\eps\,\,,
\end{eqnarray*}
so $(V_n)_{n\in\N}$ is also a Cauchy sequence and converges in operator norm to
some linear operator $U$ on $V_1$. It is easily checked that $U$ must be isometric.\qed

\begin{flemma}[Norm Inequalities]
\label{LemNormInequalities}
Let $\hr$ be a finite-dimensional Hilbert space, and
let $|\psi\rangle,|\varphi\rangle\in\hr$ with $\|\,|\psi\rangle\|=\|\,|\varphi\rangle\|=1$.
Then,
\[
   \|\,|\psi\rangle\langle\psi|-|\varphi\rangle\langle\varphi|\,\|_{\rm{Tr}}
   \leq \|\,|\psi\rangle-|\varphi\rangle\|\,\,.
\]
Moreover, if $\rho,\sigma\in\mathcal{T}_1^+(\hr)$ are density operators, then
\[
   \|\rho-\sigma\|\leq \|\rho-\sigma\|_{\rm Tr}\,\,.
\]
\end{flemma}
{\bf Proof.} Let $\Delta:=|\psi\rangle\langle\psi|-|\varphi\rangle\langle\varphi|$.
Using \cite[9.99]{NielsenChuang},
\begin{eqnarray*}
   \|\Delta\|_{\rm Tr}^2&=&1-|\langle\psi|\varphi\rangle|^2
   =\left(\strut 1-|\langle\psi|\varphi\rangle|\right)
   \underbrace{\left(\strut 1+|\langle\psi|\varphi\rangle|\right)}_{\leq 2}\\
   &\leq&2-2|\langle\psi|\varphi\rangle|\leq 2-2{\rm Re}\langle\psi|\varphi\rangle
   =\langle\psi-\varphi|\psi-\varphi\rangle=\|\,|\psi\rangle-|\varphi\rangle\|^2\,\,.
\end{eqnarray*}

Let now $\tilde\Delta:=\rho-\sigma$, then
$\tilde\Delta$ is Hermitian. We may assume that one of its eigenvalues which has largest absolute
value is positive (otherwise interchange $\rho$ and $\sigma$), thus
\begin{eqnarray*}
   \|\tilde\Delta\|&=&
   \max_{\|\,|v\rangle\|=1}\langle v|\tilde\Delta |v\rangle
   =\max_{P\mbox{ proj., } {\rm Tr } P=1}{\rm Tr}(P\tilde\Delta)
   \leq \max_{P\mbox{ proj.}} {\rm Tr}(P\tilde\Delta)=\|\tilde\Delta\|_{\rm Tr}
\end{eqnarray*}
according to \cite[9.22]{NielsenChuang}.
\qed

\begin{flemma}[Dimension Bound for Similar Subspaces]
\label{LemDimBoundSimilar}
\lineclear
Let $\hr$ be a finite-dimensional Hilbert space, and let $V,W\subset\hr$
be subspaces such that for every $|v\rangle\in V$ with $\|\,|v\rangle\|=1$
there is a vector $|w\rangle\in W$ with $\|\,|w\rangle\|=1$ which
satisfies $\|\,|v\rangle-|w\rangle\|\leq\eps$, where
$0<\eps\leq \frac 1 {4(\dim V -1)^2}$ is fixed.
Then, $\dim W\geq \dim V$. Moreover, if additionally $\eps\leq \frac 1 {36} \left(\frac 5 2\right)
^{2-2\dim V}$ holds, then
there exists an isometry $U:V\to W$ such that
$\|U-\idn\|<\frac 8 3 \sqrt{\eps}\left(\frac 5 2\right)^{\dim V}$.
\end{flemma}
{\bf Proof.} Let $\{|v_1\rangle,\ldots,|v_d\rangle\}$ be an orthonormal basis of $V$.
By assumption, there are normalized vectors $\{|w_1\rangle,\ldots,|w_d\rangle\}\subset W$ with
$\|\,|v_i\rangle-|w_i\rangle\|\leq\eps$ for every $i$.
From the definition of the trace distance for pure states (see \cite[(9.99)]{NielsenChuang}
together with Lemma~\ref{LemNormInequalities},
it follows for every $i\neq j$
\begin{eqnarray*}
   \sqrt{1-|\langle w_i|w_j\rangle|^2}&=& \|\,|w_i\rangle\langle w_i|-|w_j\rangle\langle w_j|\,\|_{\rm Tr}\\
   &\geq& \|\,|v_i\rangle\langle v_i|-|v_j\rangle\langle v_j|\,\|_{\rm Tr}
    -\|\,|v_i\rangle\langle v_i|-|w_i\rangle\langle w_i|\,\|_{\rm Tr}\\
   && -\|\,|v_j\rangle\langle v_j|-|w_j\rangle\langle w_j|\,\|_{\rm Tr}\\
   &\geq&1-\|\,|v_i\rangle-|w_i\rangle\|-\|\,|v_j\rangle-|w_j\rangle\|\\
   &\geq& 1-2\eps\,\,.
\end{eqnarray*}
Thus, $|\langle w_i|w_j\rangle| < 2 \sqrt\eps\leq \frac 1 {d-1}$, and it follows
from Lemma~\ref{LemInnerProduct} that $\dim W\geq d$. Now apply the Gram-Schmidt
orthonormalization procedure to the vectors $\{|w_i\rangle\}_{i=1}^d$:
\begin{eqnarray*}
   |\tilde e_k\rangle &:=& |w_k\rangle - \sum_{i=1}^{k-1} \langle w_k|e_i\rangle |e_i\rangle\,\,,\qquad
   |e_k\rangle := \frac{|\tilde e_k\rangle}{\|\,|\tilde e_k\rangle\|}\,\,.
\end{eqnarray*}
Use $\left|\strut \|\,|\tilde e_k\rangle\|-1\right|=\left|\strut \|\,|\tilde e_k\rangle\|
-\|\,|w_k\rangle\|\right|\leq \|\,|\tilde e_k\rangle-|w_k\rangle\|$ and calculate
\begin{eqnarray*}
   \|\,|\tilde e_k\rangle-|w_k\rangle\|&=&\left\| \sum_{i=1}^{k-1} \frac{\langle w_k|\tilde e_i\rangle
   |\tilde e_i\rangle}{\|\,|\tilde e_i\rangle\|^2}\right\|
   \leq \sum_{i=1}^{k-1}\frac{\left| \langle w_k|\tilde e_i-w_i\rangle\right| + \left|
   \langle w_k|w_i\rangle\right|}{\|\,|\tilde e_i\rangle\|}\\
   &\leq& \sum_{i=1}^{k-1} \frac{\|\,|\tilde e_i\rangle-|w_i\rangle\|+2\sqrt{\eps}}
   {1-\|\,|\tilde e_i\rangle-|w_i\rangle\|}\,\,.
\end{eqnarray*}
Let $\Delta_k:=\|\,|\tilde e_k\rangle-|w_k\rangle\|$ for every $1\leq k \leq d$. We will now
show by induction that $\Delta_k\leq 2 \sqrt{\eps}\left[ \frac 2 5\left(\frac 5 2\right)^k-1\right]$.
This is trivially true for $k=1$, since $\Delta_1=0$. Suppose it is true for every $1\leq i \leq k-1$,
then in particular, $\Delta_i\leq \frac 1 3$ by the assumptions on $\eps$ given in the statement of
this lemma, and
\begin{eqnarray*}
   \Delta_k&\leq& \sum_{i=1}^{k-1}\frac{\Delta_i+2\sqrt{\eps}}{1-\Delta_i}
   \leq  \frac 3 2 \sum_{i=1}^{k-1} \left(
      2\sqrt{\eps}\left[\frac 2 5\left(\frac 5 2 \right)^i -1\right]+2\sqrt{\eps}
   \right)\\
   &=&2 \sqrt{\eps}\left[\frac 2 5 \left(\frac 5 2\right)^k -1\right].
\end{eqnarray*}
Thus, it holds that
\begin{eqnarray*}
   \|\,|e_k\rangle-|v_k\rangle\|&\leq&\|\,|e_k\rangle-|\tilde e_k\rangle\|
   +\|\,|\tilde e_k\rangle-|w_k\rangle\|+\|\,|w_k\rangle-|v_k\rangle\|\\
   &\leq& 2\|\,|\tilde e_k\rangle-|w_k\rangle\|+\eps
   \leq 4\sqrt{\eps}\left[\frac 2 5 \left(\frac 5 2\right)^k -1\right]+\eps.
\end{eqnarray*}
Now define the linear operator $U:V\to W$ via linear extension of
$U|v_i\rangle:=|e_i\rangle$ for $1\leq i\leq d$. This map is an isometry, since it
maps an orthonormal basis onto an orthonormal basis of same dimension.
By substituting $|v\rangle=\sum_{k=1}^d \alpha_k |v_k\rangle$ and using $\eps<4\sqrt{\eps}$
and the geometric series, it easily follows that $\|\,U|v\rangle-|v\rangle\|\leq \frac 8 3
\sqrt{\eps}\left(\frac 5 2 \right)^d$ if $\|\,|v\rangle\|=1$.\qed

\begin{flemma}[Stability of the Control State]
\label{LemStability}
\lineclear
If $|\psi\rangle,|\varphi\rangle,|v\rangle\in\cn$ and $\|\,|\psi\rangle\|=\|\,|\varphi\rangle\|=1$
and $|v\rangle\neq 0$, then
it holds for every QTM $M$ and every $t\in\N_0$
\begin{eqnarray*}
   \left| \langle q_f | M_{\mathbf{C}}^t(|\psi\rangle\langle\psi|)|q_f\rangle-
   \langle q_f | M_{\mathbf{C}}^t(|\varphi\rangle\langle\varphi|)|q_f\rangle\right|
   &\leq&\left\|\strut\, |\psi\rangle\langle\psi|-|\varphi\rangle\langle\varphi|\,\right\|_{\rm Tr},\\
   \left|\strut \langle q_f|M_{\mathbf{C}}^{t}(|v\rangle\langle v|)|q_f\rangle-
   \langle q_f|M_{\mathbf{C}}^{t}(|v^0\rangle\langle v^0|)|q_f\rangle\right|
   &\leq&
   \left|\strut 1-\|\,|v\rangle\|^2\right|.
\end{eqnarray*}
\end{flemma}
{\bf Proof.} Using the Cauchy-Schwarz inequality, Lemma~\ref{LemNormInequalities} and
the contractivity of quantum operations
with respect to the trace distance (Lemma~\ref{LemContractivity}), we get
the chain of inequalities
\begin{eqnarray*}
   \Delta_t&:=&
   \left| \langle q_f | M_{\mathbf{C}}^t(|\psi\rangle\langle\psi|)|q_f\rangle-
   \langle q_f | M_{\mathbf{C}}^t(|\varphi\rangle\langle\varphi|)|q_f\rangle\right|\\
   &\leq& \left\|\,M_{\mathbf{C}}^t\left(\strut|\psi\rangle\langle\psi|\right)
   -M_{\mathbf{C}}^t\left(\strut|\varphi\rangle\langle\varphi|\right)
   \right\|\\
   &\leq& \left\|\,M_{\mathbf{C}}^t\left(\strut|\psi\rangle\langle\psi|\right)
   -M_{\mathbf{C}}^t\left(\strut|\varphi\rangle\langle\varphi|\right)
   \right\|_{\rm Tr}\\
   &\leq&\left\|\strut |\psi\rangle\langle\psi|-|\varphi\rangle\langle\varphi|\right\|_{\rm Tr}\,\,.
\end{eqnarray*}
The second inequality can be proved by an analogous calculation.
\qed

\backmatter
\chapter{Glossary of Symbols and Notation}
\begin{tabular}{|l|p{9.0cm}|c|}
\hline
{\bf \Large Notation} & {\bf \Large Meaning} & {\bf\Large Page} \\ \hline
$\mathcal{B}(\hr)$ & the set of bounded linear operators on some Hilbert space $\hr$ & \pageref{SubsecCompDecomp} \\ \hline
$\delta_{tT}$ & the Kronecker symbol:
$\delta_{tT}:=\left\{
   \begin{array}{cl}
      1 & \mbox{if }t=T,\\
      0 & \mbox{if }t\neq T.
   \end{array}
\right.$ & \pageref{EinfachSo} \\ \hline
${\rm dom}\, M$ & the domain of definition of the map (e.g. QTM) $M$ & \pageref{DefQTM} \\ \hline
$\hr_\s$ & the qubit string Hilbert space $\bigoplus_{n\in \N_0} \hr_n$ & \pageref{DefHrS} \\ \hline
$\hr_M^{(n)}(t)$ & the halting space of the QTM $M$ for time $t$ and inputs of length $n$ & \pageref{DefHaltingQubitStrings} \\ \hline
$\hr_M^{(n,\eps)}(t)$ & the approximate halting space of accuracy $\eps$ of the QTM $M$
for time $t$ and inputs of length $n$ & \pageref{DefApproxHalt} \\ \hline
$\hr_n$ & $\hr_n=\left(\C^2\right)^{\otimes n}$ with some fixed computational basis & \pageref{HK} \\ \hline
$\ell(\cdot)$ & the length of some classical finite binary string, or the base length of
some qubit string & \pageref{DefLength} \\ \hline
$\bar \ell(\rho)^{\strut}$ & the average length of some qubit string $\rho$, given by ${\rm Tr(\Lambda\rho)}$,
where $\Lambda$ is the length operator & \pageref{DefLength} \\ \hline
$M_{\mathbf C}^t(\sigma)$ & the state of the control of the QTM $M$ at time $t$, if the
input was the qubit string $\sigma$ & \pageref{MC} \\ \hline
$M_{\mathbf O}^t(\sigma)$ & the state of the output tape of the QTM $M$ at time $t$, if the
input was the qubit string $\sigma$ & \pageref{MC}\\  \hline
QTM & quantum Turing machine & \pageref{AbbrQTM} \\ \hline
$\mathcal{R}$ & ``Reading operation'': if $\sigma$ is the state of a QTM's output tape, then $\mathcal{R}(\sigma)$ is
the corresponding qubit string. & \pageref{DefReadingOp} \\ \hline
${\rm ran}\, U$ & the range of some map $U$ & \pageref{rmran} \\ \hline
$\sigma_1^n$ & the restriction of the qubit string $\sigma$ to its first $n$ qubits & \pageref{DefPrefixQTM} \\ \hline
$\mathcal{T}(\hr)$ & the set of trace-class operators on a Hilbert space $\hr$ & \pageref{DefTraceClassOp}\\ \hline
$\mathcal{T}_1^+(\hr)$ & the set of density operators, i.e. positive trace-class operators of trace $1$,
on some Hilbert space $\hr$ & \pageref{DefTraceClassOp}\\ \hline
TM & Turing machine & \pageref{AbbrQTM} \\ \hline
${\rm Tr}(A)$ & the trace of the operator $A$, if $A$ is a trace-class operator on some Hilbert space & \pageref{EqTrTrTr} \\ \hline
${\rm Tr}_{\mathbf{C}}(\rho)$ & the partial trace over the part $\mathbf C$ of the whole Hilbert space
(normally, $\mathbf C$ denotes a QTM's control) & \pageref{MC} \\
\hline
\end{tabular}

\end{document}